\documentclass[sigconf, screen, authorversion]{acmart}

\AtBeginDocument{%
  \providecommand\BibTeX{{%
    \normalfont B\kern-0.5em{\scshape i\kern-0.25em b}\kern-0.8em\TeX}}}

\copyrightyear{2024}
\acmYear{2024}
\setcopyright{acmlicensed}\acmConference[CHI '24]{Proceedings of the CHI Conference on Human Factors in Computing Systems}{May 11--16, 2024}{Honolulu, HI, USA}
\acmBooktitle{Proceedings of the CHI Conference on Human Factors in Computing Systems (CHI '24), May 11--16, 2024, Honolulu, HI, USA}
\acmDOI{10.1145/3613904.3642726}
\acmISBN{979-8-4007-0330-0/24/05}

\usepackage[most]{tcolorbox}
\usepackage{xspace}
\usepackage{multicol}
\usepackage{subcaption}
\usepackage{multirow}
\usepackage{longtable}
\usepackage{balance}

\usepackage{acmart-taps}

\usepackage{array}

\newcolumntype{P}[1]{>{\centering\arraybackslash}p{#1}}
\newcolumntype{H}{>{\setbox0=\hbox\bgroup}c<{\egroup}@{}}

\newcommand{\eg}{\textit{e.g.}}
\newcommand{\ie}{\textit{i.e.}}
\newcommand{\etal}{\textit{et~al.}}
\newcommand{\etc}{\textit{etc}}

\newcommand{\revisecr}[1]{\textcolor{black}{#1}}
\newcommand{\review}[1]{\textcolor{brown}{[#1]}\xspace}
\renewcommand{\review}[1]{}

\definecolor{mybrown}{HTML}{EEE0DA}
\definecolor{myred}{RGB}{255, 226, 221}
\definecolor{myorange}{RGB}{250, 222, 201}
\definecolor{mygreen}{RGB}{219, 237, 219}
\definecolor{mypink}{RGB}{245, 224, 233}
\definecolor{mypurple}{RGB}{232, 222, 238}
\definecolor{myblue}{RGB}{211, 229, 239}
\definecolor{myyellow}{RGB}{253, 236, 200}
\definecolor{mygrey}{RGB}{226, 226, 226}
\definecolor{mylightgrey}{RGB}{240, 240, 240}
\definecolor{navy}{RGB}{52, 85, 139}

\newcommand{\humanc}{\tcbox[on line, colback=myred, colframe=white, boxsep=0pt, left=2pt,right=2pt,top=2pt,bottom=3pt,arc=2pt,boxrule=0pt]{human-creator}\xspace}
\newcommand{\humancs}{\tcbox[on line, colback=myred, colframe=white, boxsep=0pt, left=2pt,right=2pt,top=2pt,bottom=3pt,arc=2pt,boxrule=0pt]{human-creators}\xspace}
\newcommand{\Humanc}{\tcbox[on line, colback=myred, colframe=white, boxsep=0pt, left=2pt,right=2pt,top=2pt,bottom=3pt,arc=2pt,boxrule=0pt]{Human-creator}\xspace}

\newcommand{\humancnb}{{human-creator}\xspace}
\newcommand{\humancsnb}{{human-creators}\xspace}

\newcommand{\humano}{\tcbox[on line, colback=myorange, colframe=white, boxsep=0pt, left=2pt,right=2pt,top=2pt,bottom=1pt,arc=2pt,boxrule=0pt]{human-optimizer}\xspace}
\newcommand{\humanos}{\tcbox[on line, colback=myorange, colframe=white, boxsep=0pt, left=2pt,right=2pt,top=2pt,bottom=1pt,arc=2pt,boxrule=0pt]{human-optimizers}\xspace}
\newcommand{\Humano}{\tcbox[on line, colback=myorange, colframe=white, boxsep=0pt, left=2pt,right=2pt,top=2pt,bottom=1pt,arc=2pt,boxrule=0pt]{Human-optimizer}\xspace}

\newcommand{\humanonb}{{human-optimizer}\xspace}
\newcommand{\humanosnb}{{human-optimizers}\xspace}

\newcommand{\humana}{\tcbox[on line, colback=myyellow, colframe=white, boxsep=0pt, left=2pt,right=2pt,top=2pt,bottom=3pt,arc=2pt,boxrule=0pt]{human-assistant}\xspace}
\newcommand{\humanas}{\tcbox[on line, colback=myyellow, colframe=white, boxsep=0pt, left=2pt,right=2pt,top=2pt,bottom=3pt,arc=2pt,boxrule=0pt]{human-assistants}\xspace}
\newcommand{\Humana}{\tcbox[on line, colback=myyellow, colframe=white, boxsep=0pt, left=2pt,right=2pt,top=2pt,bottom=3pt,arc=2pt,boxrule=0pt]{Human-assistant}\xspace}

\newcommand{\humanasnb}{{human-assistants}\xspace}

\newcommand{\humanr}{\tcbox[on line, colback=mypink, colframe=white, boxsep=0pt, left=2pt,right=2pt,top=2pt,bottom=3pt,arc=2pt,boxrule=0pt]{human-reviewer}\xspace}
\newcommand{\humanrs}{\tcbox[on line, colback=mypink, colframe=white, boxsep=0pt, left=2pt,right=2pt,top=2pt,bottom=3pt,arc=2pt,boxrule=0pt]{human-reviewers}\xspace}
\newcommand{\Humanr}{\tcbox[on line, colback=mypink, colframe=white, boxsep=0pt, left=2pt,right=2pt,top=2pt,bottom=3pt,arc=2pt,boxrule=0pt]{Human-reviewer}\xspace}

\newcommand{\humanrnb}{{human-reviewer}\xspace}
\newcommand{\humanrsnb}{{human-reviewers}\xspace}

\newcommand{\aia}{\tcbox[on line, colback=mypurple, colframe=white, boxsep=0pt, left=2pt,right=2pt,top=2pt,bottom=3pt,arc=2pt,boxrule=0pt]{AI-assistant}\xspace}
\newcommand{\aias}{\tcbox[on line, colback=mypurple, colframe=white, boxsep=0pt, left=2pt,right=2pt,top=2pt,bottom=3pt,arc=2pt,boxrule=0pt]{AI-assistants}\xspace}

\newcommand{\aianb}{{AI-assistant}\xspace}
\newcommand{\aiasnb}{{AI-assistants}\xspace}

\newcommand{\aic}{\tcbox[on line, colback=myblue, colframe=white, boxsep=0pt, left=2pt,right=2pt,top=2pt,bottom=3pt,arc=2pt,boxrule=0pt]{AI-creator}\xspace}
\newcommand{\aics}{\tcbox[on line, colback=myblue, colframe=white, boxsep=0pt, left=2pt,right=2pt,top=2pt,bottom=3pt,arc=2pt,boxrule=0pt]{AI-creators}\xspace}

\newcommand{\aicnb}{{AI-creator}\xspace}
\newcommand{\aicsnb}{{AI-creators}\xspace}

\newcommand{\air}{\tcbox[on line, colback=mygreen, colframe=white, boxsep=0pt, left=2pt,right=2pt,top=2pt,bottom=3pt,arc=2pt,boxrule=0pt]{AI-reviewer}\xspace}
\newcommand{\airs}{\tcbox[on line, colback=mygreen, colframe=white, boxsep=0pt, left=2pt,right=2pt,top=2pt,bottom=3pt,arc=2pt,boxrule=0pt]{AI-reviewers}\xspace}

\newcommand{\airnb}{{AI-reviewer}\xspace}
\newcommand{\airsnb}{{AI-reviewers}\xspace}

\newcommand{\aio}{\tcbox[on line, colback=mybrown, colframe=white, boxsep=0pt, left=2pt,right=2pt,top=2pt,bottom=1pt,arc=2pt,boxrule=0pt]{AI-optimizer}\xspace}
\newcommand{\aios}{\tcbox[on line, colback=mybrown, colframe=white, boxsep=0pt, left=2pt,right=2pt,top=2pt,bottom=1pt,arc=2pt,boxrule=0pt]{AI-optimizers}\xspace}

\newcommand{\aionb}{{AI-optimizer}\xspace}
\newcommand{\aiosnb}{{AI-optimizers}\xspace}

\newcommand{\na}{\tcbox[on line, colback=mylightgrey, colframe=white, boxsep=0pt, left=2pt,right=2pt,top=2pt,bottom=3pt,arc=2pt,boxrule=0pt]{N/A}\xspace}

\begin{document}

\title[Understanding Data Storytelling Tools from the Perspective of Human-AI Collaboration]{Where Are We So Far? Understanding Data Storytelling Tools from the Perspective of Human-AI Collaboration}

\author{Haotian Li}
\affiliation{%
  \institution{The Hong Kong University of Science and Technology}
  \city{Hong Kong SAR}
  \country{China}
}
\email{haotian.li@connect.ust.hk}

\author{Yun Wang}
\affiliation{%
  \institution{Microsoft Research Asia}
  \city{Beijing}
  \country{China}
}
\email{wangyun@microsoft.com}

\author{Huamin Qu}
\affiliation{%
  \institution{The Hong Kong University of Science and Technology}
  \city{Hong Kong SAR}
  \country{China}
}
\email{huamin@cse.ust.hk}

\begin{abstract}  
Data storytelling is powerful for communicating data insights, but it requires diverse skills and considerable effort from human creators. 
Recent research has widely explored the potential for artificial intelligence (AI) to support and augment humans in data storytelling. However, there lacks a systematic review to understand data storytelling tools from the perspective of human-AI collaboration, which hinders researchers from reflecting on the existing collaborative tool designs that promote humans' and AI's advantages and mitigate their shortcomings.
This paper investigated existing tools with a framework from two perspectives: the stages in the storytelling workflow where a tool serves, including analysis, planning, implementation, and communication, and the roles of humans and AI in each stage, such as creators, assistants, optimizers, and reviewers. 
Through our analysis, we recognize the common collaboration patterns in existing tools, summarize lessons learned from these patterns, and further illustrate research opportunities for human-AI collaboration in data storytelling.
\end{abstract}



\begin{teaserfigure}
  \includegraphics[width=\textwidth]{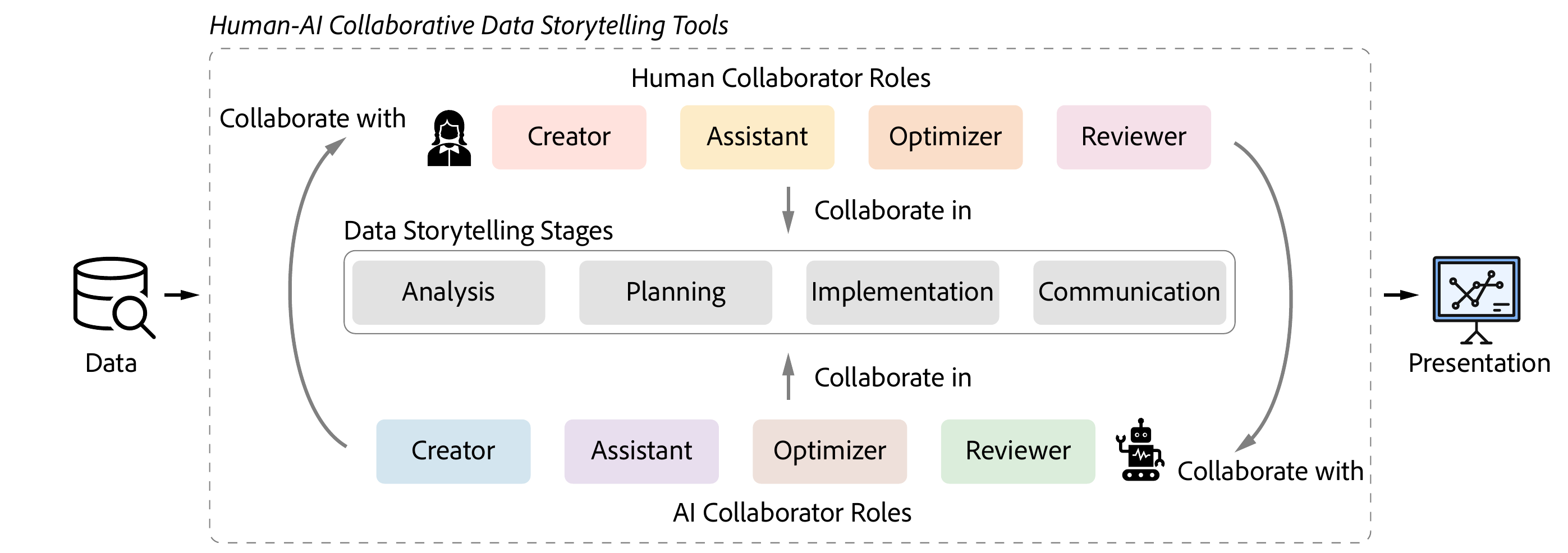}
  \caption{This figure illustrates our framework of human-AI collaboration in data storytelling tools. It characterizes human-AI collaborative data storytelling tools from (1) the stages that they cover and (2) the roles of human and AI collaborators in these stages.}
  \Description{This figure illustrates our framework of human-AI collaboration in data storytelling tools. It characterizes human-AI collaborative data storytelling tools from (1) the stages that they cover and (2) the roles of human and AI collaborators in these stages. In the whole data storytelling process, data is the input and a presentation is often the output. To transform data into presentation, data storytelling tools often facilitate four stages: analysis, planning, implementation, and communication. Humans and AI collaborate with each other in these four stages with four types of roles: creator, assistant, optimizer, and reviewer.}
  \label{fig:framework}
\end{teaserfigure}

\maketitle

\section{Introduction}\label{sec:intro}

The art of creating compelling data stories entails the integration of different aspects, such as data analysis, visualization, and presentation. However, these aspects pose significant demands for human creators, especially when they have to deal with large, complex, or dynamic data sets, manage a considerable number of data story elements, and produce visually appealing presentations. 
Moreover, crafting compelling visual data stories requires creativity, logical consideration, diverse skills from data analysis to graphic design, and a keen awareness of the audience and the context.

Researchers in the visualization (VIS) and the human-computer interaction (HCI) domains have proposed new data storytelling tools that integrate AI techniques, such as natural language processing and machine learning. 
These tools aim to enhance human-AI collaboration in data storytelling by leveraging the complementary strengths of humans and AI. For example, AI can assist humans in finding interesting data facts~\cite{wang2019datashot}, organizing story pieces~\cite{shi2020calliope},
generating visualizations~\cite{qian2020retrieve}, 
or providing feedback~\cite{fu2019visualization}, while humans can provide domain knowledge, creativity, or emotional appeal. However, designing effective human-AI collaboration in data storytelling is not trivial, as it involves many challenges and trade-offs, such as balancing the control and autonomy of the human-AI system, and facilitating the communication and coordination between the two parties. While existing surveys (\eg, \cite{ren2023re, chen2023does}) have collected a corpus of papers on data storytelling, they do not provide a
holistic understanding of how humans and AI collaborate in data storytelling tools.
To facilitate reflection on the current research progress and outlook on future opportunities, 
this paper aims to answer the research question: 
\textit{How do humans and AI collaborate in the existing data storytelling tools?}

To answer the question, we first build a paper corpus about human-AI collaborative data storytelling tools from the VIS and HCI literature.
Then the tools are coded based on our framework from two dimensions (see Figure~\ref{fig:framework}).
First, informed by the different characteristics of stages in the storytelling workflow, such as skill requirements~\cite{elmer2018organizing} and storytellers' expectations~\cite{li2023ai}, we identify \textit{the detailed stages covered by tools}, including data analysis, data story planning, data story implementation, and data communication.
Second, in each stage, the collaboration patterns are examined through \textit{human and AI collaborators' roles}, namely creators, reviewers, optimizers, and assistants. 
The roles facilitate understanding 
how humans and AI contribute to the data storytelling stage by leveraging their advantages, which is an open challenge in achieving human-AI collaboration~\cite{capel2023human}.
Combining the two perspectives, we can comprehensively understand the relationship between the stage features and the design of human-AI collaboration.

Our study results in an overview of the development of human-AI collaborative tools in data storytelling between 2010 and 2023. 
Human-AI collaborative data storytelling tools have received the most research interest since 2019, after the dominance of purely manual tools between 2016 and 2018.
We also reflect on the similarities and differences in human-AI collaboration patterns at different stages. 
Based on these findings, we further discuss the implications for the design of data storytelling systems and the future of data storytelling practice and research.

We hope our review of the current state of the art in human-AI collaborative data storytelling will be a valuable resource for researchers and designers working on human-AI collaborative tools. We also hope this paper will highlight new directions for future research. The main contributions of this paper are as follows:
\begin{itemize}
\item We conduct the first survey to characterize the human-AI collaboration in data storytelling at a stage level by building a paper corpus of human-AI collaborative data storytelling tools from the VIS and HCI literature and coding them according to the stages of the data storytelling workflow and the roles of humans and AI.
\item We report the findings from existing human-AI collaborative data storytelling tools and the design patterns of human-AI roles at different stages, identify the challenges and opportunities for the design of data storytelling systems, and further suggest future research directions from the perspective of human-AI collaboration.
\end{itemize}

\section{Related Work}
This section introduces related work about data storytelling and human-AI collaboration in visualization.

\subsection{Data Storytelling}
Data stories often present a sequence of story pieces that are supported by data facts and are accompanied by data visualizations~\cite{lee2015more}.
Data storytelling has been gaining growing interest from researchers following the seminal work by Segel and Heer~\cite{segel2010narrative}, where they outlined the narrative visualization design space by reviewing existing examples.
When creating and spreading data stories, diverse skills, including data analysis and visualization, are required in a pipeline with multiple tasks, such as planning data stories and presenting them to audiences~\cite{chevalier2018analysis, elmer2018organizing}.
To build the theory of data storytelling and enhance the practices, researchers have investigated data storytelling from multiple perspectives, including understanding real-world data stories (\eg, \cite{shu2020makes}), framing design spaces and guidelines~(\eg, \cite{hullman2013deeper}), and developing data storytelling tools (\eg, \cite{kim2019datatoon}).

To organize and review the progress of research in data storytelling, several surveys have been conducted in the past years.
Tong~\etal~\cite{tong2018storytelling} reviewed data storytelling from who, how, and why data stories are authored and spread.
Their characterization of data storytelling tools is limited to the structure of data stories, \ie, linear, parallel, or user-directed data stories.
Zhao and Elmqvist~\cite{zhao2023stories} summarized the format of data stories from six perspectives, including audience, data, media, \etc.
They do not explicitly provide a taxonomy for data storytelling tools.
Ren~\etal~\cite{ren2023re} investigated data storytelling tools leveraging narratology.
They categorize existing data storytelling tools by whether the authors of data stories introduce their own opinions and whether the audience decides the sequence of the story.

Since data storytelling often requires various skills and enormous effort from the authors~\cite{lee2015more}, applying AI in data storytelling tools to eliminate these needs has attracted researchers' attention.
To cope with the trend, a recent survey by Chen~\etal~\cite{chen2023does} classified data storytelling tools according to whether the tool leverages AI to generate data stories or support data story authoring besides data story characteristics (\ie, genre, interactivity, and structure).
Though they also attempt to characterize AI usage in tools, their taxonomy only categorizes AI-powered tools into two types: AI-supported and AI-generator tools.
According to a previous interview study~\cite{li2023ai}, data workers have various needs in different stages of data storytelling and require AI collaborators to perform multiple roles in collaboration.
Such observation reflects the need to understand a \textit{fine-grained collaboration pattern} between AI and humans \textit{in different stages} (\eg, planning and communicating the data story) of the data storytelling workflow, which is {beyond the scope of the survey by Chen~\etal~\cite{chen2023does}.}
In this paper, we review the existing human-AI collaborative data storytelling tools from the roles of humans and AI in different stages and identify potential future research opportunities.

\subsection{Human-AI Collaboration in Visualization Research}
{
Recently, there has been a growing trend of investigating how to achieve and characterize the collaboration between humans and AI 
rather than replacing humans with AI~\cite{lai2023towards, wang2020human}.
For example, in pioneering research by Endsley~\cite{endsley1987application}, the levels of automation in decision making include fully automatic decision execution, decision execution without humans' disapproval or after humans' approval, and decision recommendation to humans.
Later work by Parasuraman~\etal~\cite{parasuraman2000model} further proposed a ten-level categorization of human-AI collaboration.
Crisan~\etal~\cite{crisan2021fits} proposed a framework to characterize expected human-AI collaboration patterns in data science.
They consider humans' expertise as an additional dimension to the level of automation.
A recent research by Shi~\etal~\cite{shi2023understanding} proposed five dimensions to characterize the collaboration between designers and AI, including the skill levels of collaborators and whether humans and AI collaborate to finish the whole workflow.
The previous studies in human-AI collaboration often focus on specific domains, such as human-AI collaborative decision making (\eg,~\cite{vagia2016literature, endsley1987application, parasuraman2000model}), creativity (\eg,~\cite{shi2023understanding, lubart2005can, deterding2017mixed}), and data science~(\eg,~\cite{crisan2021fits,wang2019human}).
}

In the field of visualization research, human-AI collaboration has been investigated in the following three perspectives~\cite{wang2023vis}.
First, AI can collaborate with humans in creating and understanding visualizations (AI4VIS).
Typical tasks in this line of research include visualization recommendation~\cite{hu2019vizml, lin2023dashboard} and retrieval~\cite{li2022structure, ye2022visatlas}.
Under this category, two surveys~\cite{wang2021survey,wu2021ai4vis} were conducted to provide an overview of AI and machine learning (ML) techniques used for related tasks.
Besides, Zhu~\etal~\cite{zhu2020survey} and Qin~\etal~\cite{qin2020making} reviewed the automatic visualization recommendation techniques.
Second, visualization can serve as a tool to monitor, understand, and steer AI systems (VIS4AI) for better collaboration between humans and AI.
For example, the What-If Tool~\cite{wexler2019if} and RuleMatrix~\cite{ming2018rulematrix} apply visualizations to explain AI behaviors.
GAM Changer~\cite{wang2022interpretability} allows users to edit machine learning models with visualization interfaces.
There have been multiple papers to investigate the progress of this line of research, such as the latest ones by Subramonyam and Hullman~\cite{subramonyam2023we} and by Wang~\etal~\cite{wang2023visual}.
Lastly, visual analytics systems can serve as an interface for human-AI collaborative data analysis and decision making.
Researchers have explored human-AI collaboration in visual analytics systems for various domains, including urban analytics~\cite{liu2017smartadp}, sports~\cite{stein2017bring}, and education~\cite{li2021visual}.
Besides the reviews of visual analytics for different domains (\eg, urban~\cite{feng2022survey} and education~\cite{kui2022survey}), a recent paper by Domova and Vrotsou~\cite{domova2022model} outlined the usage of AI in these systems based on the visual analytics pipeline.

Among all three perspectives, our paper is the closest to AI4VIS research since human-AI collaborative data storytelling tools are also powered by its advances.
However, previous surveys about AI4VIS~\cite{wang2021survey, wu2021ai4vis} focus on how the AI techniques were applied in completing visualization-related tasks, including data storytelling, but 
do not specifically investigate how humans work with AI in data storytelling tools.
{To fill the gap, our study characterizes the collaboration between humans and AI from their stages and roles in collaboration, without emphasizing the techniques used in AI.
Our characterization of roles extends a previous study in human-computer collaboration by Lubart~\cite{lubart2005can}, where the author proposed four potential roles of computers in creative support, \textit{nannies}, \textit{pen-pals}, \textit{coaches}, and \textit{colleagues}.
Specifically, our framework expands the role of colleagues who work with humans in creative work to a more fine-grained level based on the work distribution between humans and AI.
The roles in our framework including creator, assistant, optimizer, and reviewer, inspired by previous literature in humans' collaborative writing~\cite{posner1992people, lowry2004building}.
Furthermore, we introduce four stages in the data storytelling workflow, \ie, analysis, planning, implementation, and communication, as another dimension to delineate where humans and AI collaborate.}

\section{Methodology}\label{sec:methodology}

This section introduces our research methodology, including how we collected and coded the corpus of tools.

\subsection{Paper Collection}

\begin{figure}
    \centering
    \includegraphics[width=\linewidth]{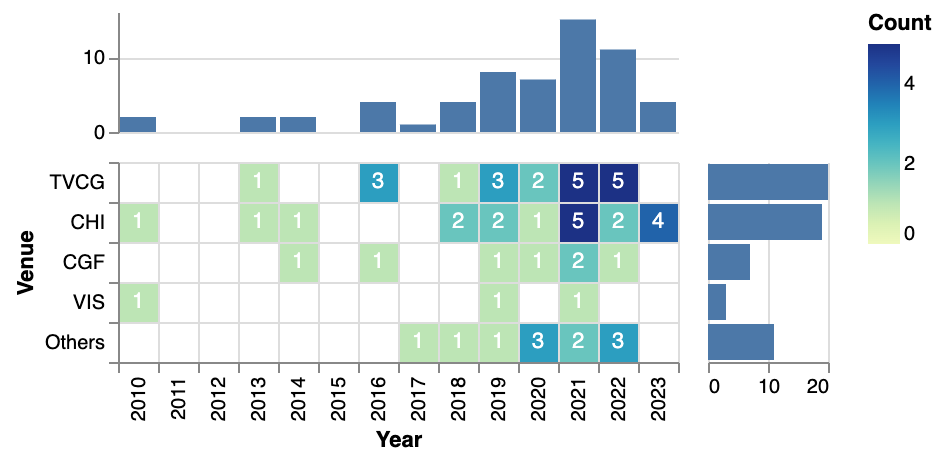}
    \caption{This figure shows an overview of our collected papers in venues and years. The venues with less than three papers are merged to ``Others.''}
    \Description{This figure shows an overview of our collected papers in venues and years with a heatmap and two histograms. It shows the number of papers between 2020 and 2023 in four major venues, i.e., TVCG, CHI, CGF, VIS, and other venues. We notice that the research in data storytelling tools receives growing interest from both the VIS and the HCI communities.}
    \label{fig:overview}
\end{figure}

To collect papers about human-AI collaborative data storytelling tools, we started our search with the data storytelling tool papers covered in the latest surveys~\cite{chen2023does, ren2023re}.
Since the two surveys only covered papers up to August 2022, we further augmented the corpus by searching recent relevant venues up to June 2023, including ACM CHI, UIST, IUI, IEEE TVCG, IEEE VIS, PacificVis, and EuroVis (papers are published in Computer Graphics Forum), with the keywords ``narrative visualization'' and ``data storytelling'' following a keyword-based paper searching strategy~\cite{mcnabb2017survey}.
Then two leading co-authors read and examined the corpus of papers individually to keep the papers describing human-AI collaborative tools for data storytelling.
We also enriched the corpus by checking the papers that cited these papers or are used as references by them in the corpus.
The conflicts in paper selection were resolved through multiple discussions between the two co-authors.
In the end, we collected 60 papers for further investigation in the paper.
Figure~\ref{fig:overview} shows an overview of the papers.
It reveals that research in data storytelling tools receives growing interest from both the VIS and the HCI communities.

When deriving the corpus of papers, we have several criteria for the papers to be included in our corpus.
First, \textit{the paper should mainly serve the purpose of data storytelling with information visualizations}.
When surveying related research, we noticed that several previous studies applied narrative visualization-related techniques for data analytics, such as VisInReport~\cite{sevastjanova2021visinreport} and NetworkNarratives~\cite{li2023networknarratives}.
Though they applied narrative techniques, they are not designed for telling data stories and are therefore not included in our corpus.
We also did not consider interactive chart recommendation tools (\eg, Voyager2~\cite{wongsuphasawat2017voyager}) or chart authoring tools (\eg, Charticulator~\cite{ren2018charticulator}) that are not specifically designed for data storytelling.
Data storytelling tools for scientific visualizations are also excluded (\eg, AniViz~\cite{akiba2009aniviz}, Molecumentary~\cite{kouril2021molecumentary}, and ScrollyVis~\cite{morth2022scrollyvis}).
Second, \textit{the paper should describe a tool for data storytelling.}
To align with the main purpose of this paper, \ie, understanding the collaboration between human and AI in tools, the scope of papers is limited to data storytelling tools rather than design spaces (\eg, narrative structure based on Freytag’s Pyramid~\cite{yang2021design}) or grammars~(\eg, Animated Vega-Lite~\cite{zong2022animated}).
To determine whether AI is applied in the tool, we followed a previous survey about AI for visualization~\cite{wu2021ai4vis} and research in human-AI collaboration~\cite{li2023ai, jiang2021supporting} to adopt an inclusive definition of AI, where heuristics-based, ML-based, and deep learning~(DL)-based AI are all included.

\subsection{Paper Coding}
Considering the goal of this paper, \ie, understanding how existing tools facilitate the collaboration between humans and AI in the data storytelling workflow, we decided to code the collaboration pattern in each step of the workflow.
To achieve this goal, we first surveyed existing research about data storytelling workflow~\cite{chevalier2018analysis,lee2015more,crisan2020passing} and human-AI collaboration patterns in data storytelling~\cite{li2023ai}.
Then we derived the preliminary framework for coding papers from previous studies.
Based on the preliminary framework, two co-authors coded all papers in the corpus individually, leveraging both papers and related presentation or demo videos if provided.
{
Then, each author checked the conflicts and adjusted their codes for apparent issues, such as missing codes.
When the conflicts were unresolved, they would be marked and brought to the weekly meeting for discussion until agreement.
If the discussion led to any update of the framework or the coding criteria, the co-authors would re-examine all their codes and update accordingly.
Finally, the co-authors reached an agreement on all codes.
}
In the following sections, we introduced the framework (Section~\ref{sec:taxonomy}) and summarized the coded tools (Section~\ref{sec:results}).
We also provide an interactive browser of the coded tools at \textcolor{navy}{\url{https://human-ai-universe.github.io/data-storytelling}}.

\section{Framework}\label{sec:taxonomy}
This section introduces the framework for characterizing the collected human-AI collaborative data storytelling tools.
A previous study~\cite{li2023ai} investigated data workers' preferred AI collaborators through an interview study.
The results reveal that data workers prefer to have different types of AI collaborators, including creators, assistants, optimizers, and reviewers, in different stages of the data storytelling workflow.
For example, AI assistants were preferred when planning data stories, such as collecting data insights, while AI creators were appreciated for making data stories, such as authoring story pieces.
However, the prior surveys on storytelling tools~\cite{chen2023does, ren2023re} do not investigate the collaboration between humans and AI at the stage level, which makes it impossible to compare the existing tools with users' expectations and identify potential future research opportunities.
To fill the gap, this paper aims to characterize the collaboration pattern in different stages.
Furthermore, the roles of human and AI collaborators can reflect how to leverage their pros and mitigate their cons in each stage, which is considered a challenge for designing a human-AI collaborative experience~\cite{capel2023human}.
As a result, our framework consists of two dimensions: (1) what \textit{stages} in the data storytelling workflow a tool serves (Section~\ref{sec:tax-stages}); and (2) what \textit{the roles of human and AI collaborators} are in each stage (Section~\ref{sec:tax-roles}).


\subsection{Stages in the Data Storytelling Workflow}\label{sec:tax-stages}
The first perspective of our framework is the tools' coverage of stages in the storytelling workflow.
To derive the categorization of stages, we first surveyed how previous research defined stages in data storytelling workflow.
{The practical workflow of data storytelling has been discussed in multiple previous papers~\cite{lee2015more, chevalier2018analysis, crisan2020passing, li2023ai}.
Among them, Lee~\etal~\cite{lee2015more} introduce a detailed workflow that consists of three stages: \textit{``explore data''}, \textit{``make a story''}, and \textit{``tell a story''}.
The workflows proposed by Chevalier~\etal~\cite{chevalier2018analysis} and Li~\etal~\cite{li2023ai} highly aligned with the one proposed by Lee~\textit{et al.}~\cite{lee2015more}.
Crisan~\etal~\cite{crisan2020passing} focused more on data analysis.
As a result, we adopted the widely recognized data storytelling workflow by Lee~\etal~\cite{lee2015more} as the preliminary version of stages in our framework.}
Furthermore, in the ``tell a story'' stage, there are two sub-stages, \textit{``build presentation''} by authors, and \textit{``share a story''} with audiences.
When reading the papers, we noticed that most tools aim to automate the sub-stage of building presentations but do not explicitly facilitate the direct communication of data stories with audiences (\eg, DataShot~\cite{wang2019datashot} and NB2Slides~\cite{zheng2022telling}).
On the other hand, some research solely provides human-AI collaborative ways for sharing data stories, such as SketchStory~\cite{lee2013sketchstory} and the interactive data story presentation approach by Hall~\etal~\cite{hall2022augmented}.
Furthermore, the potential difference between stakeholders in these two stages is mentioned by Lee~\etal~\cite{lee2015more} and Chevalier~\etal~\cite{chevalier2018analysis}.
The presentations are often built by editors, while the stories are shared among presenters and audiences.
Therefore, we decided to consider the two sub-stages in the ``tell a story'' stage as two individual stages when coding tools.

\begin{table*}
\caption{This table summarizes the four roles of human and AI collaborators.}
\Description{This table summarizes the four roles of human and AI collaborators. Human- and AI-creators finish most of the work from scratch. Human- and AI-assistants work with creators together to reduce the workload or compensate for inability. Human- and AI-optimizers automatically improve the entire or part of the data stories. Human- and AI-reviewers assess the content in data stories and provide assessment results or suggestions.}
\small
\begin{tabular}{@{}ll@{}}
\toprule
\textbf{Role} & \textbf{Definition} \\ \midrule
\Humanc, \aic   &   The collaborator who finishes most of the work from scratch         \\ \midrule
\Humana, \aia   &  The collaborator who works with creators together to reduce the workload or compensate for inability          \\ \midrule
\Humano, \aio   &   The collaborator who automatically improves the entire or part of the data stories        \\ \midrule
\Humanr, \air   &  The collaborator who assesses the content in data stories and provides assessment results or suggestions          \\ \bottomrule
\end{tabular}
\label{tab:role_def}
\end{table*}

To avoid confusion and make the names of stages simpler, inspired by Li~\etal~\cite{li2023ai}, the four stages in our framework are named as \textit{analysis}, \textit{planning}, \textit{implementation}, and \textit{communication}, which corresponds to ``explore data'', ``make a story'', ``build presentation'', and ``share a story''~\cite{lee2015more, chevalier2018analysis}, respectively.
In the \textit{analysis} stage, the users of tools often explore data to identify insights, such as the trend of data, and may summarize knowledge from the insights.
With the insights and knowledge from data, users need to \textit{plan} how the story will be told.
When planning the data story, they often determine the core message, prepare the outline, and sequence the data insights collected from the analysis.
Users next \textit{implement} data stories following their plans, such as authoring charts, writing texts, integrating story pieces, and styling the data stories.
Finally, the data stories are \textit{communicated} with the audiences on different occasions, such as team meetings or formal presentations.

With the definition above, we coded all tools in our corpus.
Notably, some tools may not only serve for one stage.
For example, Calliope~\cite{shi2020calliope} first computes the data insights, then organizes the sequence of insights, and finally creates a scrollytelling story or a poster.
According to our criteria, it covers three stages, analysis, planning, and implementation.
Furthermore, some tools may not complete all tasks in one stage, such as InfoColorizer~\cite{yuan2021infocolorizer}.
It is limited to suggesting color usage in the implementation stage.
Since we only code tools at the stage level, InfoColorizer is coded as a tool applied in the implementation stage.
{Another point is that we adopt a definition of the communication stage where a data story is directly communicated to audiences by data storytellers.
Some data story formats, such as data videos and posters, might be spread directly and asynchronously without a stage where they are explicitly presented to the audiences.
However, presenters can also communicate them to audiences, such as introducing data posters to the customers or playing data videos during the presentation.
It is not feasible for us to judge how these data stories are finally consumed.
Therefore, we consider the tools in the communication stage must facilitate direct communication between data storytellers and audiences, such as SketchStory~\cite{lee2013sketchstory}.}

\subsection{Roles of Human and AI Collaborators}\label{sec:tax-roles}
As discussed at the start of this section, we further coded how humans and AI collaborate in different stages after figuring out how tools cover different stages in the storytelling workflow.
Specifically, inspired by prior research in collaboration among humans~\cite{posner1992people, lowry2004building}, we assigned roles for human and AI collaborators in each stage covered by a data storytelling tool.
Taking Erato~\cite{sun2022erato} as an example, it facilitates the analysis, planning, and implementation stages.
In the analysis stage, humans explore the dataset with the assistance of AI.
AI suggests potential interesting data facts during exploration.
Then we coded the analysis stage of Erato as having humans as creators and AI as an assistant.
With roles like \revisecr{\textit{creator} and \textit{assistant}}, we can investigate how the existing tools distribute work among collaborators and leverage their advantages.
This section introduces how the roles are defined and applied to our corpus of papers.

We started our coding with the collaboration patterns between humans and AI in the data storytelling context by Li~\etal~\cite{li2023ai} since it is the only previous research on the topic.
However, they focus on defining the roles of AI collaborators since their purpose is to learn what assistance by AI is desired by users.
In our work, we extended their definition from AI to both AI and humans, where we had four types of AI or human collaborators: \textit{creator}, \textit{assistant}, \textit{optimizer}, and \textit{reviewer}.
For clarity and simplicity, we use ``human-'' or ``AI-'' prefixes before the roles, such as \humanc and \aia, to indicate the collaborator's identity and role in one stage. 
{The definition of the roles is introduced in Table~\ref{tab:role_def}.}

In a stage of the data storytelling workflow, an \aic or \humanc finishes most of the work from scratch and takes the highest level of control of the stage.
For example, in DataShot~\cite{wang2019datashot}, data analysis is fully completed by AI.
Then we considered AI as the creator of data analysis in DataShot.
Differently, NB2Slides~\cite{zheng2022telling} asks humans to complete the data analysis without AI's intervention.
As a result, we coded the analysis stage of NB2Slides as \humancnb.
Unlike creators, \humanas and \aias often work with creators together when authoring the data story.
They can reduce the workload of creating content in data stories or compensate for creators' inability.
As a result, they have a lower level of control.
An example of \aiasnb can be 
Chartreuse~\cite{cui2021mixed}.
In the implementation stage, AI in Chartreuse extracts reusable templates from existing infographics to simplify creating new ones.
We provide a detailed discussion about assistants in Section~\ref{sec:lesson_across}.

\textit{Optimizers} and \textit{reviewers} mostly perform after creators and assistants take action.
\aios and \humanos automatically improve the entire or part of the data stories.
Different from assistants, they do not participate in the creation process and, therefore, do not share the workload with creators directly.
They repeat the work that creators and assistants have finished to improve the quality.
For example, Notable~\cite{li2023notable} allows \humanosnb to edit the AI-created sequence of data insights.
At last, \airs and \humanrs assess the created content in data stories and provide the results or suggestions for creators to improve the data story.
Compared to optimizers, they do not take any actions to improve the data story but only provide feedback.
An example of \airnb is the approach proposed by Fan~\etal~\cite{fan2022annotating}, where AI assesses the line charts in data stories and annotates the potential deceptive issues.

\begin{figure*}
    \centering
    \includegraphics[width=\linewidth]{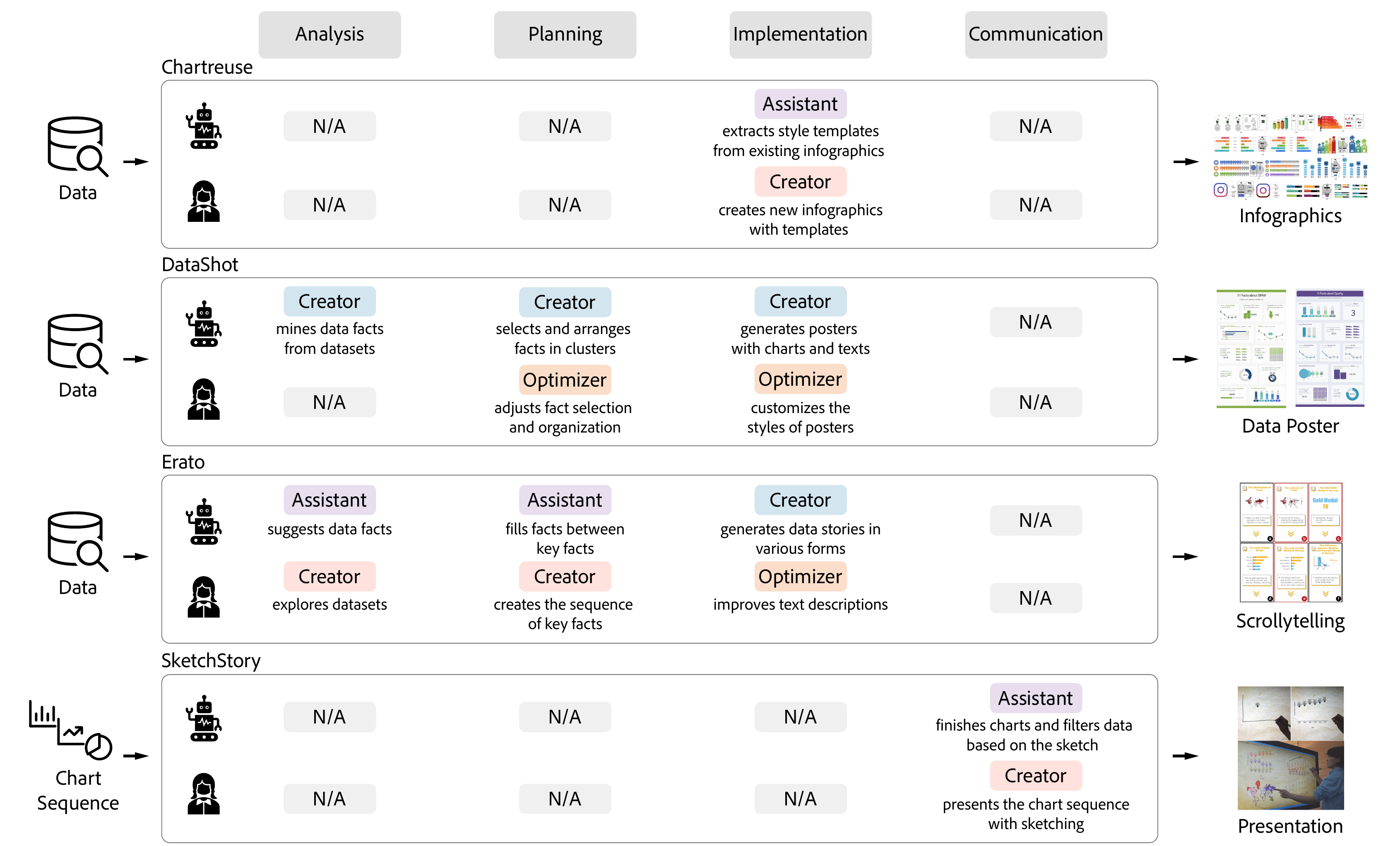}
    \caption{This figure demonstrates four example tools, including Chartreuse~\cite{cui2021mixed}, DataShot~\cite{wang2019datashot}, Erato~\cite{sun2022erato}, and SketchStory~\cite{lee2013sketchstory}, assessed by our framework.}
    \Description{This figure demonstrates four example tools assessed by our framework. We assign an AI-assistant and a Human-creator for Chartreuse in the implementation stage. DataShot has AI-creators in the analysis, planning, and implementation stages, and human-optimizers in the planning and implementation stages. Erato has AI-assistants in the analysis and planning stage, an AI-creator in the implementation stage, human-creators in the analysis and planning stages, and a human-optimizer in the implementation stage. SketchStory has an AI-assistant and an human-creator in the communication stage.}
    \label{fig:example}
\end{figure*}

We have additional considerations when coding the tools following the definition of collaborator roles.
First, we avoided assigning the \textit{creator} role to both human and AI collaborators in the same stage.
Its rationale is that we prefer to clearly distinguish whether AI or humans take control of the stage and bear more workload through the roles of creators and assistants.
Second, when a type of stakeholder takes multiple roles in one stage, we only code the primary role.
We have such consideration to simplify the coding process and the summary of common collaboration patterns.
For example, in the implementation stage of Infomages~\cite{coelho2020infomages}, after users create charts and AI merges the chart with an image, users can review the distortion of the new charts and ask AI to optimize them.
In this case, humans serve as creators and reviewers, while AI is an assistant and optimizer.
However, following the consideration, we only keep human-creator and AI-assistant since the review and optimization tasks are auxiliary and optional. 

Figure~\ref{fig:example} explains how we coded tools according to our framework by showing four tools.
The complete results of our coded human-AI collaboration patterns in different stages are available in Table~\ref{tab:tool_list}.

\section{Observations of Tools and Stages}\label{sec:results}
\begin{figure}
    \centering
    \includegraphics[width=0.9\linewidth]{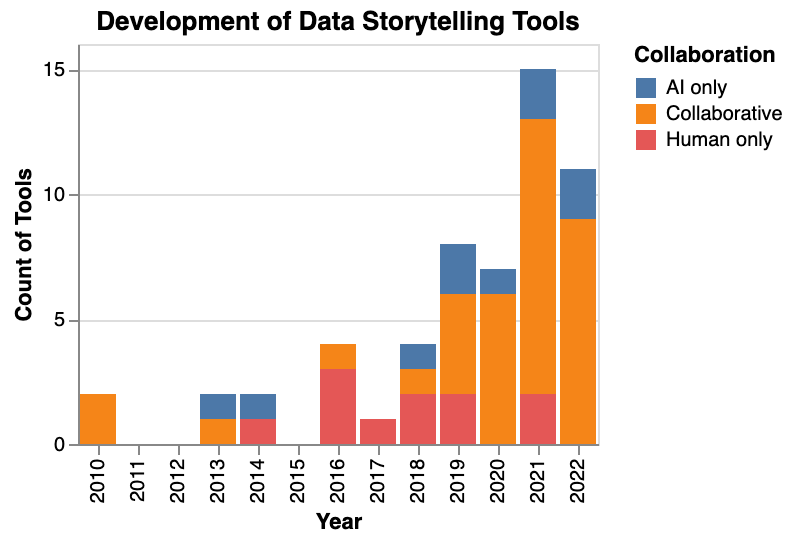}
    \caption{This figure shows the development of data storytelling tools between 2010 and 2022 with the number of tools and collaboration between humans and AI. The year 2023 is excluded since we only collect tools up to June 2023.}
    \Description{This figure includes a stacked bar chart to show the number of tools in each year between 2010 and 2022 and the collaboration between humans and AI. The chart reveals that the tools with humans only had the highest proportion between 2016 and 2018. Since 2019, most of the tools have been human-AI collaborative. The number of tools with only AI was stable after 2019. }
    \label{fig:development_collaboration}
\end{figure}

This section introduces our findings regarding tools and stages.
To provide an overview of our reviewed tools, Figure~\ref{fig:development_collaboration} shows the development of data storytelling tools.
{The chart reflects that the research of data storytelling receives increasing interest.} 
Furthermore, we notice that the purely manual tools have the highest proportion between 2016 and 2018.
{Since 2019, human-AI collaborative tools have gained interest from researchers.}

{In Table~\ref{tab:tool_list}, according to the covered stages, we find four groups of tools and separate them with horizontal lines.
The tools in the first group cover the implementation stage. The second group extends the first group to facilitate the planning stage and the implementation stage. With the tools of the third group, users can finish three stages from analysis to implementation. Finally, the fourth group of tools is designed for the communication stage only.}
{Sections~\ref{sec:group_1}-\ref{sec:group_4} will introduce the four groups in detail.
In each subsection, we organize the tools in that group based on their data story formats following prior literature~\cite{segel2010narrative, chen2023does} and discuss how they apply human-AI collaboration in tasks.}

\subsection{Group 1: Tools for the Implementation Stage} \label{sec:group_1}
The first group of 27 tools focuses on the implementation stage.
In the implementation stage, tool users build data stories with the data analysis results and story plans in the previous stages.
{Their tasks in the stage mainly include creating story pieces (\eg, charts and text), styling story pieces, adding animations, and arranging the positions of visual elements.}
According to a previous interview~\cite{li2023ai}, the implementation stage is the most desired stage where users would like to receive assistance from AI.

{Most of the early storytelling tools rely on \humancs' effort in implementation, including tools for \textbf{infographics}.}
{When implementing infographics, users' tasks include designing the visual elements, considering the positions of these elements, binding data to them, and adjusting their styles, such as colors and shapes.}
To reduce the effort of designing visual elements, DataQuilt~\cite{zhang2020dataquilt} introduces an \aia to extract glyphs from images with computer vision approaches.
Chartreuse~\cite{cui2021mixed} and Vistylist~\cite{shi2022supporting} further extend it to extracting templates (\eg, glyphs and colors) from existing infographics for styling human-created ones.
Another usage of AI-assistants is to suggest colors, as Yuan~\etal~\cite{yuan2021infocolorizer} and Liu~\etal~\cite{liu2022image} do.
To further eliminate the need to create charts,
Text-to-Viz~\cite{cui2019text} and Retrieve-then-Adapt~\cite{qian2020retrieve} explore \aics that generate infographics with users' natural language input.
The path of these tools demonstrates a trend of \textit{developing tools with higher AI automation}.
{Despite \aiasnb and \aicsnb,
applying \airs to review created charts for data storytelling has also been explored previously.
Fu~\etal~\cite{fu2019visualization} explore applying computer vision models to assess the aesthetics and memorability of infographics.
The vision models can rank the input visualizations or output the scores regarding aesthetics and memorability.}

{Besides infographics, we notice several tools are designed for creating \textbf{annotated charts}.}
{The first type of tools facilitates authors' annotating existing charts. 
The tasks assisted by these tools include adding visual embellishment and writing narrative annotations.}
Click2Annotate~\cite{chen2010click2annotate} and Touch2Annotate~\cite{chen2010touch2annotate} allow users to select their interested area and then generate related text descriptions following pre-defined templates.
Since the \aics fill the templates automatically, they may not fully understand users' intent.
Therefore, they allow \humanos to adjust the content, including the data findings and text descriptions.

{The other tools further automate annotated chart creation.
Contextifier~\cite{hullman2013contextifier} and NewsViews~\cite{gao2014newsviews} leverage \aicsnb to generate annotated charts according to data articles to help readers understand them.
The generation process does not involve humans.
Comparing these tools, we notice that \textit{the design of human-AI collaboration patterns depends on the usage scenario of the tool}.
When authoring data stories, the authors need to express their intent in data stories.
As a result, multiple tools allow \humanosnb to revise the content produced by \aicsnb to control data stories (\eg, \cite{chen2010click2annotate, chen2010touch2annotate, cui2019text, tyagi2022infographics}).
The collaboration pattern can be different when developing tools for data story consumers, like data article readers.
These tools only illustrate existing stories without considering consumers' intent.
Therefore, readers do not collaborate with AI in these tools.}

All tools discussed above are designed for static charts.
Informed by the advantages of animations, such as enhancing audiences' engagement~\cite{chevalier2016animations}, four tools transform static charts into {\textbf{animated charts or data videos}}~\cite{wang2021animated, ge2021cast, pu2021datamations, shi2021autoclips}.
{To achieve the target, the tasks conducted by authors usually involve grouping related visual elements and assigning corresponding animations to individual or grouped visual elements.
CAST~\cite{ge2021cast} provides a GUI tool for specifying animation in charts with \aiasnb.
The \aiasnb helps users group visual elements and suggests potential keyframes in the animation.
InfoMotion~\cite{wang2021animated} generates animated infographics by first deciding the group of visual elements and then assigning animations.
After generation, \humanosnb can optimize the animation further.}


\begingroup
\renewcommand{\arraystretch}{0.9}
\renewcommand{\humanc}{\tcbox[on line, colback=myred, colframe=white, boxsep=0pt, left=2pt,right=2pt,top=1pt,bottom=1pt,arc=2pt,boxrule=0pt]{H-C}\xspace}
\renewcommand{\humano}{\tcbox[on line, colback=myorange, colframe=white, boxsep=0pt, left=2pt,right=2pt,top=1pt,bottom=1pt,arc=2pt,boxrule=0pt]{H-O}\xspace}
\renewcommand{\humana}{\tcbox[on line, colback=myyellow, colframe=white, boxsep=0pt, left=2pt,right=2pt,top=1pt,bottom=1pt,arc=2pt,boxrule=0pt]{H-A}\xspace}
\renewcommand{\humanr}{\tcbox[on line, colback=mypink, colframe=white, boxsep=0pt, left=2pt,right=2pt,top=1pt,bottom=1pt,arc=2pt,boxrule=0pt]{H-R}\xspace}

\renewcommand{\aia}{\tcbox[on line, colback=mypurple, colframe=white, boxsep=0pt, left=2pt,right=2pt,top=1pt,bottom=1pt,arc=2pt,boxrule=0pt]{A-A}\xspace}
\renewcommand{\aic}{\tcbox[on line, colback=myblue, colframe=white, boxsep=0pt, left=2pt,right=2pt,top=1pt,bottom=1pt,arc=2pt,boxrule=0pt]{A-C}\xspace}
\renewcommand{\air}{\tcbox[on line, colback=mygreen, colframe=white, boxsep=0pt, left=2pt,right=2pt,top=1pt,bottom=1pt,arc=2pt,boxrule=0pt]{A-R}\xspace}
\renewcommand{\aio}{\tcbox[on line, colback=mybrown, colframe=white, boxsep=0pt, left=2pt,right=2pt,top=1pt,bottom=1pt,arc=2pt,boxrule=0pt]{A-O}\xspace}

\renewcommand{\na}{\tcbox[on line, colback=mylightgrey, colframe=white, boxsep=0pt, left=2pt,right=2pt,top=1pt,bottom=1pt,arc=2pt,boxrule=0pt]{N/A}\xspace}

\begin{table*}[]
    \caption{This table introduces our reviewed tools in detail. We coded them with the framework in Section~\ref{sec:taxonomy}. For simplicity, we shorten the names of collaborators: \aic: AI-creator, \aia: AI-assistant, \aio: AI-optimizer, \air: AI-reviewer; \humanc: Human-creator, \humana: Human-assistant, \humano: Human-optimizer, \humanr: Human-reviewer. Based on the stages, the tools can be roughly divided into four groups with horizontal lines. The first group of tools is designed particularly for the implementation stage. The second group enhances the first group by covering the planning stage. The third group considers the three stages from analysis to implementation. The last group solely serves for the communication stage.}
    \Description{This table introduces our reviewed tools in detail. We coded them with the framework in Section 4. For simplicity, we shorten the names of collaborators: A-C: AI-creator, A-A: AI-assistant, A-O: AI-optimizer, A-R: AI-reviewer; H-C: Human-creator, H-A: Human-assistant, H-O: Human-optimizer, H-R: Human-reviewer. Based on the stages, the tools can be roughly divided into four groups with horizontal lines. The lines are between Liu et al. [72] and Ellipsis [96], ChartStory [146] and CLUE [39], and Roslingifier [104] and SketchStory [58]. The first group of tools is designed particularly for the implementation stage. The second group enhances the first group by covering the planning stage. The third group considers the three stages from analysis to implementation. The last group solely serves for the communication stage.
    }
    \small
    \centering
    \begin{tabular}{lp{4cm}p{2cm}p{2cm}p{2cm}p{2cm}}
    \toprule
         \textbf{Year} & \textbf{Tool} & \textbf{Analysis} & \textbf{Planning} & \textbf{Implementation} & \textbf{Communication} \\\midrule
        
        2016 & Data-Driven Guides~\cite{kim2016data} & \na & \na & \humanc & \na \\
        2017 & ChartAccent~\cite{ren2017chartaccent} & \na & \na & \humanc & \na \\
        2018 & DataInk~\cite{xia2018dataink} & \na & \na & \humanc & \na \\  
        2019 & Timeline Storyteller~\cite{brehmer2019timeline} & \na & \na & \humanc & \na \\
        2020 & Infomages~\cite{coelho2020infomages} & \na& \na& \humanc \aia & \na \\
        2020 & DataQuilt~\cite{zhang2020dataquilt}  & \na& \na& \humanc \aia & \na \\
        2021 & Chartreuse~\cite{cui2021mixed}  & \na& \na& \humanc \aia & \na \\
        2021 & InfoColorizer~\cite{yuan2021infocolorizer}  & \na& \na& \humanc \aia & \na \\
        2021 & CAST~\cite{ge2021cast} & \na& \na& \humanc \aia & \na \\
        2022 & Vistylist~\cite{shi2022supporting}  & \na& \na& \humanc \aia & \na \\
        2013 & Contextifier~\cite{hullman2013contextifier}  & \na & \na & \aic & \na \\
        2014 & NewsViews~\cite{gao2014newsviews}  & \na & \na & \aic & \na \\
        2018 & Metoyer~\etal~\cite{metoyer2018coupling} & \na & \na & \aic & \na \\
        2019 & Chen~\etal~\cite{chen2019towards} & \na & \na & \aic & \na \\
        2020 & Retrieve-then-Adapt~\cite{qian2020retrieve} & \na & \na & \aic & \na \\
        2021 & AutoClips~\cite{shi2021autoclips} & \na & \na & \aic & \na \\
        2021 & Datamations~\cite{pu2021datamations} & \na & \na & \aic & \na \\
        2019 & DataSelfie~\cite{kim2019dataselfie} & \na & \na & \aic \humana & \na \\
        2010 & Click2Annotate~\cite{chen2010click2annotate} & \na & \na & \aic \humano & \na \\
        2010 & Touch2Annotate~\cite{chen2010touch2annotate} & \na & \na & \aic \humano & \na \\
        2018 & Voder~\cite{srinivasan2018augmenting} & \na & \na & \aic \humano & \na \\
        2019 & Text-to-Viz~\cite{cui2019text} & \na & \na & \aic \humano & \na \\
        2021 & InfoMotion~\cite{wang2021animated} & \na & \na & \aic \humano & \na \\
        2022 & Infographics Wizard~\cite{tyagi2022infographics} & \na & \na & \aic \humano & \na \\
        2019 & Fu~\etal~\cite{fu2019visualization} & \na & \na & \air & \na \\
        2022 & Fan~\etal~\cite{fan2022annotating} & \na & \na & \air & \na \\
        2022 & Liu~\etal~\cite{liu2022image} & \na & \na & \aia & \na \\ \midrule
        2014 & Ellipsis~\cite{satyanarayan2014authoring} & \na & \humanc & \humanc & \na \\
        2016 & DataClips~\cite{amini2016authoring} & \na & \humanc & \humanc & \na \\
        2021 & Idyll Studio~\cite{conlen2021idyll} & \na& \humanc & \humanc & \na \\
        2021 & VizFlow~\cite{sultanum2021leveraging} & \na& \humanc & \humanc \aia & \na \\
        2021 & Kori~\cite{latif2021kori} & \na& \humanc & \humanc \aia & \na \\
        2021 & Data Animator~\cite{thompson2021data} & \na& \humanc & \humanc \aia & \na \\
        2023 & GeoCamera~\cite{li2023geocamera} & \na& \humanc & \humanc \aia & \na \\
        2021 & VisCommentator~\cite{chen2021augmenting} & \na& \humanc & \aic \humano & \na \\
        2022 & SmartShots~\cite{tang2022smartshots} & \na& \humanc & \aic \humano & \na \\
        2022 & Sporthesia~\cite{chen2022sporthesia} & \na& \humanc & \aic \humano & \na \\
        2023 & DataParticles~\cite{cao2023dataparticles} & \na& \humanc & \aic \humano & \na \\
        2021 & Gemini$^2$~\cite{kim2021gemini} & \na& \humanc \aia & \humanc \aia & \na \\
        2020 & Gravity~\cite{obie2020authoring} & \na& \humanc \aio & \humanc & \humanc \\
        2022 & Gravity++~\cite{obie2022gravity++} & \na& \humanc \aio & \humanc & \na \\
        2021 & ChartStory~\cite{zhao2021chartstory} & \na& \aic \humano & \aic \humano & \na \\\midrule
        2016 & CLUE~\cite{gratzl2016visual} & \humanc & \humanc & \humanc & \humanc \\
        2019 & InsideInsights~\cite{mathisen2019insideinsights} & \humanc & \humanc & \humanc & \humanc \\
        2021 & ToonNote~\cite{kang2021toonnote} & \humanc & \humanc & \humanc & \na \\
        2018 & InfoNice~\cite{wang2018infonice} & \humanc & \na & \humanc & \na \\
        2023 & Slide4N~\cite{wang2023slide4n} & \humanc & \humanc \aia & \aic \humano & \na \\
        2022 & NB2Slides~\cite{zheng2022telling} & \humanc & \aic & \aic \humano & \na \\ 
        2019 & DataToon~\cite{kim2019datatoon} & \humanc \aia & \humanc &  \humanc \aia & \na \\
        2022 & Erato~\cite{sun2022erato} & \humanc \aia & \humanc \aia & \aic \humano & \na \\ 
        2023 & Notable~\cite{li2023notable} & \humanc \aia & \aic \humano & \aic \humano & \na \\
        2020 & Brexble~\cite{chotisarn2021bubble} & \aic & \na & \aic \humana & \na \\
        2019 & DataShot~\cite{wang2019datashot} & \aic & \aic \humano & \aic \humano & \na \\
        2016 & TSI~\cite{bryan2016temporal} & \aic \humana & \aic \humano & \aic \humano & \na \\
        2020 & Calliope~\cite{shi2020calliope} & \aic \humano & \aic \humano & \aic \humano & \na \\
        2020 & data2video~\cite{lu2020illustrating} & \aic \humano & \aic \humano & \aic \humana & \na \\
        2021 & Lu~\etal~\cite{lu2021automatic} & \aic \humano & \aic \humano & \aic \humano & \na \\
        2022 & Roslingifier~\cite{shin2022roslingifier} & \aic \humano & \aic \humano & \aic \humano & \na \\\midrule
        2013 & SketchStory~\cite{lee2013sketchstory} & \na & \na & \na & \humanc \aia \\
        2022 & Hall~\etal~\cite{hall2022augmented} & \na & \na & \na & \humanc \aia \\
        \bottomrule
    \end{tabular}
    \label{tab:tool_list}
\end{table*}

\endgroup

\subsection{Group 2: Tools for the Planning Stage and the Implementation Stage} \label{sec:group_2}

{Multiple tools in the previous group primarily focused on individual story pieces.}
Therefore, they do not necessarily provide functions for users to organize the story pieces.
For data stories with multiple pieces, such as data articles and data videos, creators often plan them ahead of implementation, according to Lee~\etal~\cite{lee2015more}.
The second group of 15 tools integrates the planning stage and the implementation stage for data storytelling.

{Several tools under the second category facilitate \textbf{animated chart or data video} creation.}
{In the planning stage, common tasks include arranging the sequence of keyframes in data videos~\cite{amini2016authoring, thompson2021data, li2023geocamera, tang2022smartshots, cao2023dataparticles} or picking key events in sports videos to add overlay visualizations~\cite{chen2021augmenting, chen2022sporthesia}.
The majority of tools in this group rely on \humanc to conduct the tasks in the planning stage.}
Gemini$^2$ is the only tool that applies \aias in the planning stage.
It generates intermediate charts between the start and end charts in an animated transition using GraphScape~\cite{kim2017graphscape} to minimize the cognitive load to understand the animation.
In the implementation stage, \aias and \aics are frequently applied to eliminate humans' workload, similar to the tools introduced in Section~\ref{sec:group_1}.
For example, 
DataParticles~\cite{cao2023dataparticles} parses the narration of data stories and then generates animation to illustrate the derivation of related data findings.
Besides automatic parsing, it allows \humanos to modify the animation manually, including the data and animation effect.

{Besides data videos, \textbf{data article} combines texts and multiple charts to convey data-related information.
Idyll Studio~\cite{conlen2021idyll}, Kori~\cite{latif2021kori}, and VizFlow~\cite{sultanum2021leveraging} are designed to accommodate data article authoring.
The main task of authoring data articles in the planning stage is to arrange story pieces into a coherent script.
In the implementation stage, authors often build visualizations and link visual elements to texts.
In Idyll Studio, \humancsnb write texts to organize all story pieces and prepare interactive charts in articles.
Kori and VizFlow enhance text-chart linking in the implementation stage with \aiasnb.
Kori recommends references between texts and charts.
VizFlow suggests visual elements to be annotated, such as sectors in pie charts.
}

{\textbf{Slideshow} is one of the most frequently used types of data stories~\cite{hullman2013deeper}, where multiple charts are shown sequentially to form a story.}
{It requires users to create the sequence of charts and then supplement the charts with text narrations and visual embellishment.}
Gravity~\cite{obie2020authoring} and its successor, Gravity++~\cite{obie2022gravity++}, are designed for authoring presentation slides.
They adopt \aios to optimize the human-created story piece sequence with GraphScape~\cite{kim2017graphscape}.

{
This group also has a tool for authoring \textbf{data comics}. The tasks to author data comics include organizing layouts and preparing charts with visual embellishments and narrations.
To author data comics, 
ChartStory~\cite{zhao2021chartstory} has \aics in the planning stage.}
The \aicnb in ChartStory organizes all story pieces into data comics with a minimum spanning tree.
It is the only tool to apply \aicsnb to plan data stories in this cluster.

\subsection{Group 3: Tools for the Analysis Stage to the Implementation Stage} \label{sec:group_3}

{According to Lee~\etal~\cite{lee2015more} and Chevalier~\etal~\cite{chevalier2018analysis}}, 
authors need to iterate between analysis and the other stages in the data storytelling workflow.
To facilitate such back and forth between stages, this group of 16 tools extends their covered stages to the analysis stage.
{In the analysis stage, the task of authors is to identify key data findings to be introduced in data stories from input datasets. They annotate charts or write documentations to record these findings.}

{To author data stories for analytical results, early tools mainly attempt to integrate the functionalities from analysis to implementation in the same tool without any AI support.
They save \humancs' efforts of transferring data and charts between analysis and other stages.
For example, built in Power BI, InfoNice~\cite{wang2018infonice} provides a convenient approach to customizing statistical charts into \textbf{infographics}.
Similar to DDG~\cite{kim2016data} and DataInk~\cite{xia2018dataink}, InfoNice is for individual infographic authoring.
As a result, it only supports the analysis and the implementation stage but does not support the planning stage.}

{Compared to infographics, \textbf{slideshows} and \textbf{data posters} are composed of multiple story pieces.
Therefore, the planning stage is covered by tools for these formats.
The difference between them only exists in how the story pieces are organized in the planning stage.
Slideshows focus on the sequences, while data posters emphasize the layout.
}
To create slideshows, CLUE~\cite{gratzl2016visual},
allows users to annotate explored charts and organize the sequence of charts for presentation.
Considering the long workflow, automatic end-to-end approaches were proposed to generate stories from data with \aics.
Inspired by data fact mining algorithms~\cite{tang2017extracting, ding2019quickinsights}, 
a series of data story generation approaches was proposed.
{They leverage the definition of data facts, 
to standardize the representation of data findings.}
\textit{The unified representation facilitates communicating data findings with AI and thus
enables the usage of} \aics \textit{in the first three stages of data storytelling.}
For example, DataShot~\cite{wang2019datashot} first mines all potentially interesting data facts.
Then it organizes data facts with the same focus or in the same subspace into multiple posters.
The charts and text descriptions are also generated accordingly.
Shi~\etal~\cite{shi2020calliope} and Lu~\etal~\cite{lu2021automatic} follow a similar pipeline to generate slides or scrollytelling stories.
More recently, powered by pre-trained language models, \aicsnb in NB2Slides~\cite{zheng2022telling} and Slide4N~\cite{wang2023slide4n} produce presentation slides by understanding code and documentation in notebooks.
Since these tools generate complete data stories, they often allow \humanos to optimize the output with user interfaces.

Though saving manual effort, an outstanding drawback of these end-to-end approaches is neglecting humans' intent in data analysis and follow-up communication.
{To handle the challenge, the most recent studies, Notable~\cite{li2023notable} and Erato~\cite{sun2022erato}, enhance humans' control in the analysis and planning stages with different strategies.
Notable allows \humancs to conduct data analysis in computational notebooks and applies an \aia to illustrate potentially interesting data facts for users' selection to record findings.
Then \aicnb plans the sequence of data facts and generates corresponding slides for communicating the data stories.
The strategy of Erato~\cite{sun2022erato} is to let \humancsnb explore, select, and order the key data facts in data stories in the analysis and planning stages.
Then \aianb can complete the sequence with additional data facts to make the story more coherent and smooth.}
From the tools introduced above, we observe a path of developing data storytelling tools for stages from analysis to implementation.
\textit{The tools have evolved from fully \humancnb tools, through fully \aicnb tools, and to mix-initiative tools with both \aicsnb and \humancsnb.}

{Besides the aforementioned data story formats, authoring \textbf{data comics} has also been explored. 
DataToon~\cite{kim2019datatoon} provides a platform to explore network data and author data comics to illustrate insights.
It leverages AI to suggest potential patterns in network data and add transitions to assist humans.
ToonNote~\cite{kang2021toonnote} facilitates presenting charts in notebooks as data comics to help readers focus on the analytical results rather than code.}

{Data2video~\cite{lu2020illustrating} automatically generates \textbf{data videos} for temporal data.}
A noteworthy characteristic of data2video is that it asks \humanas to annotate charts in data stories.
Besides annotation, another example task of \humanasnb is writing narrations for generated data stories~\cite{chotisarn2021bubble}.

\begin{figure}
    \centering
    \includegraphics[width=0.9\linewidth]{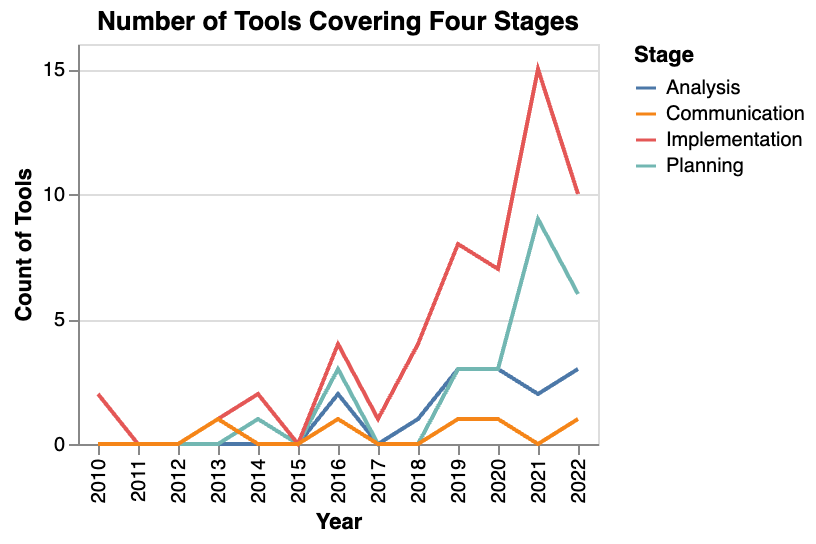}
    \vspace{-1em}
    \caption{This figure shows the number of data storytelling tools covering four stages between 2010 and 2022.}
    \Description{This figure includes a line chart showing the number of tools in all stages, i.e., analysis, planning, implementation, and communication, between 2010 and 2022. The implementation and planning stages have always been the researchers’ focus. The analysis and communication stages have not been widely considered in the past years.}
    \vspace{-1em}
    \label{fig:development_stages}
\end{figure}

\subsection{Group 4: Tools for the Communication Stage} \label{sec:group_4}

The last group of two tools applies \aias in the communication stage solely.
With a set of infographics as input, SketchStory~\cite{lee2013sketchstory} allows users to present a data story more engagingly with free-form sketching.
It recognizes users' sketches to auto-complete charts and filter data in charts.
Hall~\etal~\cite{hall2022augmented} supports mid-air gestures to help present charts in virtual meetings.
They utilize AI to recognize users' gestures and then update the chart, such as highlighting data elements or zooming in to a subset of data.

\subsection{Stages}\label{sec:results-analysis}
The previous sections illustrate an overview of surveyed tools in four clusters.
This section takes one step further to examine how the four stages, analysis, planning, implementation, and communication, are covered by those tools. 
Figure~\ref{fig:development_stages} presents the tools’ coverage of four stages over the years.
The implementation and planning stages have always been the researchers' focus.
The analysis and communication stages are not supported in most tools.

According to Table~\ref{tab:tool_list}, the \textbf{analysis} stage is covered by 16 among 60 tools in our corpus, in other words, 26.7\% of all tools.
Such results support the interview results by Li~\etal~\cite{li2023notable}, where data workers need to switch between tools for communicating data findings.
One potential reason for not supporting analysis in data storytelling tools is that building a fully functional data analysis module in data storytelling tools can involve great engineering effort and require users to adapt to a new environment.
To mitigate the gap between analysis and other stages, a potential opportunity is to integrate data storytelling tools with widely adopted data analysis platforms, such as BI tools~\cite{wang2018infonice} and computational notebooks~\cite{li2023notable, kang2021toonnote, zheng2022telling, wang2023slide4n}.

{We further notice that \textit{the coverage of other stages can be directly related to the data story format supported by tools}.
Some data story formats, such as videos, articles, and posters, can have multiple visualizations.
To facilitate authoring data stories in these formats, the tools need to support the organization of these visualizations in the  \textbf{planning} stage.
Some examples include DataClips~\cite{amini2016authoring}, Idyll Studio~\cite{conlen2021idyll}, and Data Animator~\cite{thompson2021data}.}

{The data story formats also affect whether tools are designed to facilitate direct \textbf{communication} between storytellers and the audiences.
For example, infographics and annotated charts are often used as part of data stories, such as presentation slides.
Therefore, the tools for these formats (\eg, Text-to-Viz~\cite{cui2019text} and InfoMotion~\cite{wang2021animated}) often do not cover the communication stages.
Furthermore, as discussed in Section~\ref{sec:tax-stages}, some data story formats, including data videos and posters can be directly presented to the audiences or only be viewed by the audiences asynchronously.
Therefore, these tools often do not have an explicit module for the communication stage.
On the contrary, slideshows are often used for direct communication with the audience.
As a result, several tools for slideshow authoring have embedded presentation modes, such as Gravity~\cite{obie2020authoring} and CLUE~\cite{gratzl2016visual}.
}

\begin{figure}[h!]
    \centering
    \includegraphics[width=\linewidth]{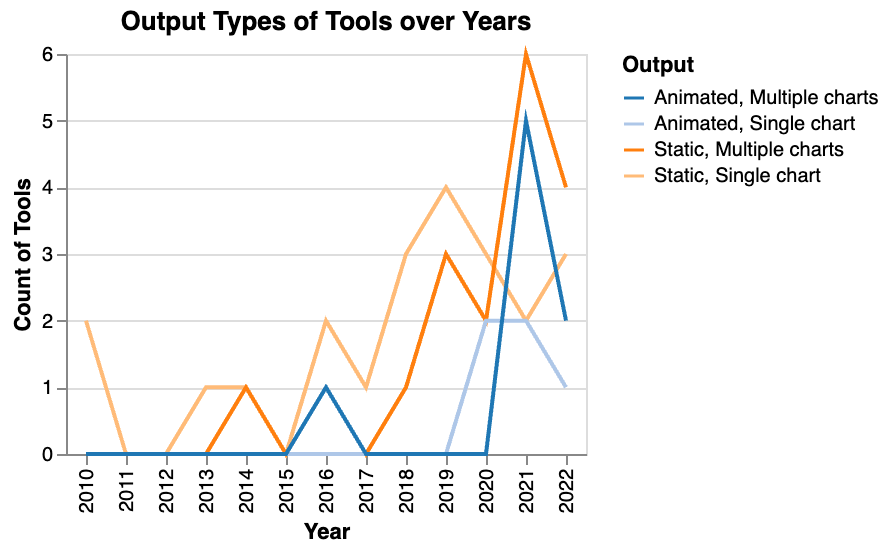}
    \caption{The figure shows the output of tools that cover the implementation stage between 2010 and 2022. The color hue encodes whether the output is animated or static. The color saturation represents whether the output includes single or multiple charts. The paper by Fu~\etal~\cite{fu2019visualization} is excluded since it outputs the scores and ranks of visualization memorability and aesthetics. }
    \Description{This figure includes a line chart showing the number of tools with four output types, i.e., animated multiple charts, animated single chart, static multiple charts, and static single chart, between 2010 and 2022 in the implementation stage. The chart shows that the number of tools for multiple charts has an increasing trend and has been greater than tools for single charts since 2021. The number of tools for animated charts also increases.}
    \label{fig:development_output}
\end{figure}

{The \textbf{implementation} stage is covered by almost all tools (58 out of 60).
We leverage the tools for the implementation stages as a lens to observe the change of data story formats supported by tools in Figure~\ref{fig:development_output}.
The chart shows that the tools are from building individual story pieces, such as single infographics, to more complex data story genres with multiple charts, \eg, data videos.
The chart also demonstrates that animated charts have gained growing interest since 2020.
Moreover, tools for animated multiple charts experienced an explosion in 2021.}

\section{Lessons Learnt about Human-AI Collaboration Patterns}\label{sec:lesson}
\begin{figure*}
        \centering
        \begin{subfigure}[b]{0.24\linewidth}
            \centering
            \includegraphics[width=\textwidth]{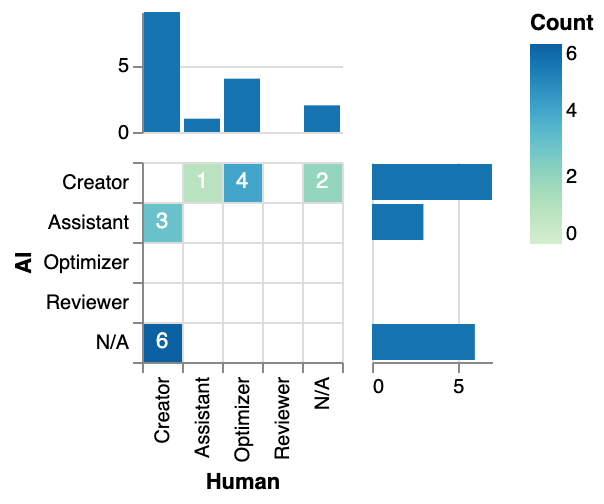}
            \caption{Analysis}  
        \end{subfigure}
        \begin{subfigure}[b]{0.24\linewidth}  
            \centering 
            \includegraphics[width=\textwidth]{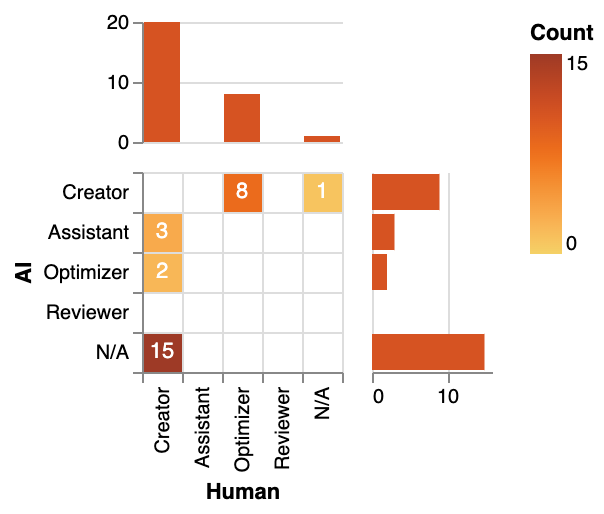}
            \caption{Planning}    
        \end{subfigure}
        \begin{subfigure}[b]{0.24\linewidth}   
            \centering 
            \includegraphics[width=\textwidth]{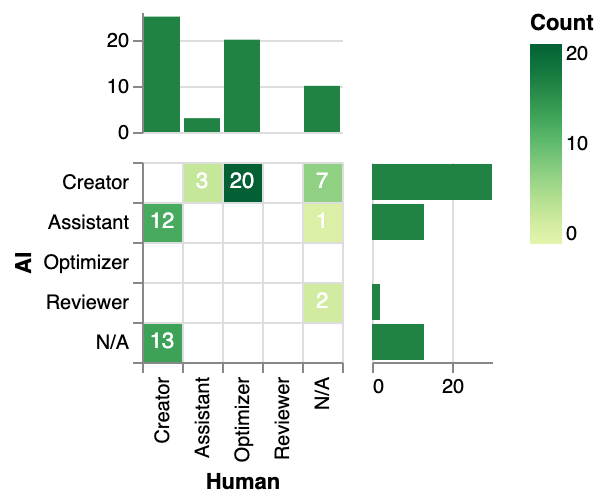}
            \caption{Implementation}    
        \end{subfigure}
        \begin{subfigure}[b]{0.24\linewidth}   
            \centering 
            \includegraphics[width=\textwidth]{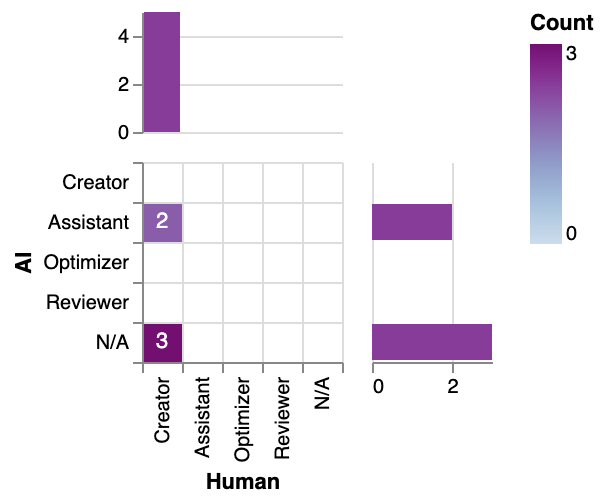}
            \caption{Communication}  
        \end{subfigure}
        \caption{This figure shows human-AI collaboration patterns in the analysis, planning, implementation, and communication stages. In all four charts, the horizontal axes represent humans' roles, and the vertical axes encode AI's roles. The colors in each chart show the frequency of collaboration patterns among all tools.}
        \Description{This figure has four subfigures to show the distribution of humans' and AI's roles in all four stages with heatmaps. The analysis and planning stages are on the top row, while the other two are on the bottom row from left to right. The figures show an imbalanced distribution of collaboration patterns, where human-creator only, AI-creator and human-optimizer, human-creator and AI-assistant, and AI-creator only are the most frequent patterns.}
        \label{fig:stage_distribution}
    \end{figure*}

This section reflects how previous research applies human-AI collaboration in data storytelling tools.
Figure~\ref{fig:stage_distribution} summarizes the human-AI collaboration patterns in four stages.

\subsection{In Individual Stages}\label{sec:lesson_individual}
We first focus on the lessons we learned from human-AI collaboration in individual stages.

\medskip
\noindent {\textbf{Analysis.}}
We realize that {\textit{data fact recommendation algorithms~\cite{tang2017extracting, ding2019quickinsights} are frequently used as} \aics \textit{or} \aias \textit{in the analysis stage}} (\eg, DataShot~\cite{wang2019datashot}, Erato~\cite{sun2022erato}).
Their advantages include mining various data findings in tabular data automatically and providing a unified representation of data findings among humans and AI~\cite{heer2019agency}.
One potential drawback is that they oversimplify the insights in data analysis, such as neglecting domain knowledge~\cite{battle2023exactly}.
We hope that there can be more research to further push the frontier of data analysis by \aicsnb.

\medskip
\noindent {\textbf{Planning.}}
{In the planning stage, we recognize that the usage of \aics is less frequent than it in the implementation stage.
The finding aligns with the results in a previous interview~\cite{li2023ai}, where data workers do not prefer AI to automate the planning stage.
A potential explanation for this phenomenon is that the way of planning data stories is directly related to the authors' intent.
Therefore, \textit{humans would like to take more control in the planning stage}.
Furthermore, planning data stories often \textit{requires understanding and designing narrative structures}, such as ``Martini Glass Structure'' and ``Drill-Down Story''~\cite{segel2010narrative}, which is challenging for AI~\cite{yang2021design}.
Another potential reason from the technical side is \textit{the challenge for \aicsnb to understand humans' intent and data findings}.
Humans' intent and findings in the data analysis may not be expressed in a standardized way~\cite{battle2022programmatic}.
For example, data findings can be expressed with arbitrary charts or texts.
It leads to the difficulties of AI's understanding.}

In the existing \aicsnb, almost all focus on planning the data story {\textit{leveraging the data relationships between story pieces}, such as placing data facts in the same subspace together~\cite{wang2019datashot, shi2020calliope}.
They often apply heuristics-based approaches to search for an optimal organization of story pieces 
based on various criteria, such as consistency~\cite{kim2017graphscape, hullman2013deeper} and logical coherence~\cite{wolf2005representing} among multiple story pieces.
However, in practice, it \textit{may not be sufficient to rely on the data relationships}}~\cite{li2023notable}.
It is essential to consider the context, users' preferences, and other semantic information when planning the data story~\cite{yang2021design}.
Towards this end, NB2Slides~\cite{zheng2022telling} summarizes the typical templates for communication between data scientists, but using fixed templates limits its flexibility.
The other approach to mitigate the problem is to leverage \humanos to further customize the data story outline.
In the future, it is possible to explore how to ask \humanas to provide background information, such as the target audiences or data analysis project information~\cite{li2023ai}, for \aicsnb to plan the data story in a context-aware manner.

On the other hand, when \humancs plan the story manually in the rest of the tools, their context and target can be considered sufficiently but the workload is high.
To reduce users' burden, \aios and \aias are applied by multiple tools, including Gravity~\cite{obie2020authoring} and Gemini$^2$~\cite{kim2021gemini}.
However, similar to previously discussed \aicsnb, {\textit{most of these \aiosnb and \aiasnb still rely on data relationships between story pieces}}.
It is possible that such AI can complement users' planning since it is expected to explore the potential organization of stories based on data relationships more thoroughly and quickly.
At the same time, there is also {\textit{a risk of breaking users' logical flows of data stories}}.
We argue that future research should examine the issue and take action to mitigate the potential problems, such as restricting AI's optimization on users' outlines.

\medskip
\noindent {\textbf{Implementation.}}
Implementation is the stage that {\textit{involves the highest proportion of AI}} in the three stages other than communication.
We exclude the communication stage from the comparison here due to the small sample size.
About 77.6\% (45 out of 58) of all tools introduce different forms of AI collaborators in the implementation stage.
For comparison, the analysis and the planning stages have 62.5\% (10 out of 16) and 48.3\% (14 out of 29) tools that apply AI.
This result {\textit{aligns with users' expectations to reduce workload in the implementation stage}}~\cite{li2023ai}.
They hoped AI could automatically generate charts or styling data stories.

\medskip
\noindent {\textbf{Communication.}}
In our corpus, 
two tools incorporate \aia for enhancing the presentation with rich interactions, including sketching~\cite{lee2013sketchstory} and gesture~\cite{hall2022augmented}.
According to a previous study, the usage of AI in communication among humans may lead to some concerns~\cite{li2023ai}.
The data workers believed that AI may have a negative effect on building relationships between presenters and audiences.
That might explain why \aicsnb are rarely researched.
{\textit{More future research is desired to understand the perception of AI when communicating data stories.}}
Furthermore, we would like to {\textit{encourage future investigations in diverse usages of AI in the stage}}.
For example, we may introduce \airsnb to assess humans' performance in communicating data stories.
Though previous research has explored visual analytics for improving presentation skills~\cite{wang2020voicecoach} and guidelines for presenting visualizations~\cite{yang2021explaining}, we have not noticed any tools for assessing or improving skills for communicating data stories.

\subsection{Across All Stages}\label{sec:lesson_across}
Figure~\ref{fig:stage_distribution} shows an imbalanced distribution of the frequencies of human-AI collaboration patterns.
Several patterns have been frequently applied, while the others are rarely explored.
Therefore, we believe it is essential to blur the boundary of different stages and examine representative patterns across stages for insights into design strategies of human-AI collaboration in storytelling tools.

\medskip
\noindent {\textbf{What are the most common human-AI collaboration patterns?}}
To identify the most common collaboration patterns, we first exclude the patterns where only humans or AI perform in individual stages due to the lack of collaboration.
Among the remaining 58 collaboration cases, we notice {\textit{that the most frequent collaboration patterns where humans and AI work together are} \humanc + \aia \footnotemark\, \textit{(20 times) and} \aic + \humano \textit{(32 times)}}.
We consider that both collaboration\footnotetext{We use the plus sign \textit{``+''} to represent the collaboration between two roles in one stage for simplicity.} patterns have their pros and cons.

\medskip

\paragraph{\Humanc + \aia.} When \humancsnb take control of one stage, they are able to {\textit{perform more free-form actions following their ideas}}.
For example, they can select any code in notebooks to organize the data story~\cite{wang2023slide4n} or pick desired glyphs for infographics in the implementation stage~\cite{zhang2020dataquilt}.
One predictable cost of the control is more manual effort from \humancsnb.
Besides, {\textit{the requirement for \aianb is also higher since they need to understand the humans' input and then take corresponding actions}}.
For example, Vistylist~\cite{shi2022supporting} leverages multiple deep learning-based vision models to extract infographic styles, such as fonts, colors, and icons.
To mitigate the problem, one common strategy in prototype tools is to limit \humancsnb' input, such as limiting chart types~\cite{sultanum2021leveraging, cui2021mixed}, restricting the input formats~\cite{li2023notable, sun2022erato}.

\medskip

\paragraph{\aic + \Humano.} When AI takes the creator role, humans' efforts can be significantly eliminated initially.
However, {\textit{the flexibility of results is greatly limited by the design of AI}}.
For example, as previously mentioned, DataShot~\cite{wang2019datashot} can only produce simple data facts pre-defined by tool developers.
Text-to-Viz~\cite{cui2019text} is limited to generating proportion-related charts.
Another issue is that \aicsnb can hardly consistently achieve satisfactory performance.
It may take considerable effort to optimize AI's output.

\paragraph{{Common characteristic}} Considering their pros and cons, there can be a {\textit{trade-off in selecting the two collaboration patterns}} when designing AI-powered data storytelling tools.
While these two patterns have pros and cons, one common characteristic of them is that {\textit{humans have sufficient agency in the overall data story creation process}}.
When humans serve as creators, they can directly decide how the story is told.
If humans are optimizers, it is possible for them to modify stories directly when \aicsnb' stories are beyond their expectations.
Such observation indicates that the tool designers often follow the common principle of enabling humans' control over AI-powered tools~\cite{amershi2019guidelines, horvitz1999principles}.

\medskip
\noindent {\textbf{What do assistants frequently perform?}}
In the previous discussion, we noticed \aiasnb commonly team up with \humancsnb.
Following prior research~\cite{li2023ai}, assistants are defined as collaborators who take supplementary tasks for leading creators during story creation.
Compared to other roles, they have a broad spectrum of usage. 
In this section, we examine the usage of \aias and \humanas to clearly understand the assistant role.

\paragraph{{The first type of assistant performs individual tasks that creators delegate to them.}}
When collaborating with these \aiasnb, \humancsnb can focus on the main tasks and save efforts in these auxiliary tasks.
For example, \humancsnb can create charts and ask \aiasnb to add styles to them~\cite{shi2022supporting, cui2021mixed}.
When \humancsnb specify keyframes in animations, \aiasnb can automatically create transitions between them~\cite{li2023geocamera, kim2021gemini}.
Similarly, \humanasnb write text descriptions for charts created by \aicsnb~\cite{chotisarn2021bubble} or add annotations~\cite{lu2020illustrating}.

\paragraph{{Another type of assistant provides suggestions for a task that creators work on.}}
For example, in the analysis stage, \aiasnb recommend interesting data facts~\cite{sun2022erato, li2023notable}.
In the planning stage, they can fill story pieces into \humancsnb' sequences to make them more coherent~\cite{kim2021gemini, sun2022erato}.
{In the implementation stage, \aiasnb in VizFlow~\cite{sultanum2021leveraging} and CAST~\cite{ge2021cast} assist in suggesting annotations and grouping similar visual elements for similar transitions.}

\medskip
\noindent{\textbf{What and why do some roles lack investigation in previous research?}}
We have discussed the common human-AI collaboration patterns previously.
Now we introduce some rarely explored collaboration patterns.

\medskip

\paragraph{\Humanc + \aio.} As introduced before, the two most common collaboration patterns are \humancsnb + \aiasnb and \aicsnb + \humanosnb.
However, it is interesting that these roles are seldom swapped between humans and AI, \ie, \humancsnb + \aiosnb and \aicsnb + \humanasnb.
Between them, \humancsnb + \aiosnb occurs less frequently only in two serial tools~\cite{obie2020authoring, obie2022gravity++} while \aiosnb are desired according to previous research~\cite{li2023ai}.
Therefore, we would like to discuss why \aiosnb may not be widely applied in existing data storytelling tools.
When \aiosnb collaborates with \humancsnb, \aiosnb act after humans' creation.
Therefore, if humans and AI have different ways of understanding the same data story, {\textit{one potential problem of letting AI optimize \humancsnb' design is to change humans' purposeful design along an undesired direction}}.
For example, \aiosnb may improve the story organization using data relationships while ignoring humans' original logic flow (see Section~\ref{sec:lesson_individual}).
What may make it even worse is an observation that there is a lack of effective communication channels between humans and AI~\cite{li2023ai}.
It may increase the risk of misunderstanding between AI and humans.

Based on these observations, we have several suggestions for applying \aiosnb with \humancsnb in data storytelling tools.
First, when applying \aiosnb, it is critical to {\textit{design appropriate strategies to incorporate humans' opinions for their control of stories}}.
For example, \aiosnb should try to follow humans' designs and sync with humans before taking action.
Second, {\textit{the usage of \aiosnb may depend on the tasks in the data storytelling workflow}}.
In our opinion, it might not be ideal to apply \aiosnb for stages where understanding data story background may affect the results considerably,
such as organizing the story.
On the other hand, \aiosnb have some opportunities to perform background-agnostic tasks, such as optimizing the layout or color usage.
Finally, {\textit{a potential alternative to \aiosnb can be \airsnb}}.
\airsnb can provide feedback on humans' designs but do not directly modify them.
In this way, humans can learn AI's suggestions and take action while considering their needs sufficiently.
\medskip

\paragraph{\air.} In our corpus, the usage of \airsnb is limited to two tools~\cite{fu2019visualization, fan2022annotating} that review individual charts' visual appearance and potential misinformation in the execution stage.
We suspect that {\textit{the reasons why \airsnb are not frequently explored correspond to the characteristics of data stories}}.
First, data stories often {\textit{involve multiple channels}}, including visualizations, animations, and narrations.
To comprehensively evaluate them, it is essential to understand all channels and the interplay between them~\cite{cheng2022investigating,wang2023wonderflow}, which poses the challenges of considering multi-modal inputs to AI systems.
Second, data stories serve as {\textit{a medium for communication between authors and audiences}}.
Therefore, when reviewing data stories, a challenge is to consider audiences' preferences and the information gap between authors and audiences~\cite{mao2019data,hou2017hacking}.
To address the challenge, recent large language models (LLMs) demonstrate their potential.
Previous research~\cite{park2023generative} reveals that LLMs can play different roles in conversations.
They may be extended to provide feedback from various audiences' perspectives for data stories.

\medskip

\paragraph{\Humanr.} Another role we would like to discuss is \humanrnb.
According to the definition, \humanrsnb provide suggestions to \aicsnb about their created content.
Then \aicsnb may update the content accordingly and ask for further human reviews.
Compared to \humanosnb, it allows users to provide feedback rather than considering and implementing solutions directly.
Such workflow {\textit{requires AI to have the ability to understand humans' feedback and perform corresponding actions iteratively}}, which may be beyond AI's ability in existing tools.
According to our observation, existing tools often apply AI to conduct given one-shot tasks, such as recommending potential animation~\cite{ge2021cast} or identifying data facts~\cite{wang2019datashot}.
These AI approaches are specialized at one assigned task and lack the ability to handle users' reviews.
Another reason is that {\textit{\humanosnb can be more efficient than \humanrsnb when handling minor tasks to be revised}}.
For example, a common task in designing infographics is to arrange the positions of the visual elements.
When AI moves a visual element to an undesired position, it might be more convenient and predictable to directly manipulate its position rather than providing feedback to AI for adjustment, as indicated by a prior discussion~\cite{shneiderman1997direct}.

\section{Suggestions and Opportunities in Human-AI Collaborative Data Storytelling}\label{sec:suggestion}

In Section~\ref{sec:taxonomy}, we introduce a framework to characterize data storytelling tools from the perspective of human-AI collaboration.
We first demonstrate how the framework can be applied to compare and group existing human-AI collaborative storytelling tools and discuss their development in Section~\ref{sec:results}.
Section~\ref{sec:lesson} further illustrates their design patterns.
This section proposes the research opportunities revealed by analyzing existing studies with the framework.

\medskip
\noindent{\textbf{Investigate the strategies to maximize automation and agency in data storytelling tools comprehensively.}}
Heer pointed out that the ultimate goal of designing human-AI collaborative tools is maximizing AI automation and human agency concurrently~\cite{heer2019agency}.
We notice an interesting trend toward this goal when analyzing the tools in our corpus.
Figure~\ref{fig:development_collaboration} reveals that the development of data storytelling tools experiences two apparent phases.
Between 2016 and 2018, most tools were manually authoring tools where \humancs take the entire control of storytelling.
Since 2019, human-AI collaborative tools have occupied the largest proportion among all tools and shown an increasing trend, 
while the number of purely automatic tools has been relatively low and stable over the years.
As the trend indicates, {\textit{researchers have widely investigated human-AI collaborative ways to automate data storytelling with consideration of humans' control}}.
This aligns with the creative nature of authoring data stories well~\cite{thudt2018exploration}, where designing compelling stories should be seen as a creative and open-ended task.
Therefore, it is hard to define a clear goal for AI to accomplish automatically.
The creation of data stories still heavily relies on humans' experience and judgment.
As a result, it is crucial to respect humans' control over the entire data storytelling workflow while attempting to improve the automation level.

\paragraph{{Individual stages}} In Section~\ref{sec:lesson_across}, we mentioned that existing work often uses two collaboration patterns to guarantee humans' control when collaborating with AI in single stages, including asking humans to author the content with AI assistance (\humanc + \aia) or allowing humans to optimize AI-created content (\aic + \humano). 
However, the two aforementioned collaboration patterns have not been systematically investigated in existing tools to decide their performance in maximizing agency and automation.
We believe future research should take one step further to {\textit{carefully examine and compare these frequent human-AI collaboration patterns}} from multiple angles, such as users' perceptions and preferences for these patterns under different scenarios.
Ultimately, these results can be transformed into design guidelines for future tool developers' references.
Moreover, along with the development of AI, other currently under-explored human-AI collaboration patterns revealed by our review (see Section~\ref{sec:lesson_across}) can also be investigated for ensuring humans' control and improving automation.
For example, \humanrs can provide feedback for AI output and save the effort of direct modification as \humanosnb do.
\aios can automatically improve humans' created content with extensive consideration of their intent.

\paragraph{{Across multiple stages}} Beyond individual stages, a more interesting observation is about the tools that cover the analysis, planning, and implementation stages experienced three phases (see Section~\ref{sec:group_3}).
The earliest tools aim to bridge the analysis and follow-up stages to reduce efforts for transferring analytical results instead of automating data storytelling directly~\cite{gratzl2016visual, mathisen2019insideinsights}.
They are still manual storytelling tools.
Later, in the second phase, researchers propose tools to generate complete data stories from datasets directly, such as DataShot~\cite{wang2019datashot} and Calliope~\cite{shi2020calliope}.
Though they allow manual optimization, humans' intent is not considered in the initial story generation.
More recently, the development of these tools reached the third phase, where humans can control data story content and structure in the analysis and planning stages, and AI takes over the implementation stage~\cite{li2023notable, sun2022erato, wang2023slide4n}.
We realized that following the trend, {\textit{tools have been closer to the expectation of tool users, where the planning stage is not preferred to be fully automated}}~\cite{li2023ai}, demonstrating the development of understanding data storytellers in our community.
Furthermore, the observation reveals a potential strategy to maximize automation and agency in human-AI collaborative tools.
When the workflow involves multiple stages, like data storytelling, {\textit{it is possible to adopt different collaboration patterns in different stages according to the stage characteristics}}.
In this way, though not maximizing the automation and agency in a specific step, they can be maximized in the entire workflow.
More research should be conducted to investigate the design strategy.
For example, the relationship between collaboration patterns and stage nature is worth more investigation.

We hope our research community can continue the investigation and innovation of maximizing agency and automation in data storytelling.
We also believe the lessons from developing human-AI collaborative data storytelling tools can potentially inspire other related fields and contribute to the future symbiosis of humans and machines~\cite{licklider1960man}.

\medskip
\noindent{\textbf{Embrace the recent advances in AI systems.}}
The past year has witnessed the emergence of powerful generative AI systems, including LLMs (\eg, GPT-3.5\footnote{\url{https://platform.openai.com/docs/models/gpt-3-5}}, GPT-4~\cite{openai2023gpt4} and Llama2~\cite{touvron2023llama}) and image generative models (\eg, Midjourney\footnote{\url{https://www.midjourney.com/}} and Stable Diffusion~\cite{rombach2022high}).
Their appearance can boost human-AI collaboration in data storytelling tools.

\paragraph{{Potential usage}} The most straightforward approach to leverage them is to {\textit{conduct generative tasks with better performance}}.
For example, recent tools\footnote{These tools are not included in our corpus since they were not officially published before June 2023.} use GPT-3.5 to create text for charts~\cite{lin2023inksight, sultanum2023datatales, ying2023reviving} instead of using template-based methods~\cite{shi2020calliope, li2023notable}.
Generating infographics from data with LLM was also explored~\cite{dibia2023lida}.
Besides performance, another outstanding advantage of using them in data storytelling is {\textit{the low requirement for training data}}.
Lu~\etal~\cite{lu2021automatic} indicated that there were no existing corpora of data stories, which resulted in the challenges of training machine learning models.
Recent LLMs show great power in zero-shot or one-shot learning tasks~\cite{wei2021finetuned}, which releases the requirement of large-scale training datasets and demonstrates their potential for generative tasks in data storytelling.

Other than generative tasks, these AI systems also enable {\textit{a large space for imagining human-AI collaborative experiences}}.
For example, Data Player animates charts based on understanding accompanying narrations with GPT-3.5~\cite{shen2023data}.
Besides, we mentioned another potential case of leveraging LLMs to review data stories from the perspectives of audiences with diverse backgrounds in Section~\ref{sec:lesson_across}.
We also anticipate that LLMs' ability to have conversations with humans may also facilitate communication between \humanrsnb and \aicsnb.

\paragraph{{Underlying risk}} Though these AI systems can bring advantages, {\textit{the risk of applying them should also be aware, such as hallucination~\cite{lee2023benefits} and bias~\cite{li2021gender}}}.
These potential problems further enhance the need for human-AI collaboration.
For example, \humanrsnb can check if the created data stories involve fake numbers and ask AI to correct them.

\begin{figure*}
    \centering
    \includegraphics[width=\linewidth]{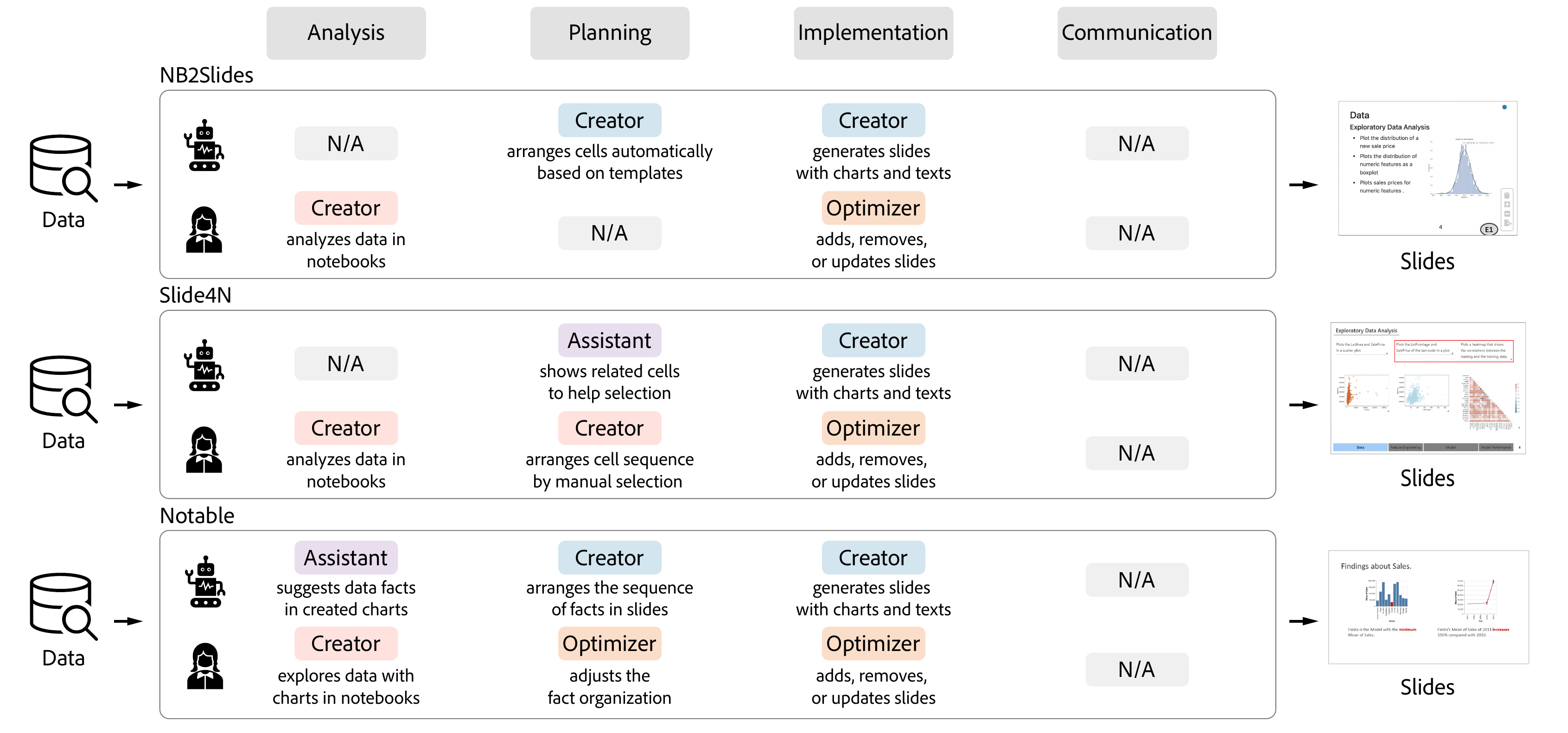}
    \caption{This figure shows the usage of our framework to compare three recent slide creation tools in computational notebooks, \ie, NB2Slides~\cite{zheng2022telling}, Slide4N~\cite{wang2023slide4n}, and Notable~\cite{li2023notable}. With our framework, their difference in human-AI collaboration is straightforward.}
    \Description{This figure shows the usage of our framework to compare three recent slide creation tools in computational notebooks, i.e., NB2Slides, Slide4N, and Notable. We code the humans' and AI's roles in all stages for these tools. NB2Slides has AI-creators in the planning and implementation stages, a human-creator in the analysis stage, and a human-optimizer in the implementation stage. Slide4N has an AI-assistant in the planning stage, an AI-creator in the implementation stage, human-creators in the analysis and planning stages, and a human-optimizer in the implementation stage. Finally, Notable applies an AI-assistant in the analysis stage, AI-creators in the planning and implementation stages, human-creator in the analysis stage, and human-optimizers in the planning and implementation stages.}
    \label{fig:comparison}
\end{figure*}

\medskip
\noindent{\textbf{Extend human-AI collaboration to multiple human and AI collaborators.}}
One observation we made during the research is that most of the surveyed storytelling tools are limited to the collaboration between one human and one AI system.
However, previous research notices that the workflow of communicating data analytical results may involve several human collaborators~\cite{lee2015more, chevalier2018analysis}, such as data analysts, story editors, and presenters.
To assist these human collaborators, different AI collaborators may also be required to facilitate their diverse backgrounds and requirements.
As a result, it is possible for future storytelling tools to {\textit{consider the collaboration among multiple stakeholders}}.
The previous research in collaborative data communication also implies the need for such tools~\cite{brehmer2021jam}.
We anticipate that the appearance of these tools has the potential to further expand the collaboration patterns identified in our paper (see Table~\ref{tab:tool_list} and Figure~\ref{fig:stage_distribution}).
{For example, we imagine that there might be a collaborative data storytelling tool that takes a slide deck as input.
Then, the slide deck can be collaboratively edited by an \aio and other \humanos, such as a product manager and a data analyst, with the tool.
To improve the slide deck, the \aionb and \humanosnb may leverage their advantages.
For example, \humanosnb can add more illustrations considering the audience's background, while the \aionb may enhance the styles.
Under such a scenario, the collaboration pattern of the tool is \humanosnb + \aionb in the implementation stage.
Our framework can also be slightly extended to distinguish the roles of different human or AI collaborators.
It is possible to distinguish them with suffixes.
For example, the data analyst (DA) can be denoted as human-optimizer-DA, and the product manager (PM) can be human-optimizer-PM.
Then, we can characterize the collaboration pattern in the implementation stage of the tool as human-optimizer-DA + human-optimizer-PM + AI-optimizer.
}

\medskip
\noindent \textbf{{Apply our human-AI collaboration framework to assess and express the similarities and differences between storytelling tools.}}
{Many recent data storytelling tools consider collaboration strategies between humans and AI explicitly as part of their research contribution, such as Erato~\cite{sun2022erato} and Slide4N~\cite{wang2023slide4n}.
However, unlike other well-established dimensions of data storytelling tools, such as story genres, it remains challenging for researchers to assess and present their new designs for human-AI collaboration in these storytelling tools.
As a result, the understanding of these tools' novelty and values may be hindered.
Our framework has the potential to address the challenge by 
providing a systematic way of characterizing human-AI collaboration in data storytelling tools.
For example, Slide4N~\cite{wang2023slide4n} and NB2Slides~\cite{zheng2022telling} are two recent tools for generating slides from computational notebooks.
As a later tool, Slide4N~\cite{wang2023slide4n} argues that adopting a human-AI collaborative approach is one of its major differences from NB2Slides~\cite{zheng2022telling}.
With our roles and stages, it is easy to notice that the difference between NB2Slides and Slide4N mainly lies in the planning stage, where Slide4N allows users to organize their story with AI assistance and NB2Slides entirely relies on AI (see Figure~\ref{fig:comparison}).
We further compare them with Notable~\cite{li2023notable}, another recent slide creation tool in notebooks, in Figure~\ref{fig:comparison}.
As the example shows, our framework can clearly expose the difference between data storytelling tools from the perspective of human-AI collaboration.
Therefore, we would like to suggest that {\textit{future researchers leverage our framework as a standard approach to demonstrate their novelty in designing human-AI collaborative storytelling tools}}.}

\section{Limitations and Future Work}
This section discusses the limitations and future directions to continue our research.

\subsection{Limitations}
{We notice three limitations in our research.}
First, \textit{the granularity of stages can be more fine-grained}.
In previous research~\cite{lee2015more, chevalier2016animations, li2023ai}, besides stages, they further identify the tasks in data storytelling workflow, such as styling data stories and answering audiences' questions.
{We did not code tools according to the human-AI collaboration patterns in these tasks since the tasks can depend on data story formats and thus may be too detailed.}
If using tasks, the clusters of tools may be less noticeable.
Furthermore, another concern is that some tasks may be mental tasks, such as connecting data findings logically~\cite{lee2015more}.
It can be hard to decide if data storytelling tools take care of these tasks.
Consequently, as a pioneer study towards human-AI collaboration in data storytelling, we provide an overview of these tools from a relatively high level, \ie, stages.
{We sincerely hope future research can further extend ours to fine-grained patterns using a systematic and data story format-agnostic definition of tasks.}

Second, \textit{our corpus of tools may not cover the latest data storytelling tools}.
As Section~\ref{sec:suggestion} indicates, the recent advance of AI has the potential to push the frontier of human-AI collaborative data storytelling tools.
Though our corpus of existing research tools may not cover the most recent ones due to the timing of the research, we believe our research can still be valuable to future tool designs with increasingly powerful AI systems. 
Our paper introduces a framework to characterize existing data storytelling tools and shows its application in comparing tools, understanding the footprints of the research community development, and identifying new research opportunities.
Future designers and researchers can follow the framework to draw inspiration, assess their tools, and compare them with prior arts.
We also plan to keep updating our online data storytelling tool browser at \textcolor{navy}{\url{https://human-ai-universe.github.io/data-storytelling}}.
It can serve as a sustainable source for the data storytelling community to reflect the progress of human-AI collaborative tools and ignite new ideas for more effective and efficient approaches.

{Third, \textit{the flexibility of our framework can be further enhanced.}
As introduced in Section~\ref{sec:tax-roles}, our framework characterizes the work distribution between humans and AI through their roles.
Though the dimension of roles reveals the common practices of human-AI collaboration patterns (see Section~\ref{sec:lesson}), it may not be flexible enough to delineate more complex collaboration patterns.
For example, our existing framework cannot reveal the transition of roles between humans and AI following specific sequences.
One potential direction is to enhance our framework with more dimensions to characterize these designs, such as a dimension to represent the role transition.
We hope the improvement in framework flexibility can be considered in future research in this line.}

\subsection{Future Work}
We further propose two directions to continue our research.
First, \textit{more design choices in human-AI collaborative data storytelling tools can be explored}, such as the communication approaches between human and AI collaborators and the detailed AI techniques.
Our paper outlines the humans' and AI's roles in human-AI collaborative data storytelling tools.
We notice that there can be diverse communication methods between humans and AI, such as programming~\cite{kim2021gemini, li2023notable}, sketching~\cite{lee2013sketchstory}, and gesture~\cite{hall2022augmented}.
Considering the page limit, we cannot include a comprehensive investigation into the communication channels in the current paper and hope to continue the research in this direction.
Furthermore, the detailed AI techniques may be worth more research.
Since data storytelling tools often cover multiple stages, they can apply several AI techniques.
A comprehensive review of AI algorithms for different stages and roles may provide a panorama of human-AI collaborative tool development from the technical perspective.
Considering our focus on roles in collaboration, we decided to blur the difference between AI techniques and leave their survey as future work.
{Lastly, an interesting future opportunity is to enhance the research into the relationship between the application of human-AI collaboration and narrative techniques, including visual narrative (\eg, the approaches of transition and highlighting) and narrative structures (\eg, monomyths)~\cite{segel2010narrative}.
Such research has the potential to guide the design of human-AI collaboration when the authors have different preferred approaches to tell data stories. }

Second, \textit{it is interesting to extend our research's scope from data storytelling tools to general visualization creation tools and creativity support tools}.
As introduced in Section~\ref{sec:methodology}, we excluded the visualization creation tools that are not specifically designed for storytelling in the paper collection phase and constrained the scope for a more focused study.
During our paper collection, we realized that existing visualization creation tools involve human-AI collaboration.
For example, VizLinter~\cite{chen2021vizlinter} applies \airsnb to identify problems in charts.
Voyager2~\cite{wongsuphasawat2017voyager} leverages \aiasnb to recommend charts based on users' intent in visual data exploration.
We hope to examine whether our framework of human and AI collaborators' roles can be applied to these visualization creation tools in the future.
Furthermore, creativity plays an important role in authoring compelling data stories~\cite{thudt2018exploration}.
Therefore, data storytelling tools are closely related to the field of creativity support.
We consider it interesting to extend our human-AI collaboration research to creativity support tools~\cite{shneiderman2007creativity}.

\section{Conclusion}
{In this paper, we systematically investigate existing human-AI collaborative data storytelling tools with our framework from two perspectives: their covered stages and the roles of humans and AI in collaboration.
Specifically, the stages include \textit{analysis}, \textit{planning}, \textit{implementation}, and \textit{communication}, and the roles of humans and AI are described by \textit{creator}, \textit{assistant}, \textit{optimizer}, and \textit{reviewer}.}

{Through reviewing these tools, we first identify that human-AI collaborative tools have \textit{received increasing research interest in the past years} and exceeded manual authoring tools since 2019.
Then we dig deeper into the design of human-AI collaborative tools and summarize the \textit{most frequent patterns of collaboration}, \ie, \humancnb + \aianb and \aicnb + \humanonb. 
We further compare their pros and cons and point out their similarity in guaranteeing humans' agency in collaborative task completion.
Besides, \textit{some roles that lack inspection are spotted and discussed}, such as \airsnb and \humanrsnb.
Based on our findings and lessons, we provide suggestions to the researchers in this field.
First, \textit{the approaches to enhance automation and agency} can be further explored and summarized to benefit future storytelling tool design.
Second, \textit{the recent advance of large-scale AI systems} has great potential in enhancing human-AI collaborative storytelling.
Moreover, it is worth more inspection to \textit{extend existing work for collaboration among multiple humans and AI collaborators.} 
Lastly, we also encourage future researchers to \textit{apply our framework in designing and presenting their human-AI collaboration strategies} in storytelling tools.}

In the future, we plan to keep track of the development of related tools and update our online browser of tools.
We hope to extend our research to other factors in human-AI collaborative data storytelling, such as communication methods between collaborators and the technical details of AI systems.
Moreover, the roles of humans and AI can be further studied and extended.
It will also be interesting to explore if our framework can be applied to other human-AI collaborative tools, \eg, visualization creation tools.


\begin{acks}
We would like to thank all reviewers and Leixian Shen for their constructive suggestions. 
This project is partially supported by HK RGC GRF grant 16210722.
\end{acks}

\balance

\bibliographystyle{ACM-Reference-Format}



\begin{thebibliography}{150}


\ifx \showCODEN    \undefined \def \showCODEN     #1{\unskip}     \fi
\ifx \showDOI      \undefined \def \showDOI       #1{#1}\fi
\ifx \showISBNx    \undefined \def \showISBNx     #1{\unskip}     \fi
\ifx \showISBNxiii \undefined \def \showISBNxiii  #1{\unskip}     \fi
\ifx \showISSN     \undefined \def \showISSN      #1{\unskip}     \fi
\ifx \showLCCN     \undefined \def \showLCCN      #1{\unskip}     \fi
\ifx \shownote     \undefined \def \shownote      #1{#1}          \fi
\ifx \showarticletitle \undefined \def \showarticletitle #1{#1}   \fi
\ifx \showURL      \undefined \def \showURL       {\relax}        \fi
\providecommand\bibfield[2]{#2}
\providecommand\bibinfo[2]{#2}
\providecommand\natexlab[1]{#1}
\providecommand\showeprint[2][]{arXiv:#2}

\bibitem[Akiba et~al\mbox{.}(2009)]%
        {akiba2009aniviz}
\bibfield{author}{\bibinfo{person}{Hiroshi Akiba}, \bibinfo{person}{Chaoli Wang}, {and} \bibinfo{person}{Kwan-Liu Ma}.} \bibinfo{year}{2009}\natexlab{}.
\newblock \showarticletitle{AniViz: A Template-Based Animation Tool for Volume Visualization}.
\newblock \bibinfo{journal}{\emph{IEEE Computer Graphics and Applications}} \bibinfo{volume}{30}, \bibinfo{number}{5} (\bibinfo{year}{2009}), \bibinfo{pages}{61--71}.
\newblock


\bibitem[Amershi et~al\mbox{.}(2019)]%
        {amershi2019guidelines}
\bibfield{author}{\bibinfo{person}{Saleema Amershi}, \bibinfo{person}{Daniel~S. Weld}, \bibinfo{person}{Mihaela Vorvoreanu}, \bibinfo{person}{Adam Fourney}, \bibinfo{person}{Besmira Nushi}, \bibinfo{person}{Penny Collisson}, \bibinfo{person}{Jina Suh}, \bibinfo{person}{Shamsi~T. Iqbal}, \bibinfo{person}{Paul~N. Bennett}, \bibinfo{person}{Kori Inkpen}, \bibinfo{person}{Jaime Teevan}, \bibinfo{person}{Ruth Kikin{-}Gil}, {and} \bibinfo{person}{Eric Horvitz}.} \bibinfo{year}{2019}\natexlab{}.
\newblock \showarticletitle{Guidelines for Human-AI Interaction}. In \bibinfo{booktitle}{\emph{Proceedings of the 2019 CHI Conference on Human Factors in Computing Systems}}. \bibinfo{pages}{1--13}.
\newblock


\bibitem[Amini et~al\mbox{.}(2016)]%
        {amini2016authoring}
\bibfield{author}{\bibinfo{person}{Fereshteh Amini}, \bibinfo{person}{Nathalie~Henry Riche}, \bibinfo{person}{Bongshin Lee}, \bibinfo{person}{Andres Monroy-Hernandez}, {and} \bibinfo{person}{Pourang Irani}.} \bibinfo{year}{2016}\natexlab{}.
\newblock \showarticletitle{Authoring Data-Driven Videos with DataClips}.
\newblock \bibinfo{journal}{\emph{IEEE Transactions on Visualization and Computer Graphics}} \bibinfo{volume}{23}, \bibinfo{number}{1} (\bibinfo{year}{2016}), \bibinfo{pages}{501--510}.
\newblock


\bibitem[Battle and Ottley(2022)]%
        {battle2022programmatic}
\bibfield{author}{\bibinfo{person}{Leilani Battle} {and} \bibinfo{person}{Alvitta Ottley}.} \bibinfo{year}{2022}\natexlab{}.
\newblock \showarticletitle{A Programmatic Definition of Visualization Insights, Objectives, and Tasks}.
\newblock \bibinfo{journal}{\emph{arXiv preprint arXiv:2206.04767}} (\bibinfo{year}{2022}).
\newblock


\bibitem[Battle and Ottley(2023)]%
        {battle2023exactly}
\bibfield{author}{\bibinfo{person}{Leilani Battle} {and} \bibinfo{person}{Alvitta Ottley}.} \bibinfo{year}{2023}\natexlab{}.
\newblock \showarticletitle{What Exactly is an Insight? A Literature Review}.
\newblock \bibinfo{journal}{\emph{arXiv preprint arXiv:2307.06551}} (\bibinfo{year}{2023}).
\newblock


\bibitem[Brehmer and Kosara(2021)]%
        {brehmer2021jam}
\bibfield{author}{\bibinfo{person}{Matthew Brehmer} {and} \bibinfo{person}{Robert Kosara}.} \bibinfo{year}{2021}\natexlab{}.
\newblock \showarticletitle{From Jam Session to Recital: Synchronous Communication and Collaboration Around Data in Organizations}.
\newblock \bibinfo{journal}{\emph{IEEE Transactions on Visualization and Computer Graphics}} \bibinfo{volume}{28}, \bibinfo{number}{1} (\bibinfo{year}{2021}), \bibinfo{pages}{1139--1149}.
\newblock


\bibitem[Brehmer et~al\mbox{.}(2019)]%
        {brehmer2019timeline}
\bibfield{author}{\bibinfo{person}{Matthew Brehmer}, \bibinfo{person}{Bongshin Lee}, \bibinfo{person}{Nathalie~Henry Riche}, \bibinfo{person}{David Tittsworth}, \bibinfo{person}{Kate Lytvynets}, \bibinfo{person}{Darren Edge}, {and} \bibinfo{person}{Christopher White}.} \bibinfo{year}{2019}\natexlab{}.
\newblock \showarticletitle{Timeline Storyteller: The Design \& Deployment of an Interactive Authoring Tool for Expressive Timeline Narratives}. In \bibinfo{booktitle}{\emph{Proceedings of the Computation+Journalism Symposium}}, Vol.~\bibinfo{volume}{2}.
\newblock


\bibitem[Bryan et~al\mbox{.}(2016)]%
        {bryan2016temporal}
\bibfield{author}{\bibinfo{person}{Chris Bryan}, \bibinfo{person}{Kwan-Liu Ma}, {and} \bibinfo{person}{Jonathan Woodring}.} \bibinfo{year}{2016}\natexlab{}.
\newblock \showarticletitle{Temporal Summary Images: An Approach to Narrative Visualization via Interactive Annotation Generation and Placement}.
\newblock \bibinfo{journal}{\emph{IEEE Transactions on Visualization and Computer Graphics}} \bibinfo{volume}{23}, \bibinfo{number}{1} (\bibinfo{year}{2016}), \bibinfo{pages}{511--520}.
\newblock


\bibitem[Cao et~al\mbox{.}(2023)]%
        {cao2023dataparticles}
\bibfield{author}{\bibinfo{person}{Yining Cao}, \bibinfo{person}{Jane~L E}, \bibinfo{person}{Zhutian Chen}, {and} \bibinfo{person}{Haijun Xia}.} \bibinfo{year}{2023}\natexlab{}.
\newblock \showarticletitle{DataParticles: Block-based and Language-oriented Authoring of Animated Unit Visualizations}. In \bibinfo{booktitle}{\emph{Proceedings of the 2023 CHI Conference on Human Factors in Computing Systems}}. \bibinfo{pages}{1--15}.
\newblock


\bibitem[Capel and Brereton(2023)]%
        {capel2023human}
\bibfield{author}{\bibinfo{person}{Tara Capel} {and} \bibinfo{person}{Margot Brereton}.} \bibinfo{year}{2023}\natexlab{}.
\newblock \showarticletitle{What is Human-Centered about Human-Centered AI? A Map of the Research Landscape}. In \bibinfo{booktitle}{\emph{Proceedings of the 2023 CHI Conference on Human Factors in Computing Systems}}. \bibinfo{pages}{1--23}.
\newblock


\bibitem[Chen et~al\mbox{.}(2023)]%
        {chen2023does}
\bibfield{author}{\bibinfo{person}{Qing Chen}, \bibinfo{person}{Shixiong Cao}, \bibinfo{person}{Jiazhe Wang}, {and} \bibinfo{person}{Nan Cao}.} \bibinfo{year}{2023}\natexlab{}.
\newblock \showarticletitle{How Does Automation Shape the Process of Narrative Visualization: A Survey of Tools}.
\newblock \bibinfo{journal}{\emph{IEEE Transactions on Visualization and Computer Graphics}} (\bibinfo{year}{2023}).
\newblock


\bibitem[Chen et~al\mbox{.}(2021a)]%
        {chen2021vizlinter}
\bibfield{author}{\bibinfo{person}{Qing Chen}, \bibinfo{person}{Fuling Sun}, \bibinfo{person}{Xinyue Xu}, \bibinfo{person}{Zui Chen}, \bibinfo{person}{Jiazhe Wang}, {and} \bibinfo{person}{Nan Cao}.} \bibinfo{year}{2021}\natexlab{a}.
\newblock \showarticletitle{VizLinter: A Linter and Fixer Framework for Data Visualization}.
\newblock \bibinfo{journal}{\emph{IEEE Transactions on Visualization and Computer Graphics}} \bibinfo{volume}{28}, \bibinfo{number}{1} (\bibinfo{year}{2021}), \bibinfo{pages}{206--216}.
\newblock


\bibitem[Chen et~al\mbox{.}(2010a)]%
        {chen2010click2annotate}
\bibfield{author}{\bibinfo{person}{Yang Chen}, \bibinfo{person}{Scott Barlowe}, {and} \bibinfo{person}{Jing Yang}.} \bibinfo{year}{2010}\natexlab{a}.
\newblock \showarticletitle{Click2Annotate: Automated Insight Externalization with Rich Semantics}. In \bibinfo{booktitle}{\emph{Proceedings of 2010 IEEE Symposium on Visual Analytics Science and Technology}}. IEEE, \bibinfo{pages}{155--162}.
\newblock


\bibitem[Chen et~al\mbox{.}(2010b)]%
        {chen2010touch2annotate}
\bibfield{author}{\bibinfo{person}{Yang Chen}, \bibinfo{person}{Jing Yang}, \bibinfo{person}{Scott Barlowe}, {and} \bibinfo{person}{Dong~H Jeong}.} \bibinfo{year}{2010}\natexlab{b}.
\newblock \showarticletitle{Touch2Annotate: Generating Better Annotations with Less Human Effort on Multi-touch Interfaces}. In \bibinfo{booktitle}{\emph{Extended Abstracts of the 2010 CHI Conference on Human Factors in Computing Systems}}. \bibinfo{pages}{3703--3708}.
\newblock


\bibitem[Chen et~al\mbox{.}(2019)]%
        {chen2019towards}
\bibfield{author}{\bibinfo{person}{Zhutian Chen}, \bibinfo{person}{Yun Wang}, \bibinfo{person}{Qianwen Wang}, \bibinfo{person}{Yong Wang}, {and} \bibinfo{person}{Huamin Qu}.} \bibinfo{year}{2019}\natexlab{}.
\newblock \showarticletitle{Towards Automated Infographic Design: Deep Learning-based Auto-Extraction of Extensible Timeline}.
\newblock \bibinfo{journal}{\emph{IEEE Transactions on Visualization and Computer Graphics}} \bibinfo{volume}{26}, \bibinfo{number}{1} (\bibinfo{year}{2019}), \bibinfo{pages}{917--926}.
\newblock


\bibitem[Chen et~al\mbox{.}(2022)]%
        {chen2022sporthesia}
\bibfield{author}{\bibinfo{person}{Zhutian Chen}, \bibinfo{person}{Qisen Yang}, \bibinfo{person}{Xiao Xie}, \bibinfo{person}{Johanna Beyer}, \bibinfo{person}{Haijun Xia}, \bibinfo{person}{Yingcai Wu}, {and} \bibinfo{person}{Hanspeter Pfister}.} \bibinfo{year}{2022}\natexlab{}.
\newblock \showarticletitle{Sporthesia: Augmenting Sports Videos Using Natural Language}.
\newblock \bibinfo{journal}{\emph{IEEE Transactions on Visualization and Computer Graphics}} \bibinfo{volume}{29}, \bibinfo{number}{1} (\bibinfo{year}{2022}), \bibinfo{pages}{918--928}.
\newblock


\bibitem[Chen et~al\mbox{.}(2021b)]%
        {chen2021augmenting}
\bibfield{author}{\bibinfo{person}{Zhutian Chen}, \bibinfo{person}{Shuainan Ye}, \bibinfo{person}{Xiangtong Chu}, \bibinfo{person}{Haijun Xia}, \bibinfo{person}{Hui Zhang}, \bibinfo{person}{Huamin Qu}, {and} \bibinfo{person}{Yingcai Wu}.} \bibinfo{year}{2021}\natexlab{b}.
\newblock \showarticletitle{Augmenting Sports Videos with VisCommentator}.
\newblock \bibinfo{journal}{\emph{IEEE Transactions on Visualization and Computer Graphics}} \bibinfo{volume}{28}, \bibinfo{number}{1} (\bibinfo{year}{2021}), \bibinfo{pages}{824--834}.
\newblock


\bibitem[Cheng et~al\mbox{.}(2022)]%
        {cheng2022investigating}
\bibfield{author}{\bibinfo{person}{Hao Cheng}, \bibinfo{person}{Junhong Wang}, \bibinfo{person}{Yun Wang}, \bibinfo{person}{Bongshin Lee}, \bibinfo{person}{Haidong Zhang}, {and} \bibinfo{person}{Dongmei Zhang}.} \bibinfo{year}{2022}\natexlab{}.
\newblock \showarticletitle{Investigating the Role and Interplay of Narrations and Animations in Data Videos}.
\newblock \bibinfo{journal}{\emph{Computer Graphics Forum}} \bibinfo{volume}{41}, \bibinfo{number}{3} (\bibinfo{year}{2022}), \bibinfo{pages}{527--539}.
\newblock


\bibitem[Chevalier et~al\mbox{.}(2016)]%
        {chevalier2016animations}
\bibfield{author}{\bibinfo{person}{Fanny Chevalier}, \bibinfo{person}{Nathalie~Henry Riche}, \bibinfo{person}{Catherine Plaisant}, \bibinfo{person}{Amira Chalbi}, {and} \bibinfo{person}{Christophe Hurter}.} \bibinfo{year}{2016}\natexlab{}.
\newblock \showarticletitle{Animations 25 Years Later: New Roles and Opportunities}. In \bibinfo{booktitle}{\emph{Proceedings of the International Working Conference on Advanced Visual Interfaces}}. \bibinfo{pages}{280--287}.
\newblock


\bibitem[Chevalier et~al\mbox{.}(2018)]%
        {chevalier2018analysis}
\bibfield{author}{\bibinfo{person}{Fanny Chevalier}, \bibinfo{person}{Melanie Tory}, \bibinfo{person}{Bongshin Lee}, \bibinfo{person}{Jarke van Wijk}, \bibinfo{person}{Giuseppe Santucci}, \bibinfo{person}{Marian D{\"o}rk}, {and} \bibinfo{person}{Jessica Hullman}.} \bibinfo{year}{2018}\natexlab{}.
\newblock \showarticletitle{From Analysis to Communication: Supporting the Lifecycle of a Story}.
\newblock In \bibinfo{booktitle}{\emph{Data-Driven Storytelling}}. \bibinfo{publisher}{AK Peters/CRC Press}, \bibinfo{pages}{151--183}.
\newblock


\bibitem[Chotisarn et~al\mbox{.}(2021)]%
        {chotisarn2021bubble}
\bibfield{author}{\bibinfo{person}{Noptanit Chotisarn}, \bibinfo{person}{Junhua Lu}, \bibinfo{person}{Libinzi Ma}, \bibinfo{person}{Jingli Xu}, \bibinfo{person}{Linhao Meng}, \bibinfo{person}{Bingru Lin}, \bibinfo{person}{Ying Xu}, \bibinfo{person}{Xiaonan Luo}, {and} \bibinfo{person}{Wei Chen}.} \bibinfo{year}{2021}\natexlab{}.
\newblock \showarticletitle{Bubble Storytelling with Automated Animation: A Brexit Hashtag Activism Case Study}.
\newblock \bibinfo{journal}{\emph{Journal of Visualization}}  \bibinfo{volume}{24} (\bibinfo{year}{2021}), \bibinfo{pages}{101--115}.
\newblock


\bibitem[Coelho and Mueller(2020)]%
        {coelho2020infomages}
\bibfield{author}{\bibinfo{person}{Darius Coelho} {and} \bibinfo{person}{Klaus Mueller}.} \bibinfo{year}{2020}\natexlab{}.
\newblock \showarticletitle{Infomages: Embedding Data into Thematic Images}.
\newblock \bibinfo{journal}{\emph{Computer Graphics Forum}} \bibinfo{volume}{39}, \bibinfo{number}{3} (\bibinfo{year}{2020}), \bibinfo{pages}{593--606}.
\newblock


\bibitem[Conlen et~al\mbox{.}(2021)]%
        {conlen2021idyll}
\bibfield{author}{\bibinfo{person}{Matthew Conlen}, \bibinfo{person}{Megan Vo}, \bibinfo{person}{Alan Tan}, {and} \bibinfo{person}{Jeffrey Heer}.} \bibinfo{year}{2021}\natexlab{}.
\newblock \showarticletitle{Idyll Studio: A Structured Editor for Authoring Interactive \& Data-Driven Articles}. In \bibinfo{booktitle}{\emph{Proceedings of the 34th Annual ACM Symposium on User Interface Software and Technology}}. \bibinfo{pages}{1--12}.
\newblock


\bibitem[Crisan and Fiore-Gartland(2021)]%
        {crisan2021fits}
\bibfield{author}{\bibinfo{person}{Anamaria Crisan} {and} \bibinfo{person}{Brittany Fiore-Gartland}.} \bibinfo{year}{2021}\natexlab{}.
\newblock \showarticletitle{Fits and Starts: Enterprise Use of {AutoML} and the Role of Humans in the Loop}. In \bibinfo{booktitle}{\emph{Proceedings of the 2021 CHI Conference on Human Factors in Computing Systems}}. \bibinfo{pages}{1--15}.
\newblock


\bibitem[Crisan et~al\mbox{.}(2020)]%
        {crisan2020passing}
\bibfield{author}{\bibinfo{person}{Anamaria Crisan}, \bibinfo{person}{Brittany Fiore-Gartland}, {and} \bibinfo{person}{Melanie Tory}.} \bibinfo{year}{2020}\natexlab{}.
\newblock \showarticletitle{Passing the Data Baton: A Retrospective Analysis on Data Science Work and Workers}.
\newblock \bibinfo{journal}{\emph{IEEE Transactions on Visualization and Computer Graphics}} \bibinfo{volume}{27}, \bibinfo{number}{2} (\bibinfo{year}{2020}), \bibinfo{pages}{1860--1870}.
\newblock


\bibitem[Cui et~al\mbox{.}(2021)]%
        {cui2021mixed}
\bibfield{author}{\bibinfo{person}{Weiwei Cui}, \bibinfo{person}{Jinpeng Wang}, \bibinfo{person}{He Huang}, \bibinfo{person}{Yun Wang}, \bibinfo{person}{Chin-Yew Lin}, \bibinfo{person}{Haidong Zhang}, {and} \bibinfo{person}{Dongmei Zhang}.} \bibinfo{year}{2021}\natexlab{}.
\newblock \showarticletitle{A Mixed-Initiative Approach to Reusing Infographic Charts}.
\newblock \bibinfo{journal}{\emph{IEEE Transactions on Visualization and Computer Graphics}} \bibinfo{volume}{28}, \bibinfo{number}{1} (\bibinfo{year}{2021}), \bibinfo{pages}{173--183}.
\newblock


\bibitem[Cui et~al\mbox{.}(2019)]%
        {cui2019text}
\bibfield{author}{\bibinfo{person}{Weiwei Cui}, \bibinfo{person}{Xiaoyu Zhang}, \bibinfo{person}{Yun Wang}, \bibinfo{person}{He Huang}, \bibinfo{person}{Bei Chen}, \bibinfo{person}{Lei Fang}, \bibinfo{person}{Haidong Zhang}, \bibinfo{person}{Jian-Guan Lou}, {and} \bibinfo{person}{Dongmei Zhang}.} \bibinfo{year}{2019}\natexlab{}.
\newblock \showarticletitle{Text-to-Viz: Automatic Generation of Infographics from Proportion-Related Natural Language Statements}.
\newblock \bibinfo{journal}{\emph{IEEE Transactions on Visualization and Computer Graphics}} \bibinfo{volume}{26}, \bibinfo{number}{1} (\bibinfo{year}{2019}), \bibinfo{pages}{906--916}.
\newblock


\bibitem[Deterding et~al\mbox{.}(2017)]%
        {deterding2017mixed}
\bibfield{author}{\bibinfo{person}{Sebastian Deterding}, \bibinfo{person}{Jonathan Hook}, \bibinfo{person}{Rebecca Fiebrink}, \bibinfo{person}{Marco Gillies}, \bibinfo{person}{Jeremy Gow}, \bibinfo{person}{Memo Akten}, \bibinfo{person}{Gillian Smith}, \bibinfo{person}{Antonios Liapis}, {and} \bibinfo{person}{Kate Compton}.} \bibinfo{year}{2017}\natexlab{}.
\newblock \showarticletitle{Mixed-Initiative Creative Interfaces}. In \bibinfo{booktitle}{\emph{Extended Abstracts of the 2017 CHI Conference on Human Factors in Computing Systems}}. \bibinfo{pages}{628--635}.
\newblock


\bibitem[Dibia(2023)]%
        {dibia2023lida}
\bibfield{author}{\bibinfo{person}{Victor Dibia}.} \bibinfo{year}{2023}\natexlab{}.
\newblock \showarticletitle{LIDA: A Tool for Automatic Generation of Grammar-Agnostic Visualizations and Infographics using Large Language Models}.
\newblock \bibinfo{journal}{\emph{arXiv preprint arXiv:2303.02927}} (\bibinfo{year}{2023}).
\newblock


\bibitem[Ding et~al\mbox{.}(2019)]%
        {ding2019quickinsights}
\bibfield{author}{\bibinfo{person}{Rui Ding}, \bibinfo{person}{Shi Han}, \bibinfo{person}{Yong Xu}, \bibinfo{person}{Haidong Zhang}, {and} \bibinfo{person}{Dongmei Zhang}.} \bibinfo{year}{2019}\natexlab{}.
\newblock \showarticletitle{QuickInsights: Quick and Automatic Discovery of Insights from Multi-Dimensional Data}. In \bibinfo{booktitle}{\emph{Proceedings of the 2019 ACM International Conference on Management of Data}}. \bibinfo{pages}{317--332}.
\newblock


\bibitem[Domova and Vrotsou(2022)]%
        {domova2022model}
\bibfield{author}{\bibinfo{person}{Veronika Domova} {and} \bibinfo{person}{Katerina Vrotsou}.} \bibinfo{year}{2022}\natexlab{}.
\newblock \showarticletitle{A Model for Types and Levels of Automation in Visual Analytics: A Survey, a Taxonomy, and Examples}.
\newblock \bibinfo{journal}{\emph{IEEE Transactions on Visualization and Computer Graphics}} (\bibinfo{year}{2022}).
\newblock


\bibitem[Elmer et~al\mbox{.}(2018)]%
        {elmer2018organizing}
\bibfield{author}{\bibinfo{person}{Christina Elmer}, \bibinfo{person}{Jonathan Schwabish}, {and} \bibinfo{person}{Benjamin Wiederkehr}.} \bibinfo{year}{2018}\natexlab{}.
\newblock \showarticletitle{Organizing the Work of Data-Driven Visual Storytelling}.
\newblock In \bibinfo{booktitle}{\emph{Data-Driven Storytelling}}. \bibinfo{publisher}{AK Peters/CRC Press}, \bibinfo{pages}{185--208}.
\newblock


\bibitem[Endsley(1987)]%
        {endsley1987application}
\bibfield{author}{\bibinfo{person}{Mica~R Endsley}.} \bibinfo{year}{1987}\natexlab{}.
\newblock \showarticletitle{The Application of Human Factors to the Development of Expert Systems for Advanced Cockpits}. In \bibinfo{booktitle}{\emph{Proceedings of the Human Factors Society Annual Meeting}}, Vol.~\bibinfo{volume}{31}. SAGE Publications Sage CA: Los Angeles, CA, \bibinfo{pages}{1388--1392}.
\newblock


\bibitem[Fan et~al\mbox{.}(2022)]%
        {fan2022annotating}
\bibfield{author}{\bibinfo{person}{Arlen Fan}, \bibinfo{person}{Yuxin Ma}, \bibinfo{person}{Michelle Mancenido}, {and} \bibinfo{person}{Ross Maciejewski}.} \bibinfo{year}{2022}\natexlab{}.
\newblock \showarticletitle{Annotating Line Charts for Addressing Deception}. In \bibinfo{booktitle}{\emph{Proceedings of the 2022 CHI Conference on Human Factors in Computing Systems}}. \bibinfo{pages}{1--12}.
\newblock


\bibitem[Feng et~al\mbox{.}(2022)]%
        {feng2022survey}
\bibfield{author}{\bibinfo{person}{Zezheng Feng}, \bibinfo{person}{Huamin Qu}, \bibinfo{person}{Shuang-Hua Yang}, \bibinfo{person}{Yulong Ding}, {and} \bibinfo{person}{Jie Song}.} \bibinfo{year}{2022}\natexlab{}.
\newblock \showarticletitle{A Survey of Visual Analytics in Urban Area}.
\newblock \bibinfo{journal}{\emph{Expert Systems}} \bibinfo{volume}{39}, \bibinfo{number}{9} (\bibinfo{year}{2022}), \bibinfo{pages}{e13065}.
\newblock


\bibitem[Fu et~al\mbox{.}(2019)]%
        {fu2019visualization}
\bibfield{author}{\bibinfo{person}{Xin Fu}, \bibinfo{person}{Yun Wang}, \bibinfo{person}{Haoyu Dong}, \bibinfo{person}{Weiwei Cui}, {and} \bibinfo{person}{Haidong Zhang}.} \bibinfo{year}{2019}\natexlab{}.
\newblock \showarticletitle{Visualization Assessment: A Machine Learning Approach}. In \bibinfo{booktitle}{\emph{Proceedings of the 2019 IEEE Visualization Conference}}. IEEE, \bibinfo{pages}{126--130}.
\newblock


\bibitem[Gao et~al\mbox{.}(2014)]%
        {gao2014newsviews}
\bibfield{author}{\bibinfo{person}{Tong Gao}, \bibinfo{person}{Jessica~R Hullman}, \bibinfo{person}{Eytan Adar}, \bibinfo{person}{Brent Hecht}, {and} \bibinfo{person}{Nicholas Diakopoulos}.} \bibinfo{year}{2014}\natexlab{}.
\newblock \showarticletitle{NewsViews: An Automated Pipeline for Creating Custom Geovisualizations for News}. In \bibinfo{booktitle}{\emph{Proceedings of the 2014 CHI Conference on Human Factors in Computing Systems}}. \bibinfo{pages}{3005--3014}.
\newblock


\bibitem[Ge et~al\mbox{.}(2021)]%
        {ge2021cast}
\bibfield{author}{\bibinfo{person}{Tong Ge}, \bibinfo{person}{Bongshin Lee}, {and} \bibinfo{person}{Yunhai Wang}.} \bibinfo{year}{2021}\natexlab{}.
\newblock \showarticletitle{CAST: Authoring Data-Driven Chart Animations}. In \bibinfo{booktitle}{\emph{Proceedings of the 2021 CHI Conference on Human Factors in Computing Systems}}. \bibinfo{pages}{1--15}.
\newblock


\bibitem[Gratzl et~al\mbox{.}(2016)]%
        {gratzl2016visual}
\bibfield{author}{\bibinfo{person}{Samuel Gratzl}, \bibinfo{person}{Alexander Lex}, \bibinfo{person}{Nils Gehlenborg}, \bibinfo{person}{Nicola Cosgrove}, {and} \bibinfo{person}{Marc Streit}.} \bibinfo{year}{2016}\natexlab{}.
\newblock \showarticletitle{From Visual Exploration to Storytelling and Back Again}.
\newblock \bibinfo{journal}{\emph{Computer Graphics Forum}} \bibinfo{volume}{35}, \bibinfo{number}{3} (\bibinfo{year}{2016}), \bibinfo{pages}{491--500}.
\newblock


\bibitem[Hall et~al\mbox{.}(2022)]%
        {hall2022augmented}
\bibfield{author}{\bibinfo{person}{Brian~D Hall}, \bibinfo{person}{Lyn Bartram}, {and} \bibinfo{person}{Matthew Brehmer}.} \bibinfo{year}{2022}\natexlab{}.
\newblock \showarticletitle{Augmented Chironomia for Presenting Data to Remote Audiences}. In \bibinfo{booktitle}{\emph{Proceedings of the 35th Annual ACM Symposium on User Interface Software and Technology}}. \bibinfo{pages}{1--14}.
\newblock


\bibitem[Heer(2019)]%
        {heer2019agency}
\bibfield{author}{\bibinfo{person}{Jeffrey Heer}.} \bibinfo{year}{2019}\natexlab{}.
\newblock \showarticletitle{Agency Plus Automation: Designing Artificial Intelligence into Interactive Systems}.
\newblock \bibinfo{journal}{\emph{Proceedings of the National Academy of Sciences}} \bibinfo{volume}{116}, \bibinfo{number}{6} (\bibinfo{year}{2019}), \bibinfo{pages}{1844--1850}.
\newblock


\bibitem[Horvitz(1999)]%
        {horvitz1999principles}
\bibfield{author}{\bibinfo{person}{Eric Horvitz}.} \bibinfo{year}{1999}\natexlab{}.
\newblock \showarticletitle{Principles of Mixed-Initiative User Interfaces}. In \bibinfo{booktitle}{\emph{Proceedings of the 1999 CHI Conference on Human Factors in Computing Systems}}. \bibinfo{pages}{159--166}.
\newblock


\bibitem[Hou and Wang(2017)]%
        {hou2017hacking}
\bibfield{author}{\bibinfo{person}{Youyang Hou} {and} \bibinfo{person}{Dakuo Wang}.} \bibinfo{year}{2017}\natexlab{}.
\newblock \showarticletitle{Hacking with NPOs: Collaborative Analytics and Broker Roles in Civic Data Hackathons}.
\newblock \bibinfo{journal}{\emph{Proceedings of the ACM on Human-Computer Interaction}} \bibinfo{volume}{1}, \bibinfo{number}{CSCW} (\bibinfo{year}{2017}), \bibinfo{pages}{1--16}.
\newblock


\bibitem[Hu et~al\mbox{.}(2019)]%
        {hu2019vizml}
\bibfield{author}{\bibinfo{person}{Kevin Hu}, \bibinfo{person}{Michiel~A Bakker}, \bibinfo{person}{Stephen Li}, \bibinfo{person}{Tim Kraska}, {and} \bibinfo{person}{C{\'e}sar Hidalgo}.} \bibinfo{year}{2019}\natexlab{}.
\newblock \showarticletitle{VizML: A Machine Learning Approach to Visualization Recommendation}. In \bibinfo{booktitle}{\emph{Proceedings of the 2019 CHI Conference on Human Factors in Computing Systems}}. \bibinfo{pages}{1--12}.
\newblock


\bibitem[Hullman et~al\mbox{.}(2013a)]%
        {hullman2013contextifier}
\bibfield{author}{\bibinfo{person}{Jessica Hullman}, \bibinfo{person}{Nicholas Diakopoulos}, {and} \bibinfo{person}{Eytan Adar}.} \bibinfo{year}{2013}\natexlab{a}.
\newblock \showarticletitle{Contextifier: Automatic Generation of Annotated Stock Visualizations}. In \bibinfo{booktitle}{\emph{Proceedings of the 2013 CHI Conference on Human Factors in Computing Systems}}. \bibinfo{pages}{2707--2716}.
\newblock


\bibitem[Hullman et~al\mbox{.}(2013b)]%
        {hullman2013deeper}
\bibfield{author}{\bibinfo{person}{Jessica Hullman}, \bibinfo{person}{Steven Drucker}, \bibinfo{person}{Nathalie~Henry Riche}, \bibinfo{person}{Bongshin Lee}, \bibinfo{person}{Danyel Fisher}, {and} \bibinfo{person}{Eytan Adar}.} \bibinfo{year}{2013}\natexlab{b}.
\newblock \showarticletitle{A Deeper Understanding of Sequence in Narrative Visualization}.
\newblock \bibinfo{journal}{\emph{IEEE Transactions on Visualization and Computer Graphics}} \bibinfo{volume}{19}, \bibinfo{number}{12} (\bibinfo{year}{2013}), \bibinfo{pages}{2406--2415}.
\newblock


\bibitem[Jiang et~al\mbox{.}(2021)]%
        {jiang2021supporting}
\bibfield{author}{\bibinfo{person}{Jialun~Aaron Jiang}, \bibinfo{person}{Kandrea Wade}, \bibinfo{person}{Casey Fiesler}, {and} \bibinfo{person}{Jed~R Brubaker}.} \bibinfo{year}{2021}\natexlab{}.
\newblock \showarticletitle{Supporting Serendipity: Opportunities and Challenges for Human-AI Collaboration in Qualitative Analysis}.
\newblock \bibinfo{journal}{\emph{Proceedings of the ACM on Human-Computer Interaction}} \bibinfo{volume}{5}, \bibinfo{number}{CSCW1} (\bibinfo{year}{2021}), \bibinfo{pages}{1--23}.
\newblock


\bibitem[Kang et~al\mbox{.}(2021)]%
        {kang2021toonnote}
\bibfield{author}{\bibinfo{person}{DaYe Kang}, \bibinfo{person}{Tony Ho}, \bibinfo{person}{Nicolai Marquardt}, \bibinfo{person}{Bilge Mutlu}, {and} \bibinfo{person}{Andrea Bianchi}.} \bibinfo{year}{2021}\natexlab{}.
\newblock \showarticletitle{ToonNote: Improving Communication in Computational Notebooks Using Interactive Data Comics}. In \bibinfo{booktitle}{\emph{Proceedings of the 2021 CHI Conference on Human Factors in Computing Systems}}. \bibinfo{pages}{1--14}.
\newblock


\bibitem[Kim et~al\mbox{.}(2019a)]%
        {kim2019datatoon}
\bibfield{author}{\bibinfo{person}{Nam~Wook Kim}, \bibinfo{person}{Nathalie Henry~Riche}, \bibinfo{person}{Benjamin Bach}, \bibinfo{person}{Guanpeng Xu}, \bibinfo{person}{Matthew Brehmer}, \bibinfo{person}{Ken Hinckley}, \bibinfo{person}{Michel Pahud}, \bibinfo{person}{Haijun Xia}, \bibinfo{person}{Michael~J McGuffin}, {and} \bibinfo{person}{Hanspeter Pfister}.} \bibinfo{year}{2019}\natexlab{a}.
\newblock \showarticletitle{DataToon: Drawing Dynamic Network Comics With Pen + Touch Interaction}. In \bibinfo{booktitle}{\emph{Proceedings of the 2019 CHI Conference on Human Factors in Computing Systems}}. \bibinfo{pages}{1--12}.
\newblock


\bibitem[Kim et~al\mbox{.}(2019b)]%
        {kim2019dataselfie}
\bibfield{author}{\bibinfo{person}{Nam~Wook Kim}, \bibinfo{person}{Hyejin Im}, \bibinfo{person}{Nathalie Henry~Riche}, \bibinfo{person}{Alicia Wang}, \bibinfo{person}{Krzysztof Gajos}, {and} \bibinfo{person}{Hanspeter Pfister}.} \bibinfo{year}{2019}\natexlab{b}.
\newblock \showarticletitle{DataSelfie: Empowering People to Design Personalized Visuals to Represent Their Data}. In \bibinfo{booktitle}{\emph{Proceedings of the 2019 CHI Conference on Human Factors in Computing Systems}}. \bibinfo{pages}{1--12}.
\newblock


\bibitem[Kim et~al\mbox{.}(2016)]%
        {kim2016data}
\bibfield{author}{\bibinfo{person}{Nam~Wook Kim}, \bibinfo{person}{Eston Schweickart}, \bibinfo{person}{Zhicheng Liu}, \bibinfo{person}{Mira Dontcheva}, \bibinfo{person}{Wilmot Li}, \bibinfo{person}{Jovan Popovic}, {and} \bibinfo{person}{Hanspeter Pfister}.} \bibinfo{year}{2016}\natexlab{}.
\newblock \showarticletitle{Data-Driven Guides: Supporting Expressive Design for Information Graphics}.
\newblock \bibinfo{journal}{\emph{IEEE Transactions on Visualization and Computer Graphics}} \bibinfo{volume}{23}, \bibinfo{number}{1} (\bibinfo{year}{2016}), \bibinfo{pages}{491--500}.
\newblock


\bibitem[Kim and Heer(2021)]%
        {kim2021gemini}
\bibfield{author}{\bibinfo{person}{Younghoon Kim} {and} \bibinfo{person}{Jeffrey Heer}.} \bibinfo{year}{2021}\natexlab{}.
\newblock \showarticletitle{Gemini$^2$: Generating Keyframe-Oriented Animated Transitions Between Statistical Graphics}. In \bibinfo{booktitle}{\emph{Proceedings of the 2021 IEEE Visualization Conference}}. IEEE, \bibinfo{pages}{201--205}.
\newblock


\bibitem[Kim et~al\mbox{.}(2017)]%
        {kim2017graphscape}
\bibfield{author}{\bibinfo{person}{Younghoon Kim}, \bibinfo{person}{Kanit Wongsuphasawat}, \bibinfo{person}{Jessica Hullman}, {and} \bibinfo{person}{Jeffrey Heer}.} \bibinfo{year}{2017}\natexlab{}.
\newblock \showarticletitle{GraphScape: A Model for Automated Reasoning about Visualization Similarity and Sequencing}. In \bibinfo{booktitle}{\emph{Proceedings of the 2017 CHI Conference on Human Factors in Computing Systems}}. \bibinfo{pages}{2628--2638}.
\newblock


\bibitem[Kouril et~al\mbox{.}(2021)]%
        {kouril2021molecumentary}
\bibfield{author}{\bibinfo{person}{David Kouril}, \bibinfo{person}{Ondrej Strnad}, \bibinfo{person}{Peter Mindek}, \bibinfo{person}{Sarkis Halladjian}, \bibinfo{person}{Tobias Isenberg}, \bibinfo{person}{Eduard Groeller}, {and} \bibinfo{person}{Ivan Viola}.} \bibinfo{year}{2021}\natexlab{}.
\newblock \showarticletitle{Molecumentary: Adaptable Narrated Documentaries Using Molecular Visualization}.
\newblock \bibinfo{journal}{\emph{IEEE Transactions on Visualization and Computer Graphics}} (\bibinfo{year}{2021}).
\newblock


\bibitem[Kui et~al\mbox{.}(2022)]%
        {kui2022survey}
\bibfield{author}{\bibinfo{person}{Xiaoyan Kui}, \bibinfo{person}{Naiming Liu}, \bibinfo{person}{Qiang Liu}, \bibinfo{person}{Jingwei Liu}, \bibinfo{person}{Xiaoqian Zeng}, {and} \bibinfo{person}{Chao Zhang}.} \bibinfo{year}{2022}\natexlab{}.
\newblock \showarticletitle{A Survey of Visual Analytics Techniques for Online Education}.
\newblock \bibinfo{journal}{\emph{Visual Informatics}} (\bibinfo{year}{2022}).
\newblock


\bibitem[Lai et~al\mbox{.}(2023)]%
        {lai2023towards}
\bibfield{author}{\bibinfo{person}{Vivian Lai}, \bibinfo{person}{Chacha Chen}, \bibinfo{person}{Alison Smith-Renner}, \bibinfo{person}{Q~Vera Liao}, {and} \bibinfo{person}{Chenhao Tan}.} \bibinfo{year}{2023}\natexlab{}.
\newblock \showarticletitle{Towards a Science of Human-AI Decision Making: An Overview of Design Space in Empirical Human-Subject Studies}. In \bibinfo{booktitle}{\emph{Proceedings of the 2023 ACM Conference on Fairness, Accountability, and Transparency}}. \bibinfo{pages}{1369--1385}.
\newblock


\bibitem[Latif et~al\mbox{.}(2021)]%
        {latif2021kori}
\bibfield{author}{\bibinfo{person}{Shahid Latif}, \bibinfo{person}{Zheng Zhou}, \bibinfo{person}{Yoon Kim}, \bibinfo{person}{Fabian Beck}, {and} \bibinfo{person}{Nam~Wook Kim}.} \bibinfo{year}{2021}\natexlab{}.
\newblock \showarticletitle{Kori: Interactive Synthesis of Text and Charts in Data Documents}.
\newblock \bibinfo{journal}{\emph{IEEE Transactions on Visualization and Computer Graphics}} \bibinfo{volume}{28}, \bibinfo{number}{1} (\bibinfo{year}{2021}), \bibinfo{pages}{184--194}.
\newblock


\bibitem[Lee et~al\mbox{.}(2013)]%
        {lee2013sketchstory}
\bibfield{author}{\bibinfo{person}{Bongshin Lee}, \bibinfo{person}{Rubaiat~Habib Kazi}, {and} \bibinfo{person}{Greg Smith}.} \bibinfo{year}{2013}\natexlab{}.
\newblock \showarticletitle{SketchStory: Telling More Engaging Stories with Data through Freeform Sketching}.
\newblock \bibinfo{journal}{\emph{IEEE Transactions on Visualization and Computer Graphics}} \bibinfo{volume}{19}, \bibinfo{number}{12} (\bibinfo{year}{2013}), \bibinfo{pages}{2416--2425}.
\newblock


\bibitem[Lee et~al\mbox{.}(2015)]%
        {lee2015more}
\bibfield{author}{\bibinfo{person}{Bongshin Lee}, \bibinfo{person}{Nathalie~Henry Riche}, \bibinfo{person}{Petra Isenberg}, {and} \bibinfo{person}{Sheelagh Carpendale}.} \bibinfo{year}{2015}\natexlab{}.
\newblock \showarticletitle{More Than Telling a Story: Transforming Data into Visually Shared Stories}.
\newblock \bibinfo{journal}{\emph{IEEE Computer Graphics and Applications}} \bibinfo{volume}{35}, \bibinfo{number}{5} (\bibinfo{year}{2015}), \bibinfo{pages}{84--90}.
\newblock


\bibitem[Lee et~al\mbox{.}(2023)]%
        {lee2023benefits}
\bibfield{author}{\bibinfo{person}{Peter Lee}, \bibinfo{person}{Sebastien Bubeck}, {and} \bibinfo{person}{Joseph Petro}.} \bibinfo{year}{2023}\natexlab{}.
\newblock \showarticletitle{Benefits, Limits, and Risks of GPT-4 as an AI Chatbot for Medicine}.
\newblock \bibinfo{journal}{\emph{New England Journal of Medicine}} \bibinfo{volume}{388}, \bibinfo{number}{13} (\bibinfo{year}{2023}), \bibinfo{pages}{1233--1239}.
\newblock


\bibitem[Li et~al\mbox{.}(2023b)]%
        {li2023ai}
\bibfield{author}{\bibinfo{person}{Haotian Li}, \bibinfo{person}{Yun Wang}, \bibinfo{person}{Q~Vera Liao}, {and} \bibinfo{person}{Huamin Qu}.} \bibinfo{year}{2023}\natexlab{b}.
\newblock \showarticletitle{Why is AI not a Panacea for Data Workers? An Interview Study on Human-AI Collaboration in Data Storytelling}.
\newblock \bibinfo{journal}{\emph{arXiv preprint arXiv:2304.08366}} (\bibinfo{year}{2023}).
\newblock


\bibitem[Li et~al\mbox{.}(2022)]%
        {li2022structure}
\bibfield{author}{\bibinfo{person}{Haotian Li}, \bibinfo{person}{Yong Wang}, \bibinfo{person}{Aoyu Wu}, \bibinfo{person}{Huan Wei}, {and} \bibinfo{person}{Huamin Qu}.} \bibinfo{year}{2022}\natexlab{}.
\newblock \showarticletitle{Structure-aware Visualization Retrieval}. In \bibinfo{booktitle}{\emph{Proceedings of the 2022 CHI Conference on Human Factors in Computing Systems}}. \bibinfo{pages}{1--14}.
\newblock


\bibitem[Li et~al\mbox{.}(2021)]%
        {li2021visual}
\bibfield{author}{\bibinfo{person}{Haotian Li}, \bibinfo{person}{Min Xu}, \bibinfo{person}{Yong Wang}, \bibinfo{person}{Huan Wei}, {and} \bibinfo{person}{Huamin Qu}.} \bibinfo{year}{2021}\natexlab{}.
\newblock \showarticletitle{A Visual Analytics Approach to Facilitate the Proctoring of Online Exams}. In \bibinfo{booktitle}{\emph{Proceedings of the 2021 CHI Conference on Human Factors in Computing Systems}}. \bibinfo{pages}{1--17}.
\newblock


\bibitem[Li et~al\mbox{.}(2023d)]%
        {li2023notable}
\bibfield{author}{\bibinfo{person}{Haotian Li}, \bibinfo{person}{Lu Ying}, \bibinfo{person}{Haidong Zhang}, \bibinfo{person}{Yingcai Wu}, \bibinfo{person}{Huamin Qu}, {and} \bibinfo{person}{Yun Wang}.} \bibinfo{year}{2023}\natexlab{d}.
\newblock \showarticletitle{Notable: On-the-fly Assistant for Data Storytelling in Computational Notebooks}. In \bibinfo{booktitle}{\emph{Proceedings of the 2023 CHI Conference on Human Factors in Computing Systems}}. \bibinfo{pages}{1--16}.
\newblock


\bibitem[Li and Bamman(2021)]%
        {li2021gender}
\bibfield{author}{\bibinfo{person}{Lucy Li} {and} \bibinfo{person}{David Bamman}.} \bibinfo{year}{2021}\natexlab{}.
\newblock \showarticletitle{Gender and Representation Bias in GPT-3 Generated Stories}. In \bibinfo{booktitle}{\emph{Proceedings of the Third Workshop on Narrative Understanding}}. \bibinfo{pages}{48--55}.
\newblock


\bibitem[Li et~al\mbox{.}(2023a)]%
        {li2023networknarratives}
\bibfield{author}{\bibinfo{person}{Wenchao Li}, \bibinfo{person}{Sarah Sch{\"o}ttler}, \bibinfo{person}{James Scott-Brown}, \bibinfo{person}{Yun Wang}, \bibinfo{person}{Siming Chen}, \bibinfo{person}{Huamin Qu}, {and} \bibinfo{person}{Benjamin Bach}.} \bibinfo{year}{2023}\natexlab{a}.
\newblock \showarticletitle{NetworkNarratives: Data Tours for Visual Network Exploration and Analysis}. In \bibinfo{booktitle}{\emph{Proceedings of the 2023 CHI Conference on Human Factors in Computing Systems}}. \bibinfo{pages}{1--15}.
\newblock


\bibitem[Li et~al\mbox{.}(2023c)]%
        {li2023geocamera}
\bibfield{author}{\bibinfo{person}{Wenchao Li}, \bibinfo{person}{Zhan Wang}, \bibinfo{person}{Yun Wang}, \bibinfo{person}{Di Weng}, \bibinfo{person}{Liwenhan Xie}, \bibinfo{person}{Siming Chen}, \bibinfo{person}{Haidong Zhang}, {and} \bibinfo{person}{Huamin Qu}.} \bibinfo{year}{2023}\natexlab{c}.
\newblock \showarticletitle{GeoCamera: Telling Stories in Geographic Visualizations with Camera Movements}. In \bibinfo{booktitle}{\emph{Proceedings of the 2023 CHI Conference on Human Factors in Computing Systems}}. \bibinfo{pages}{1--15}.
\newblock


\bibitem[Licklider(1960)]%
        {licklider1960man}
\bibfield{author}{\bibinfo{person}{Joseph~CR Licklider}.} \bibinfo{year}{1960}\natexlab{}.
\newblock \showarticletitle{Man-Computer Symbiosis}.
\newblock \bibinfo{journal}{\emph{IRE Transactions on Human Factors in Electronics}} \bibinfo{number}{1} (\bibinfo{year}{1960}), \bibinfo{pages}{4--11}.
\newblock


\bibitem[Lin et~al\mbox{.}(2023a)]%
        {lin2023dashboard}
\bibfield{author}{\bibinfo{person}{Yanna Lin}, \bibinfo{person}{Haotian Li}, \bibinfo{person}{Aoyu Wu}, \bibinfo{person}{Yong Wang}, {and} \bibinfo{person}{Huamin Qu}.} \bibinfo{year}{2023}\natexlab{a}.
\newblock \showarticletitle{DMiner: Dashboard Design Mining and Recommendation}.
\newblock \bibinfo{journal}{\emph{IEEE Transactions on Visualization and Computer Graphics}} (\bibinfo{year}{2023}).
\newblock


\bibitem[Lin et~al\mbox{.}(2023b)]%
        {lin2023inksight}
\bibfield{author}{\bibinfo{person}{Yanna Lin}, \bibinfo{person}{Haotian Li}, \bibinfo{person}{Leni Yang}, \bibinfo{person}{Aoyu Wu}, {and} \bibinfo{person}{Huamin Qu}.} \bibinfo{year}{2023}\natexlab{b}.
\newblock \showarticletitle{InkSight: Leveraging Sketch Interaction for Documenting Chart Findings in Computational Notebooks}.
\newblock \bibinfo{journal}{\emph{arXiv preprint arXiv:2307.07922}} (\bibinfo{year}{2023}).
\newblock


\bibitem[Liu et~al\mbox{.}(2017)]%
        {liu2017smartadp}
\bibfield{author}{\bibinfo{person}{Dongyu Liu}, \bibinfo{person}{Di Weng}, \bibinfo{person}{Yuhong Li}, \bibinfo{person}{Jie Bao}, \bibinfo{person}{Yu Zheng}, \bibinfo{person}{Huamin Qu}, {and} \bibinfo{person}{Yingcai Wu}.} \bibinfo{year}{2017}\natexlab{}.
\newblock \showarticletitle{SmartAdP: Visual Analytics of Large-scale Taxi Trajectories for Selecting Billboard Locations}.
\newblock \bibinfo{journal}{\emph{IEEE Transactions on Visualization and Computer Graphics}} \bibinfo{volume}{23}, \bibinfo{number}{1} (\bibinfo{year}{2017}), \bibinfo{pages}{1--10}.
\newblock


\bibitem[Liu et~al\mbox{.}(2022)]%
        {liu2022image}
\bibfield{author}{\bibinfo{person}{Shuqi Liu}, \bibinfo{person}{Mingtian Tao}, \bibinfo{person}{Yifei Huang}, \bibinfo{person}{Changbo Wang}, {and} \bibinfo{person}{Chenhui Li}.} \bibinfo{year}{2022}\natexlab{}.
\newblock \showarticletitle{Image-Driven Harmonious Color Palette Generation for Diverse Information Visualization}.
\newblock \bibinfo{journal}{\emph{IEEE Transactions on Visualization and Computer Graphics}} (\bibinfo{year}{2022}).
\newblock


\bibitem[Lowry et~al\mbox{.}(2004)]%
        {lowry2004building}
\bibfield{author}{\bibinfo{person}{Paul~Benjamin Lowry}, \bibinfo{person}{Aaron Curtis}, {and} \bibinfo{person}{Michelle~Ren{\'e} Lowry}.} \bibinfo{year}{2004}\natexlab{}.
\newblock \showarticletitle{Building a Taxonomy and Nomenclature of Collaborative Writing to Improve Interdisciplinary Research and Practice}.
\newblock \bibinfo{journal}{\emph{The Journal of Business Communication}} \bibinfo{volume}{41}, \bibinfo{number}{1} (\bibinfo{year}{2004}), \bibinfo{pages}{66--99}.
\newblock


\bibitem[Lu et~al\mbox{.}(2021)]%
        {lu2021automatic}
\bibfield{author}{\bibinfo{person}{Junhua Lu}, \bibinfo{person}{Wei Chen}, \bibinfo{person}{Hui Ye}, \bibinfo{person}{Jie Wang}, \bibinfo{person}{Honghui Mei}, \bibinfo{person}{Yuhui Gu}, \bibinfo{person}{Yingcai Wu}, \bibinfo{person}{Xiaolong~Luke Zhang}, {and} \bibinfo{person}{Kwan-Liu Ma}.} \bibinfo{year}{2021}\natexlab{}.
\newblock \showarticletitle{Automatic Generation of Unit Visualization-based Scrollytelling for Impromptu Data Facts Delivery}. In \bibinfo{booktitle}{\emph{Proceedings of the 2021 IEEE Pacific Visualization Symposium}}. IEEE, \bibinfo{pages}{21--30}.
\newblock


\bibitem[Lu et~al\mbox{.}(2020)]%
        {lu2020illustrating}
\bibfield{author}{\bibinfo{person}{Junhua Lu}, \bibinfo{person}{Jie Wang}, \bibinfo{person}{Hui Ye}, \bibinfo{person}{Yuhui Gu}, \bibinfo{person}{Zhiyu Ding}, \bibinfo{person}{Mingliang Xu}, {and} \bibinfo{person}{Wei Chen}.} \bibinfo{year}{2020}\natexlab{}.
\newblock \showarticletitle{Illustrating Changes in Time-Series Data With Data Video}.
\newblock \bibinfo{journal}{\emph{IEEE Computer Graphics and Applications}} \bibinfo{volume}{40}, \bibinfo{number}{2} (\bibinfo{year}{2020}), \bibinfo{pages}{18--31}.
\newblock


\bibitem[Lubart(2005)]%
        {lubart2005can}
\bibfield{author}{\bibinfo{person}{Todd Lubart}.} \bibinfo{year}{2005}\natexlab{}.
\newblock \showarticletitle{How can Computers be Partners in the Creative Process: Classification and Commentary on the Special Issue}.
\newblock \bibinfo{journal}{\emph{International Journal of Human-Computer Studies}} \bibinfo{volume}{63}, \bibinfo{number}{4-5} (\bibinfo{year}{2005}), \bibinfo{pages}{365--369}.
\newblock


\bibitem[Mao et~al\mbox{.}(2019)]%
        {mao2019data}
\bibfield{author}{\bibinfo{person}{Yaoli Mao}, \bibinfo{person}{Dakuo Wang}, \bibinfo{person}{Michael Muller}, \bibinfo{person}{Kush~R Varshney}, \bibinfo{person}{Ioana Baldini}, \bibinfo{person}{Casey Dugan}, {and} \bibinfo{person}{Aleksandra Mojsilovi{\'c}}.} \bibinfo{year}{2019}\natexlab{}.
\newblock \showarticletitle{How Data Scientists Work Together With Domain Experts in Scientific Collaborations: To Find The Right Answer Or To Ask The Right Question?}
\newblock \bibinfo{journal}{\emph{Proceedings of the ACM on Human-Computer Interaction}} \bibinfo{volume}{3}, \bibinfo{number}{GROUP} (\bibinfo{year}{2019}), \bibinfo{pages}{1--23}.
\newblock


\bibitem[Mathisen et~al\mbox{.}(2019)]%
        {mathisen2019insideinsights}
\bibfield{author}{\bibinfo{person}{Andreas Mathisen}, \bibinfo{person}{Tom Horak}, \bibinfo{person}{Clemens~Nylandsted Klokmose}, \bibinfo{person}{Kaj Gr{\o}nb{\ae}k}, {and} \bibinfo{person}{Niklas Elmqvist}.} \bibinfo{year}{2019}\natexlab{}.
\newblock \showarticletitle{InsideInsights: Integrating Data-Driven Reporting in Collaborative Visual Analytics}.
\newblock \bibinfo{journal}{\emph{Computer Graphics Forum}} \bibinfo{volume}{38}, \bibinfo{number}{3} (\bibinfo{year}{2019}), \bibinfo{pages}{649--661}.
\newblock


\bibitem[McNabb and Laramee(2017)]%
        {mcnabb2017survey}
\bibfield{author}{\bibinfo{person}{Liam McNabb} {and} \bibinfo{person}{Robert~S Laramee}.} \bibinfo{year}{2017}\natexlab{}.
\newblock \showarticletitle{Survey of Surveys (SoS) - Mapping The Landscape of Survey Papers in Information Visualization}.
\newblock \bibinfo{journal}{\emph{Computer Graphics Forum}} \bibinfo{volume}{36}, \bibinfo{number}{3} (\bibinfo{year}{2017}), \bibinfo{pages}{589--617}.
\newblock


\bibitem[Metoyer et~al\mbox{.}(2018)]%
        {metoyer2018coupling}
\bibfield{author}{\bibinfo{person}{Ronald Metoyer}, \bibinfo{person}{Qiyu Zhi}, \bibinfo{person}{Bart Janczuk}, {and} \bibinfo{person}{Walter Scheirer}.} \bibinfo{year}{2018}\natexlab{}.
\newblock \showarticletitle{Coupling Story to Visualization: Using Textual Analysis as a Bridge Between Data and Interpretation}. In \bibinfo{booktitle}{\emph{Proceedings of the 23rd International Conference on Intelligent User Interfaces}}. \bibinfo{pages}{503--507}.
\newblock


\bibitem[Ming et~al\mbox{.}(2018)]%
        {ming2018rulematrix}
\bibfield{author}{\bibinfo{person}{Yao Ming}, \bibinfo{person}{Huamin Qu}, {and} \bibinfo{person}{Enrico Bertini}.} \bibinfo{year}{2018}\natexlab{}.
\newblock \showarticletitle{RuleMatrix: Visualizing and Understanding Classifiers with Rules}.
\newblock \bibinfo{journal}{\emph{IEEE Transactions on Visualization and Computer Graphics}} \bibinfo{volume}{25}, \bibinfo{number}{1} (\bibinfo{year}{2018}), \bibinfo{pages}{342--352}.
\newblock


\bibitem[M{\"o}rth et~al\mbox{.}(2022)]%
        {morth2022scrollyvis}
\bibfield{author}{\bibinfo{person}{Eric M{\"o}rth}, \bibinfo{person}{Stefan Bruckner}, {and} \bibinfo{person}{Noeska~N Smit}.} \bibinfo{year}{2022}\natexlab{}.
\newblock \showarticletitle{ScrollyVis: Interactive Visual Authoring of Guided Dynamic Narratives for Scientific Scrollytelling}.
\newblock \bibinfo{journal}{\emph{IEEE Transactions on Visualization and Computer Graphics}} (\bibinfo{year}{2022}).
\newblock


\bibitem[Obie et~al\mbox{.}(2020)]%
        {obie2020authoring}
\bibfield{author}{\bibinfo{person}{Humphrey~O Obie}, \bibinfo{person}{Caslon Chua}, \bibinfo{person}{Iman Avazpour}, \bibinfo{person}{Mohamed Abdelrazek}, \bibinfo{person}{John Grundy}, {and} \bibinfo{person}{Tomasz Bednarz}.} \bibinfo{year}{2020}\natexlab{}.
\newblock \showarticletitle{Authoring Logically Sequenced Visual Data Stories with Gravity}.
\newblock \bibinfo{journal}{\emph{Journal of Computer Languages}}  \bibinfo{volume}{58} (\bibinfo{year}{2020}), \bibinfo{pages}{100961}.
\newblock


\bibitem[Obie et~al\mbox{.}(2022)]%
        {obie2022gravity++}
\bibfield{author}{\bibinfo{person}{Humphrey~O Obie}, \bibinfo{person}{Dac Thanh~Chuong Ho}, \bibinfo{person}{Iman Avazpour}, \bibinfo{person}{John Grundy}, \bibinfo{person}{Mohamed Abdelrazek}, \bibinfo{person}{Tomasz Bednarz}, {and} \bibinfo{person}{Caslon Chua}.} \bibinfo{year}{2022}\natexlab{}.
\newblock \showarticletitle{Gravity++: A Graph-based Framework for Constructing Interactive Visualization Narratives}.
\newblock \bibinfo{journal}{\emph{Journal of Computer Languages}}  \bibinfo{volume}{71} (\bibinfo{year}{2022}), \bibinfo{pages}{101125}.
\newblock


\bibitem[OpenAI(2023)]%
        {openai2023gpt4}
\bibfield{author}{\bibinfo{person}{OpenAI}.} \bibinfo{year}{2023}\natexlab{}.
\newblock \showarticletitle{{GPT}-4 Technical Report}.
\newblock \bibinfo{journal}{\emph{arXiv preprint arXiv:2303.08774}} (\bibinfo{year}{2023}).
\newblock


\bibitem[Parasuraman et~al\mbox{.}(2000)]%
        {parasuraman2000model}
\bibfield{author}{\bibinfo{person}{Raja Parasuraman}, \bibinfo{person}{Thomas~B Sheridan}, {and} \bibinfo{person}{Christopher~D Wickens}.} \bibinfo{year}{2000}\natexlab{}.
\newblock \showarticletitle{A Model for Types and Levels of Human Interaction with Automation}.
\newblock \bibinfo{journal}{\emph{IEEE Transactions on Systems, Man, and Cybernetics-Part A: Systems and Humans}} \bibinfo{volume}{30}, \bibinfo{number}{3} (\bibinfo{year}{2000}), \bibinfo{pages}{286--297}.
\newblock


\bibitem[Park et~al\mbox{.}(2023)]%
        {park2023generative}
\bibfield{author}{\bibinfo{person}{Joon~Sung Park}, \bibinfo{person}{Joseph~C O'Brien}, \bibinfo{person}{Carrie~J Cai}, \bibinfo{person}{Meredith~Ringel Morris}, \bibinfo{person}{Percy Liang}, {and} \bibinfo{person}{Michael~S Bernstein}.} \bibinfo{year}{2023}\natexlab{}.
\newblock \showarticletitle{Generative Agents: Interactive Simulacra of Human Behavior}.
\newblock \bibinfo{journal}{\emph{arXiv preprint arXiv:2304.03442}} (\bibinfo{year}{2023}).
\newblock


\bibitem[Posner and Baecker(1992)]%
        {posner1992people}
\bibfield{author}{\bibinfo{person}{Ilona~R Posner} {and} \bibinfo{person}{Ronald~M Baecker}.} \bibinfo{year}{1992}\natexlab{}.
\newblock \showarticletitle{How People Write Together (Groupware)}. In \bibinfo{booktitle}{\emph{Proceedings of the 25th Hawaii International Conference on System Sciences}}, Vol.~\bibinfo{volume}{4}. IEEE, \bibinfo{pages}{127--138}.
\newblock


\bibitem[Pu et~al\mbox{.}(2021)]%
        {pu2021datamations}
\bibfield{author}{\bibinfo{person}{Xiaoying Pu}, \bibinfo{person}{Sean Kross}, \bibinfo{person}{Jake~M Hofman}, {and} \bibinfo{person}{Daniel~G Goldstein}.} \bibinfo{year}{2021}\natexlab{}.
\newblock \showarticletitle{Datamations: Animated Explanations of Data Analysis Pipelines}. In \bibinfo{booktitle}{\emph{Proceedings of the 2021 CHI Conference on Human Factors in Computing Systems}}. \bibinfo{pages}{1--14}.
\newblock


\bibitem[Qian et~al\mbox{.}(2020)]%
        {qian2020retrieve}
\bibfield{author}{\bibinfo{person}{Chunyao Qian}, \bibinfo{person}{Shizhao Sun}, \bibinfo{person}{Weiwei Cui}, \bibinfo{person}{Jian-Guang Lou}, \bibinfo{person}{Haidong Zhang}, {and} \bibinfo{person}{Dongmei Zhang}.} \bibinfo{year}{2020}\natexlab{}.
\newblock \showarticletitle{Retrieve-Then-Adapt: Example-based Automatic Generation for Proportion-related Infographics}.
\newblock \bibinfo{journal}{\emph{IEEE Transactions on Visualization and Computer Graphics}} \bibinfo{volume}{27}, \bibinfo{number}{2} (\bibinfo{year}{2020}), \bibinfo{pages}{443--452}.
\newblock


\bibitem[Qin et~al\mbox{.}(2020)]%
        {qin2020making}
\bibfield{author}{\bibinfo{person}{Xuedi Qin}, \bibinfo{person}{Yuyu Luo}, \bibinfo{person}{Nan Tang}, {and} \bibinfo{person}{Guoliang Li}.} \bibinfo{year}{2020}\natexlab{}.
\newblock \showarticletitle{Making Data Visualization More Efficient and Effective: A Survey}.
\newblock \bibinfo{journal}{\emph{The VLDB Journal}}  \bibinfo{volume}{29} (\bibinfo{year}{2020}), \bibinfo{pages}{93--117}.
\newblock


\bibitem[Ren et~al\mbox{.}(2017)]%
        {ren2017chartaccent}
\bibfield{author}{\bibinfo{person}{Donghao Ren}, \bibinfo{person}{Matthew Brehmer}, \bibinfo{person}{Bongshin Lee}, \bibinfo{person}{Tobias H{\"o}llerer}, {and} \bibinfo{person}{Eun~Kyoung Choe}.} \bibinfo{year}{2017}\natexlab{}.
\newblock \showarticletitle{ChartAccent: Annotation for Data-Driven Storytelling}. In \bibinfo{booktitle}{\emph{Proceedings of the 2017 IEEE Pacific Visualization Symposium}}. \bibinfo{pages}{230--239}.
\newblock


\bibitem[Ren et~al\mbox{.}(2018)]%
        {ren2018charticulator}
\bibfield{author}{\bibinfo{person}{Donghao Ren}, \bibinfo{person}{Bongshin Lee}, {and} \bibinfo{person}{Matthew Brehmer}.} \bibinfo{year}{2018}\natexlab{}.
\newblock \showarticletitle{Charticulator: Interactive Construction of Bespoke Chart Layouts}.
\newblock \bibinfo{journal}{\emph{IEEE Transactions on Visualization and Computer Graphics}} \bibinfo{volume}{25}, \bibinfo{number}{1} (\bibinfo{year}{2018}), \bibinfo{pages}{789--799}.
\newblock


\bibitem[Ren et~al\mbox{.}(2023)]%
        {ren2023re}
\bibfield{author}{\bibinfo{person}{Pengkun Ren}, \bibinfo{person}{Yi Wang}, {and} \bibinfo{person}{Fan Zhao}.} \bibinfo{year}{2023}\natexlab{}.
\newblock \showarticletitle{Re-understanding of Data Storytelling Tools from a Narrative Perspective}.
\newblock \bibinfo{journal}{\emph{Visual Intelligence}} \bibinfo{volume}{1}, \bibinfo{number}{1} (\bibinfo{year}{2023}), \bibinfo{pages}{11}.
\newblock


\bibitem[Rombach et~al\mbox{.}(2022)]%
        {rombach2022high}
\bibfield{author}{\bibinfo{person}{Robin Rombach}, \bibinfo{person}{Andreas Blattmann}, \bibinfo{person}{Dominik Lorenz}, \bibinfo{person}{Patrick Esser}, {and} \bibinfo{person}{Bj{\"o}rn Ommer}.} \bibinfo{year}{2022}\natexlab{}.
\newblock \showarticletitle{High-Resolution Image Synthesis with Latent Diffusion Models}. In \bibinfo{booktitle}{\emph{Proceedings of the IEEE/CVF Conference on Computer Vision and Pattern Recognition}}. \bibinfo{pages}{10684--10695}.
\newblock


\bibitem[Satyanarayan and Heer(2014)]%
        {satyanarayan2014authoring}
\bibfield{author}{\bibinfo{person}{Arvind Satyanarayan} {and} \bibinfo{person}{Jeffrey Heer}.} \bibinfo{year}{2014}\natexlab{}.
\newblock \showarticletitle{Authoring Narrative Visualizations with Ellipsis}.
\newblock \bibinfo{journal}{\emph{Computer Graphics Forum}} \bibinfo{volume}{33}, \bibinfo{number}{3} (\bibinfo{year}{2014}), \bibinfo{pages}{361--370}.
\newblock


\bibitem[Segel and Heer(2010)]%
        {segel2010narrative}
\bibfield{author}{\bibinfo{person}{Edward Segel} {and} \bibinfo{person}{Jeffrey Heer}.} \bibinfo{year}{2010}\natexlab{}.
\newblock \showarticletitle{Narrative Visualization: Telling Stories with Data}.
\newblock \bibinfo{journal}{\emph{IEEE Transactions on Visualization and Computer Graphics}} \bibinfo{volume}{16}, \bibinfo{number}{6} (\bibinfo{year}{2010}), \bibinfo{pages}{1139--1148}.
\newblock


\bibitem[Sevastjanova et~al\mbox{.}(2021)]%
        {sevastjanova2021visinreport}
\bibfield{author}{\bibinfo{person}{Rita Sevastjanova}, \bibinfo{person}{Mennatallah El-Assady}, \bibinfo{person}{Adam Bradley}, \bibinfo{person}{Christopher Collins}, \bibinfo{person}{Miriam Butt}, {and} \bibinfo{person}{Daniel Keim}.} \bibinfo{year}{2021}\natexlab{}.
\newblock \showarticletitle{VisInReport: Complementing Visual Discourse Analytics Through Personalized Insight Reports}.
\newblock \bibinfo{journal}{\emph{IEEE Transactions on Visualization and Computer Graphics}} \bibinfo{volume}{28}, \bibinfo{number}{12} (\bibinfo{year}{2021}), \bibinfo{pages}{4757--4769}.
\newblock


\bibitem[Shen et~al\mbox{.}(2023)]%
        {shen2023data}
\bibfield{author}{\bibinfo{person}{Leixian Shen}, \bibinfo{person}{Yizhi Zhang}, \bibinfo{person}{Haidong Zhang}, {and} \bibinfo{person}{Yun Wang}.} \bibinfo{year}{2023}\natexlab{}.
\newblock \showarticletitle{Data Player: Automatic Generation of Data Videos with Narration-Animation Interplay}.
\newblock \bibinfo{journal}{\emph{arXiv preprint arXiv:2308.04703}} (\bibinfo{year}{2023}).
\newblock


\bibitem[Shi et~al\mbox{.}(2021)]%
        {shi2021autoclips}
\bibfield{author}{\bibinfo{person}{Danqing Shi}, \bibinfo{person}{Fuling Sun}, \bibinfo{person}{Xinyue Xu}, \bibinfo{person}{Xingyu Lan}, \bibinfo{person}{David Gotz}, {and} \bibinfo{person}{Nan Cao}.} \bibinfo{year}{2021}\natexlab{}.
\newblock \showarticletitle{AutoClips: An Automatic Approach to Video Generation from Data Facts}.
\newblock \bibinfo{journal}{\emph{Computer Graphics Forum}} \bibinfo{volume}{40}, \bibinfo{number}{3} (\bibinfo{year}{2021}), \bibinfo{pages}{495--505}.
\newblock


\bibitem[Shi et~al\mbox{.}(2020)]%
        {shi2020calliope}
\bibfield{author}{\bibinfo{person}{Danqing Shi}, \bibinfo{person}{Xinyue Xu}, \bibinfo{person}{Fuling Sun}, \bibinfo{person}{Yang Shi}, {and} \bibinfo{person}{Nan Cao}.} \bibinfo{year}{2020}\natexlab{}.
\newblock \showarticletitle{Calliope: Automatic Visual Data Story Generation from a Spreadsheet}.
\newblock \bibinfo{journal}{\emph{IEEE Transactions on Visualization and Computer Graphics}} \bibinfo{volume}{27}, \bibinfo{number}{2} (\bibinfo{year}{2020}), \bibinfo{pages}{453--463}.
\newblock


\bibitem[Shi et~al\mbox{.}(2023)]%
        {shi2023understanding}
\bibfield{author}{\bibinfo{person}{Yang Shi}, \bibinfo{person}{Tian Gao}, \bibinfo{person}{Xiaohan Jiao}, {and} \bibinfo{person}{Nan Cao}.} \bibinfo{year}{2023}\natexlab{}.
\newblock \showarticletitle{Understanding Design Collaboration Between Designers and Artificial Intelligence: A Systematic Literature Review}.
\newblock \bibinfo{journal}{\emph{Proceedings of the ACM on Human-Computer Interaction}} \bibinfo{volume}{7}, \bibinfo{number}{CSCW2} (\bibinfo{year}{2023}), \bibinfo{pages}{1--35}.
\newblock


\bibitem[Shi et~al\mbox{.}(2022)]%
        {shi2022supporting}
\bibfield{author}{\bibinfo{person}{Yang Shi}, \bibinfo{person}{Pei Liu}, \bibinfo{person}{Siji Chen}, \bibinfo{person}{Mengdi Sun}, {and} \bibinfo{person}{Nan Cao}.} \bibinfo{year}{2022}\natexlab{}.
\newblock \showarticletitle{Supporting Expressive and Faithful Pictorial Visualization Design with Visual Style Transfer}.
\newblock \bibinfo{journal}{\emph{IEEE Transactions on Visualization and Computer Graphics}} \bibinfo{volume}{29}, \bibinfo{number}{1} (\bibinfo{year}{2022}), \bibinfo{pages}{236--246}.
\newblock


\bibitem[Shin et~al\mbox{.}(2022)]%
        {shin2022roslingifier}
\bibfield{author}{\bibinfo{person}{Minjeong Shin}, \bibinfo{person}{Joohee Kim}, \bibinfo{person}{Yunha Han}, \bibinfo{person}{Lexing Xie}, \bibinfo{person}{Mitchell Whitelaw}, \bibinfo{person}{Bum~Chul Kwon}, \bibinfo{person}{Sungahn Ko}, {and} \bibinfo{person}{Niklas Elmqvist}.} \bibinfo{year}{2022}\natexlab{}.
\newblock \showarticletitle{Roslingifier: Semi-Automated Storytelling for Animated Scatterplots}.
\newblock \bibinfo{journal}{\emph{IEEE Transactions on Visualization and Computer Graphics}} (\bibinfo{year}{2022}).
\newblock


\bibitem[Shneiderman(2007)]%
        {shneiderman2007creativity}
\bibfield{author}{\bibinfo{person}{Ben Shneiderman}.} \bibinfo{year}{2007}\natexlab{}.
\newblock \showarticletitle{Creativity Support Tools: Accelerating Discovery and Innovation}.
\newblock \bibinfo{journal}{\emph{Commun. ACM}} \bibinfo{volume}{50}, \bibinfo{number}{12} (\bibinfo{year}{2007}), \bibinfo{pages}{20--32}.
\newblock


\bibitem[Shneiderman and Maes(1997)]%
        {shneiderman1997direct}
\bibfield{author}{\bibinfo{person}{Ben Shneiderman} {and} \bibinfo{person}{Pattie Maes}.} \bibinfo{year}{1997}\natexlab{}.
\newblock \showarticletitle{Direct Manipulation vs. Interface Agents}.
\newblock \bibinfo{journal}{\emph{Interactions}} \bibinfo{volume}{4}, \bibinfo{number}{6} (\bibinfo{year}{1997}), \bibinfo{pages}{42--61}.
\newblock


\bibitem[Shu et~al\mbox{.}(2020)]%
        {shu2020makes}
\bibfield{author}{\bibinfo{person}{Xinhuan Shu}, \bibinfo{person}{Aoyu Wu}, \bibinfo{person}{Junxiu Tang}, \bibinfo{person}{Benjamin Bach}, \bibinfo{person}{Yingcai Wu}, {and} \bibinfo{person}{Huamin Qu}.} \bibinfo{year}{2020}\natexlab{}.
\newblock \showarticletitle{What Makes a Data-GIF Understandable?}
\newblock \bibinfo{journal}{\emph{IEEE Transactions on Visualization and Computer Graphics}} \bibinfo{volume}{27}, \bibinfo{number}{2} (\bibinfo{year}{2020}), \bibinfo{pages}{1492--1502}.
\newblock


\bibitem[Srinivasan et~al\mbox{.}(2018)]%
        {srinivasan2018augmenting}
\bibfield{author}{\bibinfo{person}{Arjun Srinivasan}, \bibinfo{person}{Steven~M Drucker}, \bibinfo{person}{Alex Endert}, {and} \bibinfo{person}{John Stasko}.} \bibinfo{year}{2018}\natexlab{}.
\newblock \showarticletitle{Augmenting Visualizations with Interactive Data Facts to Facilitate Interpretation and Communication}.
\newblock \bibinfo{journal}{\emph{IEEE Transactions on Visualization and Computer Graphics}} \bibinfo{volume}{25}, \bibinfo{number}{1} (\bibinfo{year}{2018}), \bibinfo{pages}{672--681}.
\newblock


\bibitem[Stein et~al\mbox{.}(2017)]%
        {stein2017bring}
\bibfield{author}{\bibinfo{person}{Manuel Stein}, \bibinfo{person}{Halldor Janetzko}, \bibinfo{person}{Andreas Lamprecht}, \bibinfo{person}{Thorsten Breitkreutz}, \bibinfo{person}{Philipp Zimmermann}, \bibinfo{person}{Bastian Goldl{\"u}cke}, \bibinfo{person}{Tobias Schreck}, \bibinfo{person}{Gennady Andrienko}, \bibinfo{person}{Michael Grossniklaus}, {and} \bibinfo{person}{Daniel~A Keim}.} \bibinfo{year}{2017}\natexlab{}.
\newblock \showarticletitle{Bring It to the Pitch: Combining Video and Movement Data to Enhance Team Sport Analysis}.
\newblock \bibinfo{journal}{\emph{IEEE Transactions on Visualization and Computer Graphics}} \bibinfo{volume}{24}, \bibinfo{number}{1} (\bibinfo{year}{2017}), \bibinfo{pages}{13--22}.
\newblock


\bibitem[Subramonyam and Hullman(2023)]%
        {subramonyam2023we}
\bibfield{author}{\bibinfo{person}{Hariharan Subramonyam} {and} \bibinfo{person}{Jessica Hullman}.} \bibinfo{year}{2023}\natexlab{}.
\newblock \showarticletitle{Are We Closing the Loop Yet? Gaps in the Generalizability of VIS4ML Research}.
\newblock \bibinfo{journal}{\emph{arXiv preprint arXiv:2308.06290}} (\bibinfo{year}{2023}).
\newblock


\bibitem[Sultanum et~al\mbox{.}(2021)]%
        {sultanum2021leveraging}
\bibfield{author}{\bibinfo{person}{Nicole Sultanum}, \bibinfo{person}{Fanny Chevalier}, \bibinfo{person}{Zoya Bylinskii}, {and} \bibinfo{person}{Zhicheng Liu}.} \bibinfo{year}{2021}\natexlab{}.
\newblock \showarticletitle{Leveraging Text-Chart Links to Support Authoring of Data-Driven Articles with VizFlow}. In \bibinfo{booktitle}{\emph{Proceedings of the 2021 CHI Conference on Human Factors in Computing Systems}}. \bibinfo{pages}{1--17}.
\newblock


\bibitem[Sultanum and Srinivasan(2023)]%
        {sultanum2023datatales}
\bibfield{author}{\bibinfo{person}{Nicole Sultanum} {and} \bibinfo{person}{Arjun Srinivasan}.} \bibinfo{year}{2023}\natexlab{}.
\newblock \showarticletitle{DataTales: Investigating the use of Large Language Models for Authoring Data-Driven Articles}.
\newblock \bibinfo{journal}{\emph{arXiv preprint arXiv:2308.04076}} (\bibinfo{year}{2023}).
\newblock


\bibitem[Sun et~al\mbox{.}(2022)]%
        {sun2022erato}
\bibfield{author}{\bibinfo{person}{Mengdi Sun}, \bibinfo{person}{Ligan Cai}, \bibinfo{person}{Weiwei Cui}, \bibinfo{person}{Yanqiu Wu}, \bibinfo{person}{Yang Shi}, {and} \bibinfo{person}{Nan Cao}.} \bibinfo{year}{2022}\natexlab{}.
\newblock \showarticletitle{Erato: Cooperative Data Story Editing via Fact Interpolation}.
\newblock \bibinfo{journal}{\emph{IEEE Transactions on Visualization and Computer Graphics}} \bibinfo{volume}{29}, \bibinfo{number}{1} (\bibinfo{year}{2022}), \bibinfo{pages}{983--993}.
\newblock


\bibitem[Tang et~al\mbox{.}(2017)]%
        {tang2017extracting}
\bibfield{author}{\bibinfo{person}{Bo Tang}, \bibinfo{person}{Shi Han}, \bibinfo{person}{Man~Lung Yiu}, \bibinfo{person}{Rui Ding}, {and} \bibinfo{person}{Dongmei Zhang}.} \bibinfo{year}{2017}\natexlab{}.
\newblock \showarticletitle{Extracting Top-K Insights from Multi-dimensional Data}. In \bibinfo{booktitle}{\emph{Proceedings of the 2017 ACM International Conference on Management of Data}}. \bibinfo{pages}{1509--1524}.
\newblock


\bibitem[Tang et~al\mbox{.}(2022)]%
        {tang2022smartshots}
\bibfield{author}{\bibinfo{person}{Tan Tang}, \bibinfo{person}{Junxiu Tang}, \bibinfo{person}{Jiewen Lai}, \bibinfo{person}{Lu Ying}, \bibinfo{person}{Yingcai Wu}, \bibinfo{person}{Lingyun Yu}, {and} \bibinfo{person}{Peiran Ren}.} \bibinfo{year}{2022}\natexlab{}.
\newblock \showarticletitle{SmartShots: An Optimization Approach for Generating Videos with Data Visualizations Embedded}.
\newblock \bibinfo{journal}{\emph{ACM Transactions on Interactive Intelligent Systems}} \bibinfo{volume}{12}, \bibinfo{number}{1} (\bibinfo{year}{2022}), \bibinfo{pages}{1--21}.
\newblock


\bibitem[Thompson et~al\mbox{.}(2021)]%
        {thompson2021data}
\bibfield{author}{\bibinfo{person}{John~R Thompson}, \bibinfo{person}{Zhicheng Liu}, {and} \bibinfo{person}{John Stasko}.} \bibinfo{year}{2021}\natexlab{}.
\newblock \showarticletitle{Data Animator: Authoring Expressive Animated Data Graphics}. In \bibinfo{booktitle}{\emph{Proceedings of the 2021 CHI Conference on Human Factors in Computing Systems}}. \bibinfo{pages}{1--18}.
\newblock


\bibitem[Thudt et~al\mbox{.}(2018)]%
        {thudt2018exploration}
\bibfield{author}{\bibinfo{person}{Alice Thudt}, \bibinfo{person}{Jagoda Walny}, \bibinfo{person}{Theresia Gschwandtner}, \bibinfo{person}{Jason Dykes}, {and} \bibinfo{person}{John Stasko}.} \bibinfo{year}{2018}\natexlab{}.
\newblock \showarticletitle{Exploration and Explanation in Data-Driven Storytelling}.
\newblock In \bibinfo{booktitle}{\emph{Data-Driven Storytelling}}. \bibinfo{publisher}{AK Peters/CRC Press}, \bibinfo{pages}{59--83}.
\newblock


\bibitem[Tong et~al\mbox{.}(2018)]%
        {tong2018storytelling}
\bibfield{author}{\bibinfo{person}{Chao Tong}, \bibinfo{person}{Richard~C. Roberts}, \bibinfo{person}{Rita Borgo}, \bibinfo{person}{Sean~P. Walton}, \bibinfo{person}{Robert~S. Laramee}, \bibinfo{person}{Kodzo Wegba}, \bibinfo{person}{Aidong Lu}, \bibinfo{person}{Yun Wang}, \bibinfo{person}{Huamin Qu}, \bibinfo{person}{Qiong Luo}, {and} \bibinfo{person}{Xiaojuan Ma}.} \bibinfo{year}{2018}\natexlab{}.
\newblock \showarticletitle{Storytelling and Visualization: An Extended Survey}.
\newblock \bibinfo{journal}{\emph{Information}} \bibinfo{volume}{9}, \bibinfo{number}{3} (\bibinfo{year}{2018}), \bibinfo{pages}{65}.
\newblock


\bibitem[Touvron et~al\mbox{.}(2023)]%
        {touvron2023llama}
\bibfield{author}{\bibinfo{person}{Hugo Touvron}, \bibinfo{person}{Louis Martin}, \bibinfo{person}{Kevin Stone}, \bibinfo{person}{Peter Albert}, \bibinfo{person}{Amjad Almahairi}, \bibinfo{person}{Yasmine Babaei}, \bibinfo{person}{Nikolay Bashlykov}, \bibinfo{person}{Soumya Batra}, \bibinfo{person}{Prajjwal Bhargava}, \bibinfo{person}{Shruti Bhosale}, {et~al\mbox{.}}} \bibinfo{year}{2023}\natexlab{}.
\newblock \showarticletitle{Llama 2: Open Foundation and Fine-Tuned Chat Models}.
\newblock \bibinfo{journal}{\emph{arXiv preprint arXiv:2307.09288}} (\bibinfo{year}{2023}).
\newblock


\bibitem[Tyagi et~al\mbox{.}(2022)]%
        {tyagi2022infographics}
\bibfield{author}{\bibinfo{person}{Anjul Tyagi}, \bibinfo{person}{Jian Zhao}, \bibinfo{person}{Pushkar Patel}, \bibinfo{person}{Swasti Khurana}, {and} \bibinfo{person}{Klaus Mueller}.} \bibinfo{year}{2022}\natexlab{}.
\newblock \showarticletitle{Infographics Wizard: Flexible Infographics Authoring and Design Exploration}.
\newblock \bibinfo{journal}{\emph{Computer Graphics Forum}} \bibinfo{volume}{41}, \bibinfo{number}{3} (\bibinfo{year}{2022}), \bibinfo{pages}{121--132}.
\newblock


\bibitem[Vagia et~al\mbox{.}(2016)]%
        {vagia2016literature}
\bibfield{author}{\bibinfo{person}{Marialena Vagia}, \bibinfo{person}{Aksel~A Transeth}, {and} \bibinfo{person}{Sigurd~A Fjerdingen}.} \bibinfo{year}{2016}\natexlab{}.
\newblock \showarticletitle{A Literature Review on the Levels of Automation during the Years. What are the Different Taxonomies that have been Proposed?}
\newblock \bibinfo{journal}{\emph{Applied Ergonomics}}  \bibinfo{volume}{53} (\bibinfo{year}{2016}), \bibinfo{pages}{190--202}.
\newblock


\bibitem[Wang et~al\mbox{.}(2020a)]%
        {wang2020human}
\bibfield{author}{\bibinfo{person}{Dakuo Wang}, \bibinfo{person}{Elizabeth Churchill}, \bibinfo{person}{Pattie Maes}, \bibinfo{person}{Xiangmin Fan}, \bibinfo{person}{Ben Shneiderman}, \bibinfo{person}{Yuanchun Shi}, {and} \bibinfo{person}{Qianying Wang}.} \bibinfo{year}{2020}\natexlab{a}.
\newblock \showarticletitle{From Human-Human Collaboration to Human-AI Collaboration: Designing AI Systems That Can Work Together with People}. In \bibinfo{booktitle}{\emph{Extended Abstracts of the 2020 CHI Conference on Human Factors in Computing Systems}}. \bibinfo{pages}{1--6}.
\newblock


\bibitem[Wang et~al\mbox{.}(2019b)]%
        {wang2019human}
\bibfield{author}{\bibinfo{person}{Dakuo Wang}, \bibinfo{person}{Justin~D Weisz}, \bibinfo{person}{Michael Muller}, \bibinfo{person}{Parikshit Ram}, \bibinfo{person}{Werner Geyer}, \bibinfo{person}{Casey Dugan}, \bibinfo{person}{Yla Tausczik}, \bibinfo{person}{Horst Samulowitz}, {and} \bibinfo{person}{Alexander Gray}.} \bibinfo{year}{2019}\natexlab{b}.
\newblock \showarticletitle{Human-AI Collaboration in Data Science: Exploring Data Scientists' Perceptions of Automated AI}.
\newblock \bibinfo{journal}{\emph{Proceedings of the ACM on Human-Computer Interaction}} \bibinfo{volume}{3}, \bibinfo{number}{CSCW} (\bibinfo{year}{2019}), \bibinfo{pages}{1--24}.
\newblock


\bibitem[Wang et~al\mbox{.}(2023b)]%
        {wang2023slide4n}
\bibfield{author}{\bibinfo{person}{Fengjie Wang}, \bibinfo{person}{Xuye Liu}, \bibinfo{person}{Oujing Liu}, \bibinfo{person}{Ali Neshati}, \bibinfo{person}{Tengfei Ma}, \bibinfo{person}{Min Zhu}, {and} \bibinfo{person}{Jian Zhao}.} \bibinfo{year}{2023}\natexlab{b}.
\newblock \showarticletitle{Slide4N: Creating Presentation Slides from Computational Notebooks with Human-AI Collaboration}. In \bibinfo{booktitle}{\emph{Proceedings of the 2023 CHI Conference on Human Factors in Computing Systems}}. \bibinfo{pages}{1--18}.
\newblock


\bibitem[Wang et~al\mbox{.}(2023a)]%
        {wang2023visual}
\bibfield{author}{\bibinfo{person}{Junpeng Wang}, \bibinfo{person}{Shixia Liu}, {and} \bibinfo{person}{Wei Zhang}.} \bibinfo{year}{2023}\natexlab{a}.
\newblock \showarticletitle{Visual Analytics For Machine Learning: A Data Perspective Survey}.
\newblock \bibinfo{journal}{\emph{arXiv preprint arXiv:2307.07712}} (\bibinfo{year}{2023}).
\newblock


\bibitem[Wang et~al\mbox{.}(2021a)]%
        {wang2021survey}
\bibfield{author}{\bibinfo{person}{Qianwen Wang}, \bibinfo{person}{Zhutian Chen}, \bibinfo{person}{Yong Wang}, {and} \bibinfo{person}{Huamin Qu}.} \bibinfo{year}{2021}\natexlab{a}.
\newblock \showarticletitle{A Survey on ML4VIS: Applying Machine Learning Advances to Data Visualization}.
\newblock \bibinfo{journal}{\emph{IEEE Transactions on Visualization and Computer Graphics}} \bibinfo{volume}{28}, \bibinfo{number}{12} (\bibinfo{year}{2021}), \bibinfo{pages}{5134--5153}.
\newblock


\bibitem[Wang et~al\mbox{.}(2023d)]%
        {wang2023vis}
\bibfield{author}{\bibinfo{person}{Xumeng Wang}, \bibinfo{person}{Ziliang Wu}, \bibinfo{person}{Wenqi Huang}, \bibinfo{person}{Yating Wei}, \bibinfo{person}{Zhaosong Huang}, \bibinfo{person}{Mingliang Xu}, {and} \bibinfo{person}{Wei Chen}.} \bibinfo{year}{2023}\natexlab{d}.
\newblock \showarticletitle{VIS+AI: Integrating Visualization with Artificial Intelligence for Efficient Data Analysis}.
\newblock \bibinfo{journal}{\emph{Frontiers of Computer Science}} \bibinfo{volume}{17}, \bibinfo{number}{6} (\bibinfo{year}{2023}), \bibinfo{pages}{1--12}.
\newblock


\bibitem[Wang et~al\mbox{.}(2020b)]%
        {wang2020voicecoach}
\bibfield{author}{\bibinfo{person}{Xingbo Wang}, \bibinfo{person}{Haipeng Zeng}, \bibinfo{person}{Yong Wang}, \bibinfo{person}{Aoyu Wu}, \bibinfo{person}{Zhida Sun}, \bibinfo{person}{Xiaojuan Ma}, {and} \bibinfo{person}{Huamin Qu}.} \bibinfo{year}{2020}\natexlab{b}.
\newblock \showarticletitle{VoiceCoach: Interactive Evidence-based Training for Voice Modulation Skills in Public Speaking}. In \bibinfo{booktitle}{\emph{Proceedings of the 2020 CHI Conference on Human Factors in Computing Systems}}. \bibinfo{pages}{1--12}.
\newblock


\bibitem[Wang et~al\mbox{.}(2021b)]%
        {wang2021animated}
\bibfield{author}{\bibinfo{person}{Yun Wang}, \bibinfo{person}{Yi Gao}, \bibinfo{person}{Ray Huang}, \bibinfo{person}{Weiwei Cui}, \bibinfo{person}{Haidong Zhang}, {and} \bibinfo{person}{Dongmei Zhang}.} \bibinfo{year}{2021}\natexlab{b}.
\newblock \showarticletitle{Animated Presentation of Static Infographics with InfoMotion}.
\newblock \bibinfo{journal}{\emph{Computer Graphics Forum}} \bibinfo{volume}{40}, \bibinfo{number}{3} (\bibinfo{year}{2021}), \bibinfo{pages}{507--518}.
\newblock


\bibitem[Wang et~al\mbox{.}(2023c)]%
        {wang2023wonderflow}
\bibfield{author}{\bibinfo{person}{Yun Wang}, \bibinfo{person}{Leixian Shen}, \bibinfo{person}{Zhengxin You}, \bibinfo{person}{Xinhuan Shu}, \bibinfo{person}{Bongshin Lee}, \bibinfo{person}{John Thompson}, \bibinfo{person}{Haidong Zhang}, {and} \bibinfo{person}{Dongmei Zhang}.} \bibinfo{year}{2023}\natexlab{c}.
\newblock \showarticletitle{WonderFlow: Narration-Centric Design of Animated Data Videos}.
\newblock \bibinfo{journal}{\emph{arXiv preprint arXiv:2308.04040}} (\bibinfo{year}{2023}).
\newblock


\bibitem[Wang et~al\mbox{.}(2019a)]%
        {wang2019datashot}
\bibfield{author}{\bibinfo{person}{Yun Wang}, \bibinfo{person}{Zhida Sun}, \bibinfo{person}{Haidong Zhang}, \bibinfo{person}{Weiwei Cui}, \bibinfo{person}{Ke Xu}, \bibinfo{person}{Xiaojuan Ma}, {and} \bibinfo{person}{Dongmei Zhang}.} \bibinfo{year}{2019}\natexlab{a}.
\newblock \showarticletitle{DataShot: Automatic Generation of Fact Sheets from Tabular Data}.
\newblock \bibinfo{journal}{\emph{IEEE Transactions on Visualization and Computer Graphics}} \bibinfo{volume}{26}, \bibinfo{number}{1} (\bibinfo{year}{2019}), \bibinfo{pages}{895--905}.
\newblock


\bibitem[Wang et~al\mbox{.}(2018)]%
        {wang2018infonice}
\bibfield{author}{\bibinfo{person}{Yun Wang}, \bibinfo{person}{Haidong Zhang}, \bibinfo{person}{He Huang}, \bibinfo{person}{Xi Chen}, \bibinfo{person}{Qiufeng Yin}, \bibinfo{person}{Zhitao Hou}, \bibinfo{person}{Dongmei Zhang}, \bibinfo{person}{Qiong Luo}, {and} \bibinfo{person}{Huamin Qu}.} \bibinfo{year}{2018}\natexlab{}.
\newblock \showarticletitle{InfoNice: Easy Creation of Information Graphics}. In \bibinfo{booktitle}{\emph{Proceedings of the 2018 CHI Conference on Human Factors in Computing Systems}}. \bibinfo{pages}{1--12}.
\newblock


\bibitem[Wang et~al\mbox{.}(2022)]%
        {wang2022interpretability}
\bibfield{author}{\bibinfo{person}{Zijie~J Wang}, \bibinfo{person}{Alex Kale}, \bibinfo{person}{Harsha Nori}, \bibinfo{person}{Peter Stella}, \bibinfo{person}{Mark~E Nunnally}, \bibinfo{person}{Duen~Horng Chau}, \bibinfo{person}{Mihaela Vorvoreanu}, \bibinfo{person}{Jennifer Wortman~Vaughan}, {and} \bibinfo{person}{Rich Caruana}.} \bibinfo{year}{2022}\natexlab{}.
\newblock \showarticletitle{Interpretability, Then What? Editing Machine Learning Models to Reflect Human Knowledge and Values}. In \bibinfo{booktitle}{\emph{Proceedings of the 28th ACM SIGKDD Conference on Knowledge Discovery and Data Mining}}. \bibinfo{pages}{4132--4142}.
\newblock


\bibitem[Wei et~al\mbox{.}(2021)]%
        {wei2021finetuned}
\bibfield{author}{\bibinfo{person}{Jason Wei}, \bibinfo{person}{Maarten Bosma}, \bibinfo{person}{Vincent Zhao}, \bibinfo{person}{Kelvin Guu}, \bibinfo{person}{Adams~Wei Yu}, \bibinfo{person}{Brian Lester}, \bibinfo{person}{Nan Du}, \bibinfo{person}{Andrew~M Dai}, {and} \bibinfo{person}{Quoc~V Le}.} \bibinfo{year}{2021}\natexlab{}.
\newblock \showarticletitle{Finetuned Language Models are Zero-Shot Learners}. In \bibinfo{booktitle}{\emph{Proceedings of the International Conference on Learning Representations}}.
\newblock


\bibitem[Wexler et~al\mbox{.}(2019)]%
        {wexler2019if}
\bibfield{author}{\bibinfo{person}{James Wexler}, \bibinfo{person}{Mahima Pushkarna}, \bibinfo{person}{Tolga Bolukbasi}, \bibinfo{person}{Martin Wattenberg}, \bibinfo{person}{Fernanda Vi{\'e}gas}, {and} \bibinfo{person}{Jimbo Wilson}.} \bibinfo{year}{2019}\natexlab{}.
\newblock \showarticletitle{The What-If Tool: Interactive Probing of Machine Learning Models}.
\newblock \bibinfo{journal}{\emph{IEEE Transactions on Visualization and Computer Graphics}} \bibinfo{volume}{26}, \bibinfo{number}{1} (\bibinfo{year}{2019}), \bibinfo{pages}{56--65}.
\newblock


\bibitem[Wolf and Gibson(2005)]%
        {wolf2005representing}
\bibfield{author}{\bibinfo{person}{Florian Wolf} {and} \bibinfo{person}{Edward Gibson}.} \bibinfo{year}{2005}\natexlab{}.
\newblock \showarticletitle{Representing Discourse Coherence: A Corpus-Based Study}.
\newblock \bibinfo{journal}{\emph{Computational Linguistics}} \bibinfo{volume}{31}, \bibinfo{number}{2} (\bibinfo{year}{2005}), \bibinfo{pages}{249--287}.
\newblock


\bibitem[Wongsuphasawat et~al\mbox{.}(2017)]%
        {wongsuphasawat2017voyager}
\bibfield{author}{\bibinfo{person}{Kanit Wongsuphasawat}, \bibinfo{person}{Zening Qu}, \bibinfo{person}{Dominik Moritz}, \bibinfo{person}{Riley Chang}, \bibinfo{person}{Felix Ouk}, \bibinfo{person}{Anushka Anand}, \bibinfo{person}{Jock Mackinlay}, \bibinfo{person}{Bill Howe}, {and} \bibinfo{person}{Jeffrey Heer}.} \bibinfo{year}{2017}\natexlab{}.
\newblock \showarticletitle{Voyager 2: Augmenting Visual Analysis with Partial View Specifications}. In \bibinfo{booktitle}{\emph{Proceedings of the 2017 CHI Conference on Human Factors in Computing Systems}}. \bibinfo{pages}{2648--2659}.
\newblock


\bibitem[Wu et~al\mbox{.}(2021)]%
        {wu2021ai4vis}
\bibfield{author}{\bibinfo{person}{Aoyu Wu}, \bibinfo{person}{Yun Wang}, \bibinfo{person}{Xinhuan Shu}, \bibinfo{person}{Dominik Moritz}, \bibinfo{person}{Weiwei Cui}, \bibinfo{person}{Haidong Zhang}, \bibinfo{person}{Dongmei Zhang}, {and} \bibinfo{person}{Huamin Qu}.} \bibinfo{year}{2021}\natexlab{}.
\newblock \showarticletitle{AI4VIS: Survey on Artificial Intelligence Approaches for Data Visualization}.
\newblock \bibinfo{journal}{\emph{IEEE Transactions on Visualization and Computer Graphics}} (\bibinfo{year}{2021}).
\newblock


\bibitem[Xia et~al\mbox{.}(2018)]%
        {xia2018dataink}
\bibfield{author}{\bibinfo{person}{Haijun Xia}, \bibinfo{person}{Nathalie Henry~Riche}, \bibinfo{person}{Fanny Chevalier}, \bibinfo{person}{Bruno De~Araujo}, {and} \bibinfo{person}{Daniel Wigdor}.} \bibinfo{year}{2018}\natexlab{}.
\newblock \showarticletitle{DataInk: Direct and Creative Data-Oriented Drawing}. In \bibinfo{booktitle}{\emph{Proceedings of the 2018 CHI Conference on Human Factors in Computing Systems}}. \bibinfo{pages}{1--13}.
\newblock


\bibitem[Yang et~al\mbox{.}(2021a)]%
        {yang2021explaining}
\bibfield{author}{\bibinfo{person}{Leni Yang}, \bibinfo{person}{Cindy Xiong}, \bibinfo{person}{Jason~K Wong}, \bibinfo{person}{Aoyu Wu}, {and} \bibinfo{person}{Huamin Qu}.} \bibinfo{year}{2021}\natexlab{a}.
\newblock \showarticletitle{Explaining with Examples: Lessons Learned from Crowdsourced Introductory Description of Information Visualizations}.
\newblock \bibinfo{journal}{\emph{IEEE Transactions on Visualization and Computer Graphics}} (\bibinfo{year}{2021}).
\newblock


\bibitem[Yang et~al\mbox{.}(2021b)]%
        {yang2021design}
\bibfield{author}{\bibinfo{person}{Leni Yang}, \bibinfo{person}{Xian Xu}, \bibinfo{person}{XingYu Lan}, \bibinfo{person}{Ziyan Liu}, \bibinfo{person}{Shunan Guo}, \bibinfo{person}{Yang Shi}, \bibinfo{person}{Huamin Qu}, {and} \bibinfo{person}{Nan Cao}.} \bibinfo{year}{2021}\natexlab{b}.
\newblock \showarticletitle{A Design Space for Applying the Freytag's Pyramid Structure to Data Stories}.
\newblock \bibinfo{journal}{\emph{IEEE Transactions on Visualization and Computer Graphics}} \bibinfo{volume}{28}, \bibinfo{number}{1} (\bibinfo{year}{2021}), \bibinfo{pages}{922--932}.
\newblock


\bibitem[Ye et~al\mbox{.}(2022)]%
        {ye2022visatlas}
\bibfield{author}{\bibinfo{person}{Yilin Ye}, \bibinfo{person}{Rong Huang}, {and} \bibinfo{person}{Wei Zeng}.} \bibinfo{year}{2022}\natexlab{}.
\newblock \showarticletitle{VISAtlas: An Image-based Exploration and Query System for Large Visualization Collections via Neural Image Embedding}.
\newblock \bibinfo{journal}{\emph{IEEE Transactions on Visualization and Computer Graphics}} (\bibinfo{year}{2022}).
\newblock


\bibitem[Ying et~al\mbox{.}(2023)]%
        {ying2023reviving}
\bibfield{author}{\bibinfo{person}{Lu Ying}, \bibinfo{person}{Yun Wang}, \bibinfo{person}{Haotian Li}, \bibinfo{person}{Shuguang Dou}, \bibinfo{person}{Haidong Zhang}, \bibinfo{person}{Xinyang Jiang}, \bibinfo{person}{Huamin Qu}, {and} \bibinfo{person}{Yingcai Wu}.} \bibinfo{year}{2023}\natexlab{}.
\newblock \showarticletitle{Reviving Static Charts into Live Charts}.
\newblock \bibinfo{journal}{\emph{arXiv preprint arXiv:2309.02967}} (\bibinfo{year}{2023}).
\newblock


\bibitem[Yuan et~al\mbox{.}(2021)]%
        {yuan2021infocolorizer}
\bibfield{author}{\bibinfo{person}{Lin-Ping Yuan}, \bibinfo{person}{Ziqi Zhou}, \bibinfo{person}{Jian Zhao}, \bibinfo{person}{Yiqiu Guo}, \bibinfo{person}{Fan Du}, {and} \bibinfo{person}{Huamin Qu}.} \bibinfo{year}{2021}\natexlab{}.
\newblock \showarticletitle{InfoColorizer: Interactive Recommendation of Color Palettes for Infographics}.
\newblock \bibinfo{journal}{\emph{IEEE Transactions on Visualization and Computer Graphics}} \bibinfo{volume}{28}, \bibinfo{number}{12} (\bibinfo{year}{2021}), \bibinfo{pages}{4252--4266}.
\newblock


\bibitem[Zhang et~al\mbox{.}(2020)]%
        {zhang2020dataquilt}
\bibfield{author}{\bibinfo{person}{Jiayi~Eris Zhang}, \bibinfo{person}{Nicole Sultanum}, \bibinfo{person}{Anastasia Bezerianos}, {and} \bibinfo{person}{Fanny Chevalier}.} \bibinfo{year}{2020}\natexlab{}.
\newblock \showarticletitle{DataQuilt: Extracting Visual Elements from Images to Craft Pictorial Visualizations}. In \bibinfo{booktitle}{\emph{Proceedings of the 2020 CHI Conference on Human Factors in Computing Systems}}. \bibinfo{pages}{1--13}.
\newblock


\bibitem[Zhao et~al\mbox{.}(2021)]%
        {zhao2021chartstory}
\bibfield{author}{\bibinfo{person}{Jian Zhao}, \bibinfo{person}{Shenyu Xu}, \bibinfo{person}{Senthil Chandrasegaran}, \bibinfo{person}{Chris Bryan}, \bibinfo{person}{Fan Du}, \bibinfo{person}{Aditi Mishra}, \bibinfo{person}{Xin Qian}, \bibinfo{person}{Yiran Li}, {and} \bibinfo{person}{Kwan-Liu Ma}.} \bibinfo{year}{2021}\natexlab{}.
\newblock \showarticletitle{ChartStory: Automated Partitioning, Layout, and Captioning of Charts into Comic-Style Narratives}.
\newblock \bibinfo{journal}{\emph{IEEE Transactions on Visualization and Computer Graphics}} \bibinfo{volume}{29}, \bibinfo{number}{2} (\bibinfo{year}{2021}), \bibinfo{pages}{1384--1399}.
\newblock


\bibitem[Zhao and Elmqvist(2023)]%
        {zhao2023stories}
\bibfield{author}{\bibinfo{person}{Zhenpeng Zhao} {and} \bibinfo{person}{Niklas Elmqvist}.} \bibinfo{year}{2023}\natexlab{}.
\newblock \showarticletitle{The Stories We Tell About Data: Surveying Data-Driven Storytelling Using Visualization}.
\newblock \bibinfo{journal}{\emph{IEEE Computer Graphics and Applications}} (\bibinfo{year}{2023}).
\newblock


\bibitem[Zheng et~al\mbox{.}(2022)]%
        {zheng2022telling}
\bibfield{author}{\bibinfo{person}{Chengbo Zheng}, \bibinfo{person}{Dakuo Wang}, \bibinfo{person}{April~Yi Wang}, {and} \bibinfo{person}{Xiaojuan Ma}.} \bibinfo{year}{2022}\natexlab{}.
\newblock \showarticletitle{Telling Stories from Computational Notebooks: AI-Assisted Presentation Slides Creation for Presenting Data Science Work}. In \bibinfo{booktitle}{\emph{Proceedings of the 2022 CHI Conference on Human Factors in Computing Systems}}. \bibinfo{pages}{1--20}.
\newblock


\bibitem[Zhu et~al\mbox{.}(2020)]%
        {zhu2020survey}
\bibfield{author}{\bibinfo{person}{Sujia Zhu}, \bibinfo{person}{Guodao Sun}, \bibinfo{person}{Qi Jiang}, \bibinfo{person}{Meng Zha}, {and} \bibinfo{person}{Ronghua Liang}.} \bibinfo{year}{2020}\natexlab{}.
\newblock \showarticletitle{A Survey on Automatic Infographics and Visualization Recommendations}.
\newblock \bibinfo{journal}{\emph{Visual Informatics}} \bibinfo{volume}{4}, \bibinfo{number}{3} (\bibinfo{year}{2020}), \bibinfo{pages}{24--40}.
\newblock


\bibitem[Zong et~al\mbox{.}(2022)]%
        {zong2022animated}
\bibfield{author}{\bibinfo{person}{Jonathan Zong}, \bibinfo{person}{Josh Pollock}, \bibinfo{person}{Dylan Wootton}, {and} \bibinfo{person}{Arvind Satyanarayan}.} \bibinfo{year}{2022}\natexlab{}.
\newblock \showarticletitle{Animated Vega-Lite: Unifying Animation with a Grammar of Interactive Graphics}.
\newblock \bibinfo{journal}{\emph{IEEE Transactions on Visualization and Computer Graphics}} \bibinfo{volume}{29}, \bibinfo{number}{1} (\bibinfo{year}{2022}), \bibinfo{pages}{149--159}.
\newblock


\end{thebibliography}

\end{document}